\documentclass[iop, numberedappendix]{emulateapj}
\usepackage{natbib}
\usepackage{multirow}

\shorttitle{UltraVISTA SMF at $4<z<7$}
\shortauthors{Stefanon et al.}

\begin{document}

\title{Stellar mass functions of galaxies at $4<\MakeLowercase{z}<7$ from an IRAC-selected Sample in COSMOS/UltraVISTA: Limits on the abundance of very massive galaxies}

\author{Mauro Stefanon\altaffilmark{1}, Danilo Marchesini\altaffilmark{2}, Adam Muzzin\altaffilmark{3}, Gabriel Brammer\altaffilmark{4}, James S. Dunlop\altaffilmark{5}, Marijin Franx\altaffilmark{3}, Johan P. U. Fynbo\altaffilmark{6}, Ivo Labb\'e\altaffilmark{3}, Bo Milvang-Jensen\altaffilmark{6}, Pieter G. van Dokkum\altaffilmark{7}}

\email{Email: stefanonm@missouri.edu}

\altaffiltext{1}{Physics and Astronomy Department, University of Missouri, Columbia, MO 65211, USA}
\altaffiltext{2}{Physics and Astronomy Department, Tufts University, Robinson Hall, Room 257, Medford, MA, 02155, USA}
\altaffiltext{3}{Leiden Observatory, Leiden University, P.O. Box 9513, 2300 RA Leiden, The Netherlands}
\altaffiltext{4}{Space Telescope Science Institute, Baltimore, MD 21218, USA}
\altaffiltext{5}{SUPA - Scottish Universities Physics Alliance - Institute for Astronomy, University of Edinburgh, Royal Observatory, Edinburgh EH9 3HJ, UK}
\altaffiltext{6}{Dark Cosmology Centre, Niels Bohr Institute, Copenhagen University, Juliane Maries Vej 30, DK-2100 Copenhagen O, Denmark}
\altaffiltext{7}{Department of Astronomy, Yale University, New Haven, CT 06511, USA}

\begin{abstract}

We build a Spitzer IRAC complete catalog of objects, obtained by complementing the $K_\mathrm{s}$-band selected UltraVISTA catalog with objects detected in IRAC only. With the aim of identifying massive (i.e., $\log(M_*/M_\odot)>11$)  galaxies at $4<z<7$, we consider the systematic effects on the measured photometric redshifts from the introduction of an old and dusty SED template and from the introduction of  a bayesian prior taking into account the brightness of the objects,  as well as the systematic effects from different star formation histories (SFHs) and from nebular emission lines in the recovery of stellar population parameters. We show that our results are most affected by the bayesian luminosity prior, while nebular emission lines and SFHs only introduce a small dispersion in the measurements. Specifically, the number of $4<z<7$  galaxies ranges from 52 to 382 depending on the adopted configuration.  Using these results we investigate, for the first time, the evolution of the massive end of the stellar mass functions (SMFs) at $4<z<7$. Given the rarity of very massive galaxies in the early universe, major contributions to the total error budget come from cosmic variance and poisson noise. The SMF obtained without the introduction of the bayesian luminosity prior does not show any evolution from $z\sim6.5$ to $z\sim 3.5$, implying that massive galaxies could already be present when the Universe was $\sim0.9$~Gyr old. 
However, the introduction of the bayesian luminosity prior reduces the number of $z>4$ galaxies with best fit masses $\log(M_*/M_\odot)>11$ by 83\%, implying a rapid growth of very massive galaxies in the first 1.5 Gyr of cosmic history. From the stellar-mass complete sample,  we identify one candidate of a very massive ($\log(M_*/M_\odot)\sim11.5$), quiescent galaxy at $z\sim5.4$, with MIPS $24\mu$m detection suggesting the presence of a powerful obscured AGN. Finally, we show that the number of massive galaxies at $4<z<7$ measured in this work matches the number of massive galaxies at $3<z<6$ predicted by current models of galaxy formation.

\end{abstract}


\section{Introduction}

Understanding the formation and evolution of galaxies has remained a central topic in astrophysics for many decades.
Since the pioneering works (e.g., \citealt{hoyle1953}), this field has seen a substantial step forward, especially in the last decade,
thanks to the developments in data acquisition and analysis techniques. 
Specifically, this has allowed the community to study a large
number of galaxies up to $z\gtrsim4$, with the frontier gradually shifting to $z\sim8-10$ (see e.g., \citealt{fontana2010, pentericci2011,mclure2011,yan2012,caruana2012,ono2012,bouwens2013, curtis-lake2013,labbe2013,finkelstein2013, coe2014,ellis2013,bradley2014,zitrin2014,zheng2014}).

Different measurements of galaxy masses provide information on different physical aspects.  
Dynamical masses measure the associated gravitational potential allowing for a closer insight into the dark matter halo properties and their evolution across cosmic time (see e.g., \citealt{gerhard2001,thomas2011,cappellari2013,beifiori2014}). However, recovering dynamical masses requires spectroscopy,  more difficult to obtain for galaxies at high redshift, given their lower brightness. Stellar masses, on the other hand, reflect the information on the luminous matter content. They do not require spectroscopy, as they can also be recovered from modelling of the multi-wavelength photometry, and are thus available for a much larger sample of high redshift galaxies. The observed profile of the stellar mass function (SMF) across cosmic time is the result not only of the effects of the gravitational potential governing the assembly of the dark matter halos, but  also of the physical processes which govern the baryonic matter, such as the formation of new stars or the quenching of star formation.

The SMF is then one of the statistical tools that are most commonly used to trace the evolution of the galaxy populations across cosmic time and is one 
of the main observables whose reproduction is a necessary step for the validation of galaxy formation models.  Several measurements of the SMF exist up to $z\lesssim4$ 
(see  e.g., \citealt{muzzin2013b} and \citealt{ilbert2013} and references therein). The emerging picture is that some of the very massive galaxies were already in place at $z\sim4$ and their associated number density has rapidly evolved between $z\sim4$ and $z\sim1.5$. From a complementary perspective, so-called archeological studies have shown that local most massive galaxies formed most of their stars during a short burst at $z\gtrsim4$ (e.g., \citealt{thomas2010}). This picture is however complicated by the fact that the evolution of dark matter halos follows a rigid hierarchical structure, with more 
massive haloes forming at later times only (see e.g., \citealt{springel2005,baugh2006} and references therein). It is therefore important to be able to track the formation and evolution of the most massive galaxies at earlier cosmic times.

The SMF measurements at $z\lesssim3.5$ are typically constructed from samples selected in the near-infrared (NIR) $K$ band. The advantage of the single-band photometric selection is that it allows 
for the assembly of stellar mass complete samples of galaxies. However, the rest-frame optical Balmer/4000\AA~break begins to enter the K band at $z\sim4$ and it is completely included by $z\sim5$. Because of this effect,  galaxies at $z>4$ with a given stellar mass will then be characterized by even fainter K-band fluxes than galaxies at $z<4$ with the same stellar mass and mass-to-light ratio (M/L). The dimmed K-band fluxes drop below the detection threshold, preventing these objects from being detected.  Conversely, detected objects will be characterized by K-band fluxes equal or higher than the detection threshold, and thus correspond to higher values of stellar mass (for the same M/L value) required for the selection of stellar-mass-complete samples of galaxies.

Furthermore, the K-band selection becomes more and more sensitive to the extinction by dust with increasing redshift, as the rest-frame optical and ultra-violet (UV) enter the K-band. Given the current photometric depth of K-band selected samples, these factors thus limit our ability to perform statistical studies at higher redshifts. For this 
reason, SMF at $z>4$ are generally computed using samples of galaxies selected in the rest-frame UV via color selection criteria, through the so-called \emph{dropout} technique (exceptions exist, such as \citealt{caputi2011} who measure the SMF up to $z\sim 5$ using photometric redshifts from an IRAC $4.5\mu$m-selected sample in the UDS field). However, by 
construction, dropout selections are biased against evolved (i.e., quiescent) galaxies and/or dust-extincted systems as they preferentially pick those galaxies with brighter rest-frame UV luminosities, indicative of those systems with a recent burst or still ongoing star-formation and low dust 
content. This is even more important for massive galaxies, as recent studies have shown an increase of dust extinction with redshift (see e.g., \citealt{whitaker2010,marchesini2010,marchesini2014}). Specifically, this  prevents from searching for massive and massive and quiescent galaxies at $z>4$ (see e.g., \citealt{mobasher2005,mclure2006,dunlop2007,lundgren2014}), potentially introducing a bias in our knowledge of the high-mass end population of galaxies and, in general, in the first $\sim $2~Gyr of cosmic history.

On the other side, as we will show in Sect. \ref{sect:smf_completeness}, the stellar mass completeness limit in bands at wavelength $\gtrsim3-5\mu$m (e.g., IRAC $3.6\mu$m and $4.5\mu$m bands) is roughly constant for $4<z<8$, and, with current imaging 
depth, corresponds to $M_*\sim10^{11}M_\odot$ (see also e.g., \citealt{fontana2006,ilbert2010,caputi2011}). However, the larger point-spread functions (PSF) of IRAC bands, 
compared to those in the optical/NIR imaging, enhance the problem of source blending  in the measurements of the fluxes; these effects become even more important when IRAC bands are considered for the detection of sources. The most commonly adopted solution consists in performing the photometry using positional and morphological information from higher resolution imaging, usually in bands different from that of interest, and under the assumption that the morphological properties of the objects in the lower resolution band do not significantly differ from those in the higher resolution one (see e.g., \citealt{fernandez-soto1999,labbe2005,laidler2007}). If this approach solves the problem of contamination in the flux measurement, it can not be directly adopted for the detection of sources in IRAC bands, as it relies on an already existing catalog.

The aim of this work is to search for a population of massive ($\log(M_*/M_\odot)>11$) galaxies at $4<z<7$ and to study their evolution through the analysis of the  SMF  in three redshift bins: $4<z<5$, $5<z<6$ and $6<z<7$.  To this aim, we used an IRAC $4.5\mu$m-complete sample of galaxies assembled complementing the UltraVISTA DR1 $K_\mathrm{s}$-band selected catalog by \citet{muzzin2013a} with detections on the residual maps of IRAC $3.6\mu$m and $4.5\mu$m bands. Indeed, the IRAC flux measurements included in the UltraVISTA $K_\mathrm{s}$-band selected catalog were obtained using the technique described above, which relies on position and morphological information from the K-band map. These measurements are less sensitive to the contamination from neighbours than those performed directly on the IRAC maps. Performing the detections on the IRAC residual maps allows for the inclusion in the catalog of all those sources with K-band fluxes below the K-band detection threshold. The advantage of an IRAC-based sample is that it is possible to detect galaxies at $4<z<7$ with stellar masses lower than with the current UltraVISTA $K_\mathrm{s}$-band based sample, for the same M/L value, since $z<7$ galaxies in the IRAC $3.6\mu$m and  $4.5\mu$m do not suffer from the dimming due to the Balmer/4000\AA~break, which instead affects the K-band data. This approach then allowed us to exploit the deeper completeness in stellar mass associated to the IRAC bands, while solving the issues in the detection and flux measurements from source blending in purely IRAC-detected catalogs.

The still large uncertainties in the knowledge of the SEDs of galaxies in this range of redshift can introduce systematic effects which potentially bias the measurements of photometric redshifts and stellar population parameters. We approach this problem by exploring different configurations for the measurements of photometric redshifts and stellar population parameters and treat the results as systematic effects. In order to make the presentation more organized, we assume the sample obtained from the most relaxed configuration to be the default sample and consider the statistical differences arising from the other configurations as systematic effects. One of the main results of this approach, presented in Section \ref{sect:robust_massive}, is a sample of galaxies at $z>4$ and with $\log(M_*/M_\odot)>11$ irrespective of the configuration adopted for the measurement of photometric redshifts and stellar population parameters.

This paper is organized as follows.  In Section 2 we describe how we build the IRAC-complete sample, complementing the UltraVISTA DR1 $K_\mathrm{s}$-band selected catalog by \citet{muzzin2013a} with detections on the residual maps of IRAC $3.6\mu$m and $4.5\mu$m bands. In Section 3 we present the step-by-step process we adopted to obtain a clean sample of massive ($\log(M_*/M_\odot)>11$) galaxies at $4<z<7$, removing galaxies potentially contaminated by AGN and potential lower redshift interlopers, and addressing different factors which can affect the measurements of photometric redshifts and stellar population parameters.  In Section 4 we present the robust sample of galaxies  with $\log(M_*/M_\odot)>11$ at $z>4$ irrespectively of the systematic effects discussed in Section 3. Specifically, in Section 4.3 we present the SMF measurements in the redshift range $4<z<7$ and show the systematic effects arising from the factors presented in Section 3 on the SMF measurements. Our results are discussed in Section 5, while we summarize and conclude in Section 6.

Throughout this work we adopt a concordance 
cosmology, with $H_0=70$~km/s/Mpc, $\Omega_m=0.3$ and $\Omega_\Lambda=0.7$. Unless otherwise specified, all magnitudes are referred to the AB system, while stellar masses were computed using the \citet{chabrier2003} initial mass function (IMF).

\section{The photometric sample}
\label{sect:sample}

For this work, we complemented the public catalog from \citet{muzzin2013a} with detections on the IRAC residual maps from S-COSMOS \citep{sanders2007,ilbert2010}.  In the following paragraphs we present the details of the procedure we followed to construct the IRAC-complete sample from the $K_\mathrm{s}$-band selected catalog of \citet{muzzin2013a}. The $K_\mathrm{s}$-band selected catalog of \citet{muzzin2013a} is based on the first data release (DR1) of the UltraVISTA survey \citep{mccracken2012} and delivers 30-bands flux information for 262615 objects.  The DR1 of the UltraVISTA survey
is characterized by deep (90\% point source completeness $K_\mathrm{s}=23.4$AB) imaging in four broad-band NIR filters ($YJHK_\mathrm{s}$) as well as
one narrow-band filter centered on H$\alpha$ at $z = 0.8$ (NB118). Images in each band cover an area of $\sim1.6$ degrees$^2$ centered on the COSMOS field (\citealt{scoville2007}). This field is characterized by extensive deep multi-wavelength coverage, ranging from the X-rays (\citealt{hasinger2007,elvis2009}) to the radio (\citealt{schinnerer2007}). Specifically, Spitzer IRAC imaging data were collected in the framework of the S-COSMOS project  \citep{sanders2007,ilbert2010}.

\begin{figure}
\includegraphics[width=8.5cm, bb=80 0 412 500, clip]{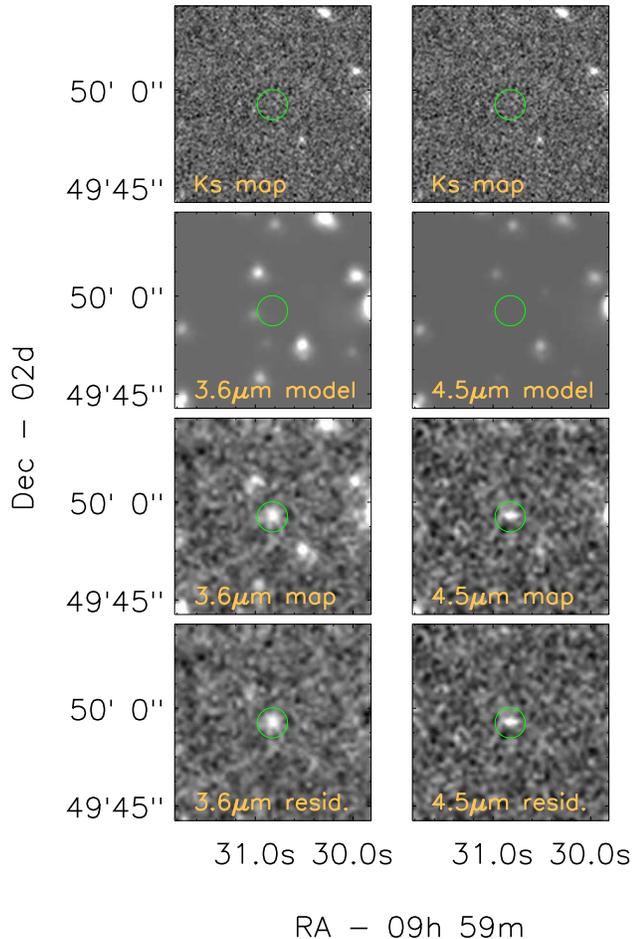} 
\caption{Example of the steps involved in the detection of sources in the IRAC bands. The left and right columns refer to the same procedure in the IRAC $3.6\mu$m  and $4.5\mu$m, respectively. In each column, the top panel shows the cutout of the UltraVISTA $K_\mathrm{s}$-band frame centered on the position of an IRAC source randomly picked among those originally not included in the $K_\mathrm{s}$-band catalog of \citet{muzzin2013a}. Its position is also marked across all panels by the green circle to facilitate its identification. The second panel reproduces the brightness profiles of all the objects, for the same region of sky, from the $K_\mathrm{s}$-band catalog, after convolving them with the kernel required to match the PSF of the corresponding IRAC channel and after applying to each source the flux scale factor from the best fit process. This image, then, constitutes the best-fit \emph{model} of the science frame based on the information available from the $K_\mathrm{s}$ band. Sources not included in the $K_\mathrm{s}$-based catalog will not be reproduced in the model image. The third panel panel shows the original IRAC science frame. The panel at the bottom presents the \emph{residual} image obtained subtracting the model from the science frame. The additional IRAC source is still present, cleaned from neighbors. We adopted the residual images as input for the detection with SExtractor. \label{fig:dropout}}
\end{figure}

The flux measurements in the S-COSMOS IRAC  for the UltraVISTA $K_\mathrm{s}$-band selected catalog of \citet{muzzin2013a} were obtained using a source-fitting code specifically developed to recover fluxes in heavily confused images \citep{labbe2005}. The code is based on the assumption that the morphology of each object in the higher-resolution band ($K_\mathrm{s}$ for our case) does not differ too much from the intrinsic morphology in the lower resolution image (i.e., IRAC). It is then possible to use the position and brightness profile of the source on the high resolution image as a prior for the same object in the lower resolution image. The brightness profile of the source in the high-resolution image is convolved with the kernel required to match the low-resolution image PSF. The result is then a template of the object in the low-resolution image modulo its total flux which is obtained via best fit. The actual flux measurements for each object is however performed in apertures on a per-object basis, after all the neighbouring objects have been removed using the information from the fitting procedure. One of the diagnostic outputs from this procedure is the image resulting from subtracting all the fitted sources from the input science frame, i.e., a residual image. Since no new detection is performed during the source-fitting measurements, the residual image will contain those sources not detected in the high-resolution image. An example of this process is shown in Figure \ref{fig:dropout}. We complemented the UltraVISTA DR1 $K_\mathrm{s}$-band selected photometric catalog of \citet{muzzin2013a} with objects detected on the IRAC-band residual images. These objects are too faint to be robustly detected in the DR1 K-band image and therefore were not included in the K-selected UltraVISTA DR1 catalogIRAC bands, whose $K_\mathrm{s}$-band counterparts  are too faint to be included in the original $K_\mathrm{s}$-band selected UltraVISTA catalog. This was achieved by performing an additional detection on the residual images resulting from the template fitting procedure as described below. 

At first, we run SExtractor (\citealt{bertin1996}) on S-COSMOS IRAC $3.6\mu$m and $4.5\mu$m bands residual images independently. Only those objects with a matching position within a radius of 5 pixels (equivalent to $0.75''$), and with a SNR $>5$ in the two IRAC bands were kept. This approach limits the number of spurious sources, although it introduces a completeness in stellar mass shifted towards higher stellar masses, given the shallower depth of the IRAC $4.5\mu$m map compared to the IRAC $3.6\mu$m one. 
Objects within $\sim100''$ from the border of the UltraVISTA $K_\mathrm{s}$-band map were purged to remove detection in those regions with lower signal-to-noise ratio. In total, 408 new sources over an area of $1.5$ square degree were added to the existing UltraVISTA catalog, representing an increase up to 39\% (depending on the configuration adopted for the recovery of photometric redshifts - see Section 3) to the number of galaxies at $z>4$ in the sample.

Spectral energy distributions (SEDs) were then built using matched aperture photometry from CFHTLS \citep{cuillandre2012}, Subaru \citep{taniguchi2007}, UltraVISTA \citep{mccracken2012} and S-COSMOS \citep{sanders2007}. Photometry was performed with SExtractor in dual mode, with IRAC $3.6\mu$m residual image as the detection image, to take advantage of its slightly better image quality compared to IRAC $4.5\mu$m map. We adopted apertures of $2.1''$ diameter for CFHTLS, Subaru and UltraVISTA frames, $3.0''$ for IRAC bands and $5.0''$ for MIPS 24$\mu$m as a compromise between the highest signal-to-noise ratio measurements and the possible loss of flux due to the uncertainty in the position of the source on the IRAC frame. The sizes of the aperture adopted for the photometry were chosen taking into consideration potential contamination from neighboring objects. The FWHM of the broadest PSF in ground-based data is about $1''$, so that the chosen aperture corresponds to $2\times$FWHM. The fraction of pairs closer than the adopted aperture radius is about $2\%$,  a value which still ensures a limited contamination fraction from neighbors. Finally, objects close to bright stars and those with contamination by bright saturated nearby objects in more than three filters were excluded from the selection (flag \texttt{use=1} in the UltraVISTA catalog).   For IRAC and MIPS data, instead, the flux measurement is done for each object after cleaning from neighbors, minimizing potential contamination effects before the actual measurement is done. 

Total fluxes were finally computed applying an aperture correction obtained from the curve of growth of a sample of bright and isolated point sources in each band.

This sample was finally cleaned from potential brown dwarfs (BD) and variable objects. Current observations of BD SEDs do not yet cover the wavelength region redder than $\sim2\mu$m. The molecular bands in the photosphere of the cool stars result in peaks and plateaux which largely overlap with the NIR broad bands (see e.g., Figure 5 in \citealt{bowler2012}). These features do not allow for a reliable removal of the degeneracy between BD and red/high-z galaxies. The identification of candidate brown dwarfs was therefore carried out in two steps: at first we pre-selected the sample of candidate BDs from color-color plots; the PSF of each candidate was then measured on the public HST/ACS frames (\citealt{koekemoer2007} - where ACS was not available, we considered those bands with the best compromise between resolution and S/N). We thus excluded from our sample those pre-selected objects with a full-width-half-maximum (FWHM) smaller than $0.11''$ for HST ($\sim0.8''$ for ground-based), being consistent with point sources. We further removed sources which showed clear signs of temporal variability. Through this selection we removed 54 objects from the original sample, leading to a sample of 502 galaxies at $z>4$.

\begin{figure}
\includegraphics[width=8.5cm]{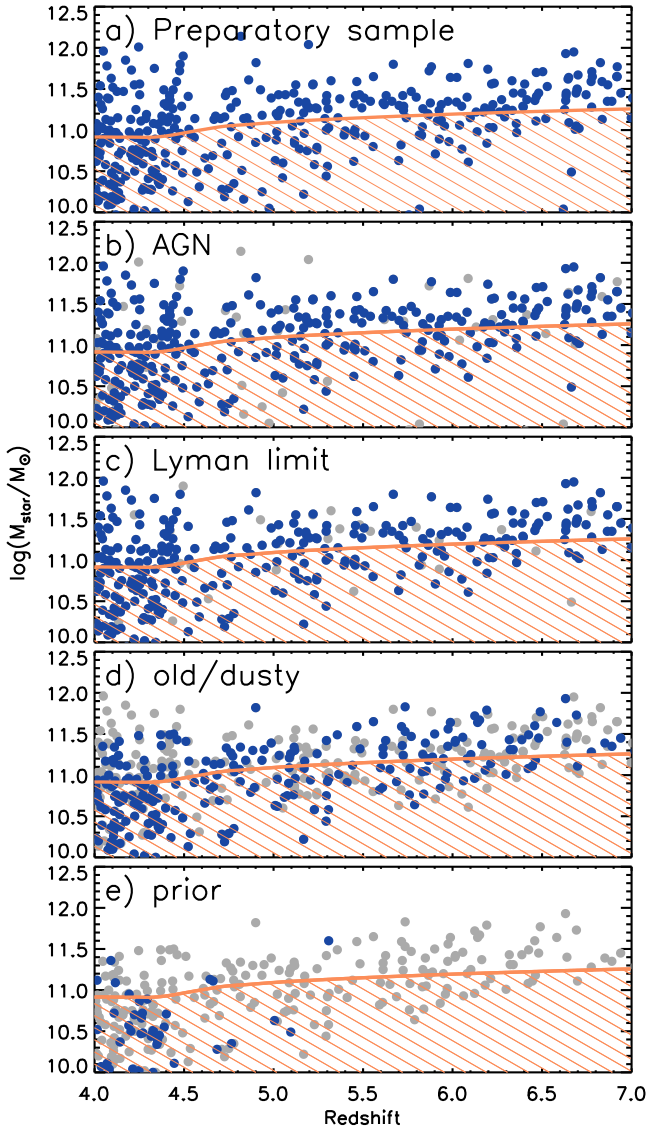} 
\caption{The five panels track the effect of the selections applied to the sample and of the different configurations adopted for the measurements of photometric redshifts. From top to bottom, the panels refer to (a) the initial sample, obtained after cleaning from brown dwarfs, (b) the initial sample after eliminating AGN candidates; (c) the AGN-cleaned sample after removing those objects with an emission blueward of the Lyman limit; (d) the sample obtained after introducing an old and dusty template for the measurements of photometric redshift and (e) the sample obtained after applying the bayesian luminosity prior in the measurements of the photometric redshifts. In each panel, the filled blue circles correspond to the object properties after applying the corresponding selection, while filled grey circles represent the objects from the progenitor sample, i.e., the sample identified by filled blue circles in the previous panel. The orange curve delimits the stellar-mass complete sample (see Sect. \ref{sect:smf_completeness}). The percentages of objects excluded from the sample because potential AGN (18\%) or because showing a detection blue-ward of the Lyman limit  (8\%) or after introducing the dusty template (19\%) are relatively small, the percentage of objects excluded from the $z>4$ sample when the luminosity prior is introduced in the measurements of photometric redshifts is very high, reaching 83\%. \label{fig:z_mstar}}
\end{figure}

\section{Photometric redshifts and stellar population parameters}
\label{sect:photoz}

\begin{figure*} 
\begin{tabular}{cc}
\hspace{-0.1cm}\includegraphics[width=8.5cm]{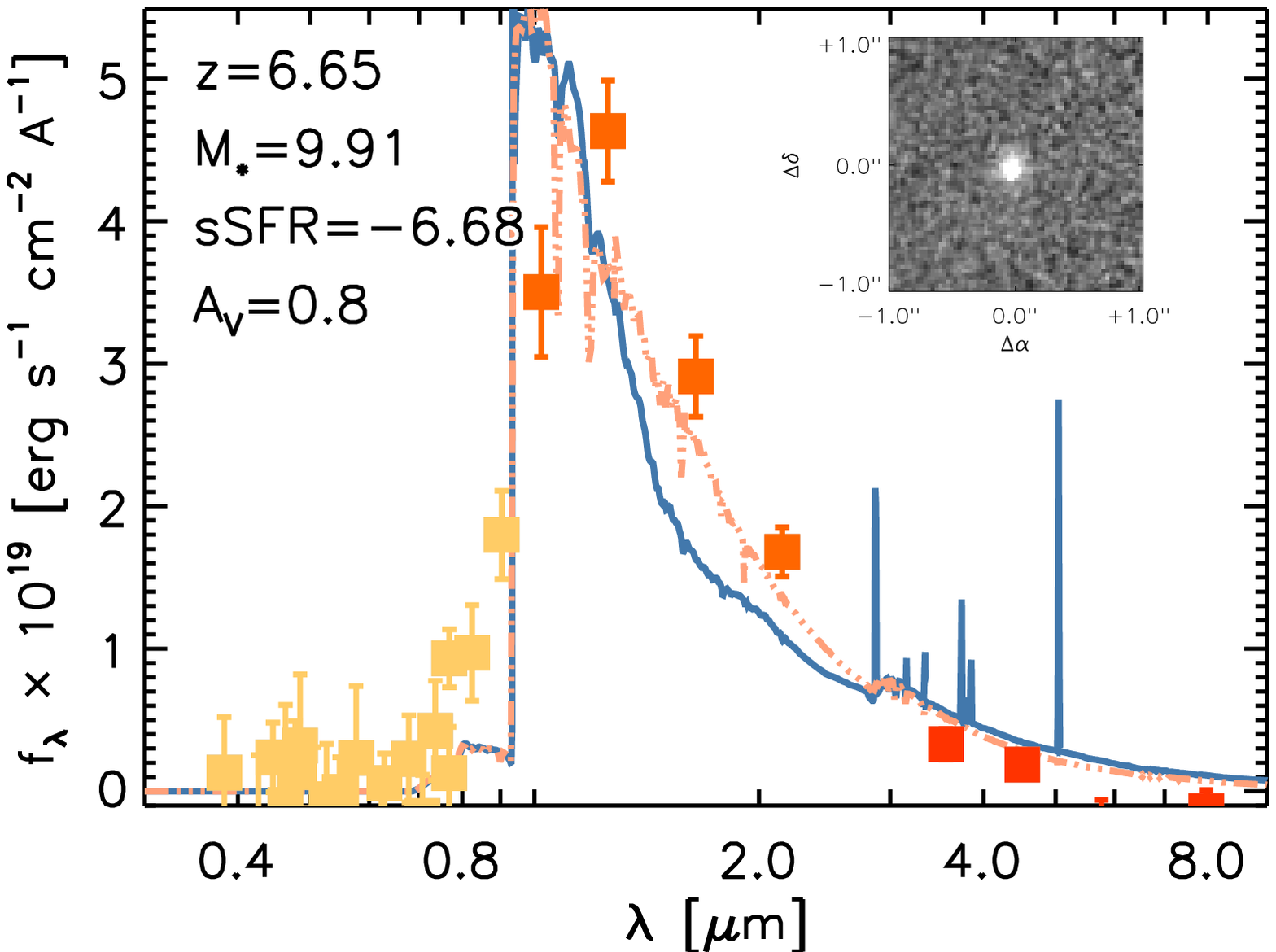}  & \hspace{-0.1cm}\includegraphics[width=8.5cm]{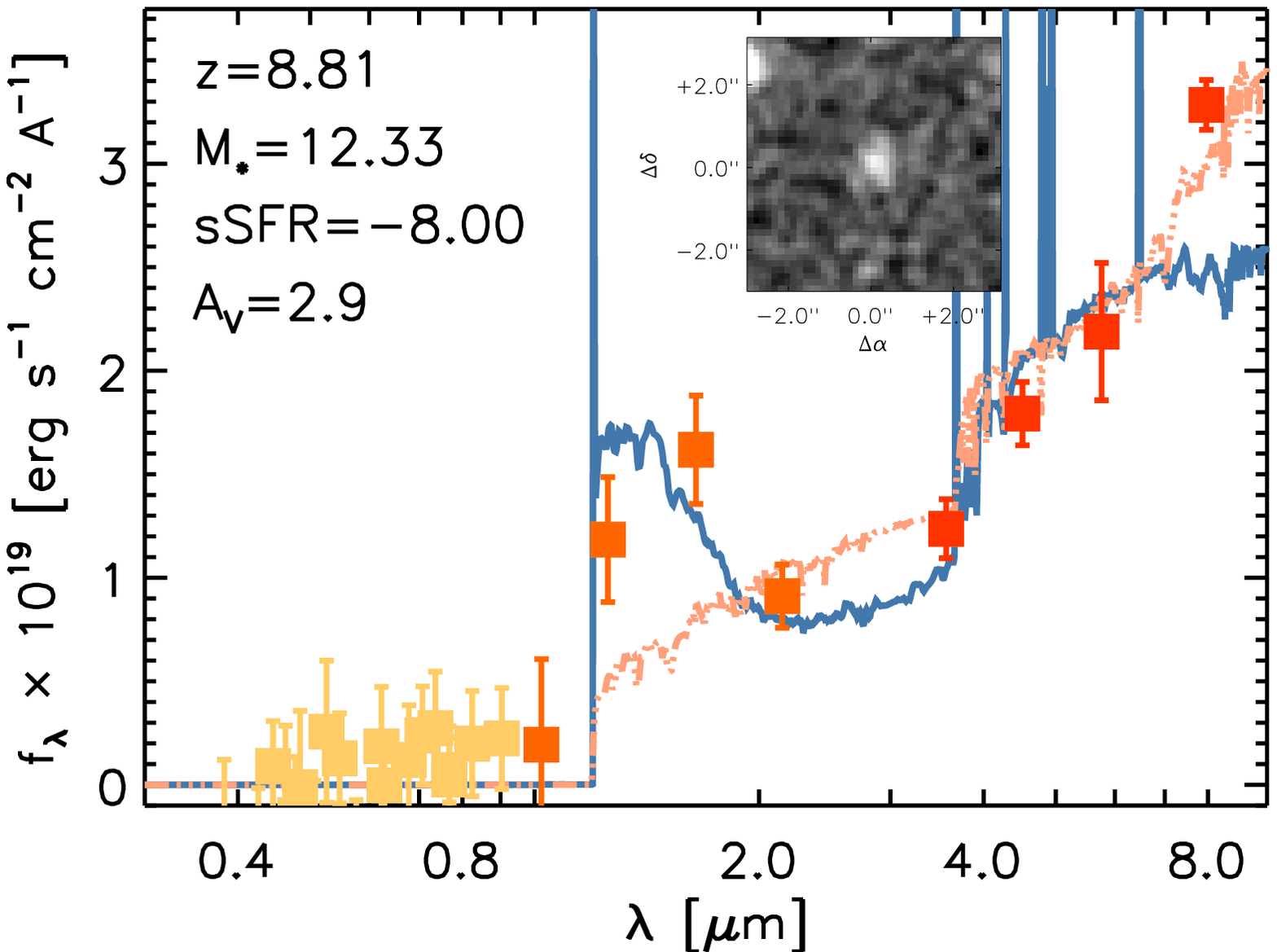}
\end{tabular}
\caption{Examples of two AGN candidates. The measured photometry in observer frame is represented by the filled colored squares with $1\sigma$ error bars (yellow for optical bands, orange for UltraVISTA $Y,J,H$ and $K_\mathrm{s}$ bands, red for IRAC $3.6\mu$ to $8.0\mu$m). The best fitting SED from EAzY and from FAST are shown as the blue and pink curves respectively.  The main physical parameters are also listed at the top-left corner. Top to bottom they are: the photometric redshift ($z$), the  $\log(M_*/M_\odot)$, the $\log(\mathrm{sSFR/yr}^{-1})$ and the extinction expressed in magnitudes. The inset shows a cutout of the object in the HST ACS F814W  band and UltraVISTA $K_\mathrm{s}$ for the object on the left and right respectively. The HST ACS F814W cutout shows a point-source morphology.  Objects like the two presented here were excluded from the sample. \label{fig:sed_AGN} }
\end{figure*}

The selection of a sample of reliable (massive) $z>4$ galaxies from a photometric catalog requires taking into account many different aspects that can possibly taint the sample and/or introduce systematic effects in the measurements of their physical parameters (see e.g., \citealt{dahlen2010,marchesini2010,lee2012}). Specifically,  when the measurements of photometric redshifts are involved, the lack of extensive spectroscopic data on $z>4$ galaxies limits our knowledge on the SEDs of such objects, making the choice of suitable SED templates less straightforward, thus affecting the reliability of the measurements of photometric redshifts. A second source of uncertainty in the estimation of photometric redshifts comes from the adoption of a flux- and redshift-based bayesian prior.  As originally shown in \citet{benitez2000}, the adoption of prior information through the Bayesian formalism can drastically limit the fraction of outliers and reduce systematic biases in the measurement of photometric redshifts.  Specifically, the prior adopted in this work consists in the distribution of galaxies as a function of redshift and of flux density as it was recovered from semi-analytic models (see Section \ref{sect:prior}). Furthermore, the presence, in a galaxy core, of an Active Galactic Nucleus (AGN) can bias the measurements of the stellar population parameters; finally, as it has recently been shown, biases in the estimation of the stellar masses can also be introduced by nebular emission lines. In the following sections we will systematically address these uncertainties. The involved steps are briefly summarized hereafter. Figures \ref{fig:z_mstar} through \ref{fig:nl99}  track the discussed systematic effects as this sample is polished.
We start from the full sample of $z>4$ galaxies obtained after cleaning the composite sample from point-source objects and potential brown dwarfs (the \emph{preparatory} sample - Section \ref{sect:initial_sample}). We then identify and remove from this sample galaxies potentially contaminated by AGN (Section \ref{sect:AGN}). Successively, we purge those objects with a non-zero flux in those bands bluer than the Lyman limit at the redshift measured for each galaxy, as such objects are inconsistent with $z>4$ galaxies (Section \ref{sect:default_sample}). The sample obtained after this multi-step polishing process constitutes our \emph{default} sample of $z>4$ galaxies.  Panels a) through c) in Figure \ref{fig:z_mstar} give a graphical representation of these steps.  Successively, we study the systematic effects on the redshift measurement of the inclusion in the set of SED templates of a maximally red SED template (Section \ref{sect:dusty}) and of the bayesian luminosity prior (Section \ref{sect:prior}) - see also panels d) and e) in Figure \ref{fig:z_mstar}, respectively. Finally we analyze the systematic effects of contamination by nebular emission lines and of different star-formation histories (SFHs) in the recovery of the stellar population parameters (Sections \ref{sect:neblines} and \ref{sect:sfh}).

In total, from the combination of the above systematic effect analysis, this approach generated 32 samples of massive $4<z<7$ galaxies, each one corresponding to a different configuration: two configurations are the result of the introduction/exclusion of the old and dusty template; two configurations result from the activation or not of the bayesian prior on the flux, four configurations proceed from the different treatment of the contamination by nebular lines and two more configurations derive from the considered SFHs. Given the current uncertainties in the measurement of the physical parameters of $z>4$ galaxies from broad-band photometry, we would like to stress here that each one of these samples is potentially a consistent measurement of the properties of $z>4$ galaxies. In order to make the presentation of the results clearer, in this work we assume as a reference the sample obtained from the configuration with the smaller set of assumptions, i.e., photometric redshifts computed without the inclusion of the old and dusty template and without activating the bayesian prior on the flux, and stellar population parameters measured with the delayed exponential SFH and without applying any correction for the contamination by nebular lines. We therefore consider all the other configurations as systematic effects.

\subsection{The preparatory sample}
\label{sect:initial_sample}

Photometric redshifts were initially computed for all the 502 objects in the sample presented on Sect. \ref{sect:sample}, using EAzY \citep{brammer2008}. EAzY performs the redshift measurements by fitting the observed spectral energy distributions to linear combinations of a number of templates. The template set adopted for the preparatory sample is constituted by six templates from the PEGASE models \citep{fioc1999}  which also include emission lines and was presented in \citet{muzzin2013a}. Although EAzY allows for the use of a bayesian prior, we did not activate this option at this stage. The systematic effects of the introduction of the bayesian prior will be discussed in Sect. \ref{sect:prior}

Since EAzY does not deliver information on the stellar population parameters, the photometric redshifts were then used as an input to FAST \citep{kriek2009} for the measurements of stellar masses, star-formation rates and ages.  Indeed FAST can measure photometric redshifts using the $\chi^2$ minimization procedure, although it does not provide the possibility of introducing any bayesian luminosity prior, as instead is the case for EAzY. The uncertainties associated to the stellar population parameters are natively computed by FAST through Monte Carlo simulations. The errors on the stellar population parameters quoted in the figures of this paper refer to the upper and lower 68\% confidence intervals produced by FAST, unless otherwise specified. The synergy between EAzY and FAST allows then for an optimal measurement of photometric redshifts and stellar population parameters.

We adopted the  \citet{bc03} models, a \citet{chabrier2003} initial mass function, solar metallicity, $0<A_V<10$ and  a delayed-exponential ($\tau$-model) SFH.  We also note that the set of templates adopted for the measurements of the stellar population parameters does not include any SED template with AGN emission. These templates were not included since a robust fit of such models to the observed data for $z>4$ galaxies would also require coverage in bands red-ward of IRAC/MIPS, unavailable for this work. The distribution of the stellar mass with redshift for this sample is presented in panel a) of Figure \ref{fig:z_mstar}.

\subsection{AGN contamination}
\label{sect:AGN}

The presence, in a galaxy core, of an active galactic nucleus (AGN) can bias the measurements of the stellar population parameters obtained thorough SED modelling when an AGN component is not properly included in the set of SED templates. However, a reliable estimation of the contribution from the AGN requires the modelling of the rest-frame infra-red region of the SED, which would provide information on the dusty torus around the central black hole. Spectroscopy would certainly be the preferred tool, as it allows to recover information on the main emission lines characteristic of AGNs, such as Ly$\alpha$, Ly$\beta$, H$\alpha$, H$\beta$, N\textsc{V} $\lambda1242$\AA, Si\textsc{IV}$\lambda1393,1402$\AA, C\textsc{III}] $\lambda 1909$\AA, C\textsc{IV} $\lambda1549$\AA, N\textsc{III}] $\lambda 1750$\AA, O\textsc{III}] $\lambda 1663$\AA, Mg\textsc{II} $\lambda2798$\AA~and Ca\textsc{II} $\lambda 8498,8542,8662$\AA. However, this kind of information is extremely difficult to obtain for high-z galaxies, as the expected flux is very low.

Given the above uncertainties, we opted for removing from the sample all potential Type-1 AGNs. When viewed face-on, AGNs are characterized by extremely compact or point-source morphologies, corresponding to the compact region around the central black hole.  On the other side, emission from the AGN can strongly contaminate the rest-frame optical region of the SED resulting in a characteristic excess compared to the SEDs of galaxies not hosting an AGN.

In order to select all potential AGNs, we visually inspected both the image stamps and the observed SEDs of all the objects in the preparatory sample and removed those showing a point-source morphology and/or whose SED presented the characteristic excess in the observed SED (occurring in the IRAC bands, given the redshift range considered here), signature of  potential contamination by emission from the AGN.  The right panel of Figure \ref{fig:sed_AGN} shows the SED and image stamp of an object randomly chosen among those whose SED presents an excess in the rest-frame optical (corresponding to the IRAC bands in the observer frame). The visual inspection on morphology was performed on  the ACS F814W band, and adopting optical/NIR bands where no ACS coverage was available.  In the left panel of Figure \ref{fig:sed_AGN} we show an example of such objects, with an extremely compact morphology and a strong and broad emission possibly corresponding to the rest-frame Lyman $\alpha$. We further improved this selection by matching our sample to public XMM-Newton \citep{cappelluti2009}, Chandra/ACIS catalogs \citep{elvis2009} and to the SDSS Quasar Catalog \citep{paris2012}.

With this procedure, we flagged as potential AGNs  91 sources, corresponding to 18\% of the preparatory sample; of these, 12 sources were already detected in X-ray data, while 79 are candidate new AGN. The objects flagged as AGN are visible as grey filled circles in panel b) of Figure \ref{fig:z_mstar}. As it can be seen from panel b), several objects with high stellar mass turn out to actually be potential AGNs. The high stellar mass recovered from the SED fitting for  theses objects  is likely the result of the fact that when the stellar population parameters are measured adopting an SED template set which does not include any AGN-specific SED, the excess in flux introduced by the existing AGN component biases towards higher values the stellar mass measurements. This, then, emphasizes the importance of this kind of selection in such works, as it can significantly bias the statistical analysis. All the 91 sources flagged as potential AGN were removed from the preparatory sample.

\subsection{Galaxies at $z>4$: the default sample}
\label{sect:default_sample}

The neutral hydrogen clouds which constitute the inter-galactic medium absorb the light emitted by distant sources when  the wavelength of the photon is shorter than that corresponding to the 912\AA~Lyman limit at the redshift of the emitting source measured from the rest-frame of the HI cloud. Although the amount of absorption depends on the redshift of the emitting object, for objects at $z\gtrsim 3$, the absorption can be considered total (see e.g.,  \citealt{moller1990}, \citealt{madau1995}). From an observational point of view, this physical effect manifests itself as an absence of flux in bands covering the wavelength region blueward of the 912\AA~Lyman limit at the observer-frame. Specifically, this means that any object with a non-zero flux measurements in bands covering the wavelength region $\lambda_{obs}<(1+z)\times 912$\AA~(with $z$ the presumed redshift of the source) is instead likely to be at lower redshift. We note here that the Lyman forest for sources at $z\sim4.5 $ still transmits $\sim30\%$ of the photons (see e.g., \citealt{madau1995}), meaning that we can only consider as effective limit the Lyman 912\AA~limit, but not the 1216\AA~Lyman-$\alpha$. Although by $z\sim6.5$ the transmission has dropped to 2\%, for consistency we apply the same selection criteria over the full redshift range.

\begin{figure*}
\begin{tabular}{cc}
\hspace{-0.1cm}\includegraphics[width=8.5cm]{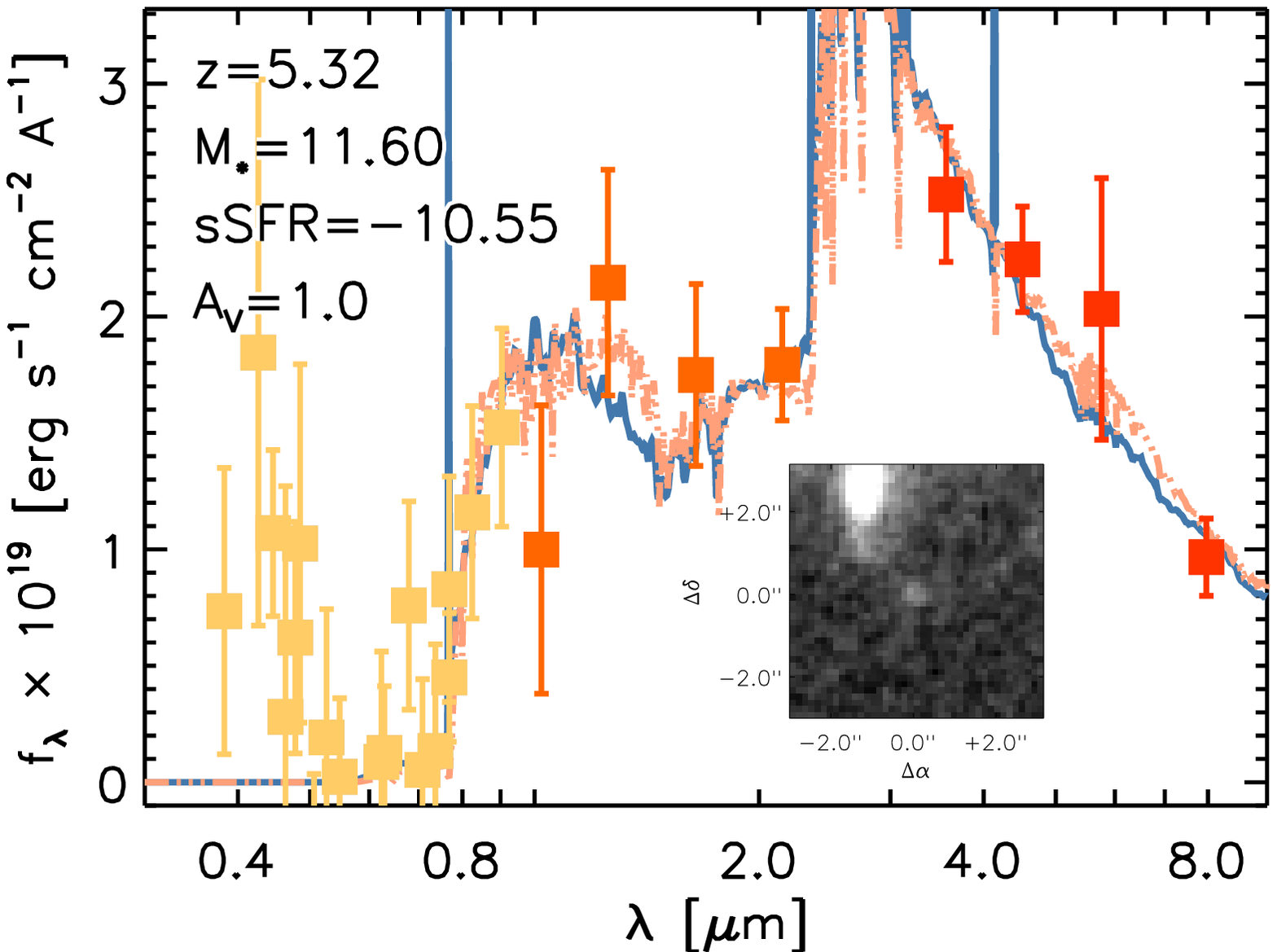}  & \hspace{-0.1cm}\includegraphics[width=8.5cm]{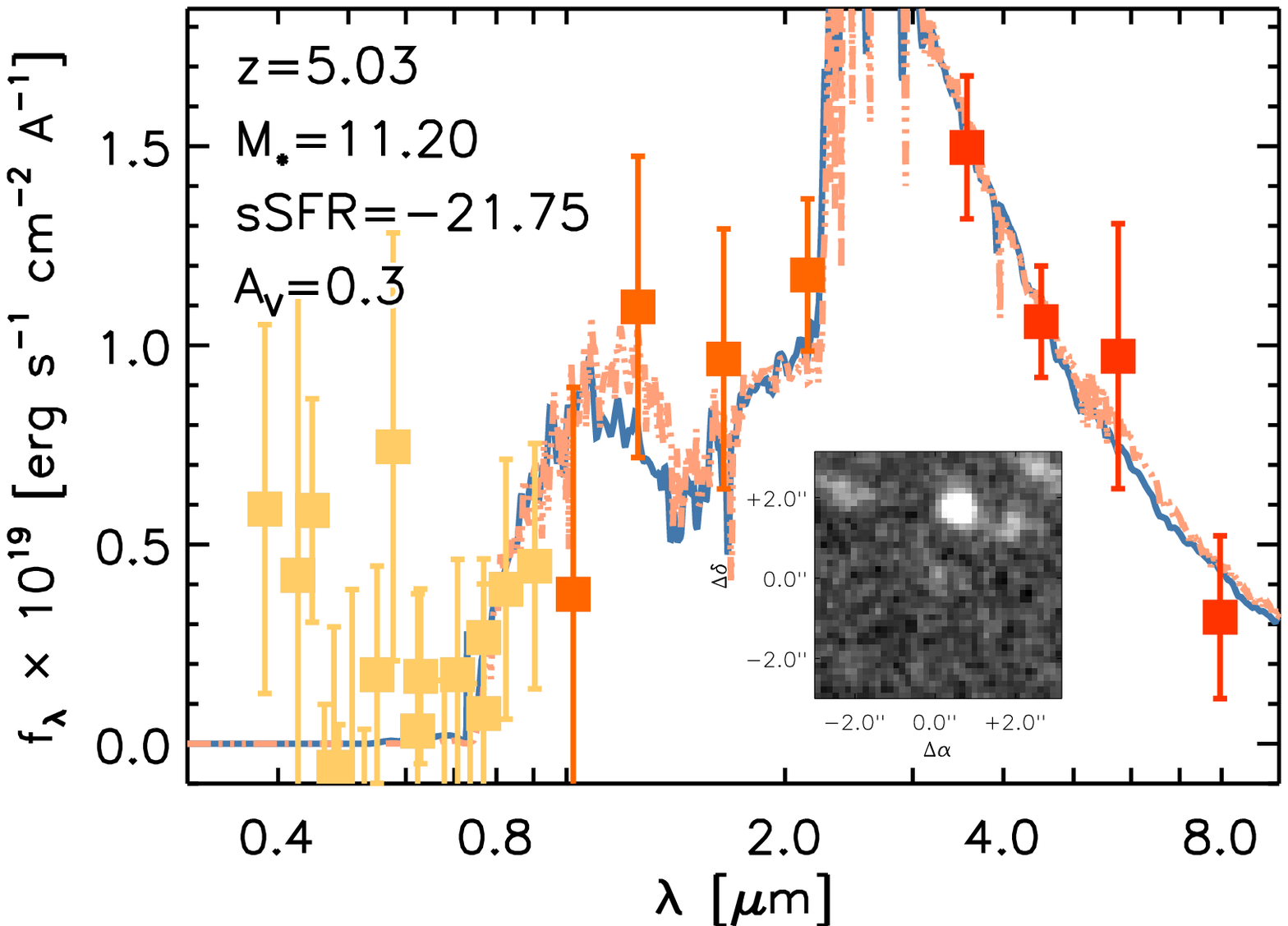}
\end{tabular}
\caption{Examples of objects excluded from the sample of $z>4$ galaxies because presenting an excess in the flux measured in bands blue-ward of the observer-frame Lyman limit. The colored squares mark the flux measurements in observer frame with the associated error bars; the blue and pink curves represent the best-fit SED from EAzY and FAST respectively.  The main physical properties are listed at the top-left corner (see Figure \ref{fig:sed_AGN} for further details). The inset reproduces the B-band cutout centered on the object position. The photometry in some of the bands bluer than the Lyman limit present an excess of flux. The presence of emission associated to the object is confirmed by the cutout. Objects like the two shown here were excluded from the sample of $z>4$ galaxies.  \label{fig:sed_lya} }
\end{figure*}

As a consistency check, for each object we stacked the cutouts in those bands bluer than the observer frame Lyman limit and excluded from the sample those objects whose stacked image showed a clear excess at visual inspection. We also visually inspected the SEDs of the AGN-purged sample to identify those objects with a non-zero flux  (at 1$\sigma$) in those bands bluer than the Lyman limit. The image stamps of the selected objects in the bands blueward of the observer-frame Lyman limit were successively visually inspected in order to disentangle whether the observed flux excess was the result of contamination from bright and/or nearby objects or it was a genuine emission from the object. In this latter case, the object was removed from the sample. Following this procedure, we removed from the AGN-cleaned sample 33 sources with non-zero flux measurements blueward of the Lyman limit; these sources are marked by grey filled circle in panel c) of Figure \ref{fig:z_mstar}, while in Figure \ref{fig:sed_lya} we show the SEDs and cutouts for two  objects randomly chosen among those removed from the sample. The low number of objects excluded from the sample in this step supports the robustness of our photometric redshift measurements.

In our analysis, we considered as default sample the preparatory sample cleaned from AGN and from sources with clear detection in bands bluer than the Lyman limit at the redshift of each source. This sample includes 382 galaxies, which correspond to a fraction of 75\% with respect to the preparatory sample, or 92\% of the AGN-cleaned sample.

\subsection{SED templates: the maximally red SED}
\label{sect:dusty}
\begin{figure} 
\hspace{-0.1cm}\includegraphics[width=8.5cm]{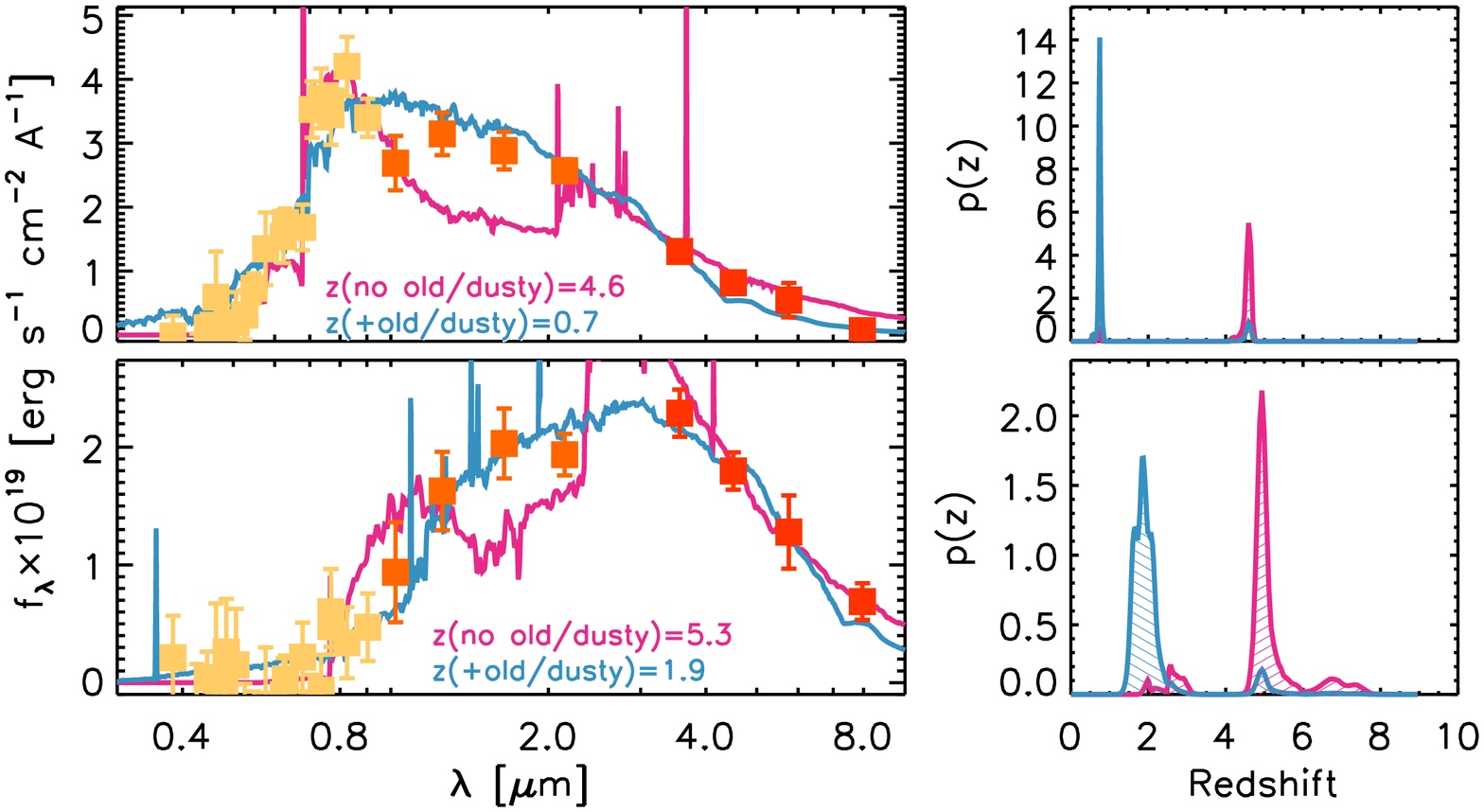} 
\caption{Examples of photometric redshift measurements with and without including the old and dusty template into the template set. Each one of the two rows refers to a distinct object, randomly chosen among those whose photometric redshift was $z>4$ without the inclusion of the old and dusty template, but became $z<4$ after including it in the template set.  For each object, the panel on the left shows the measured photometry in observer frame (colored points) together with the high-redshift and low-redshift best-fit templates (magenta and blue curves respectively). The panel on the right shown the probability distribution of the redshift ($p(z)$) for the two cases of no old and dusty template (magenta region) and with the old and dusty template (blue region). The introduction of the dusty template favors the lower redshift solutions. \label{fig:sed_dusty} }
\end{figure}

We investigated the {systematic} effects on the measurement of photometric redshifts of the inclusion of a \emph{maximally red} template, i. e. a passively evolving 1.5~Gyr old and dusty ($A_V=2.5$ mag)  galaxy from \citet{muzzin2013a}. A non-negligible fraction of objects whose redshift was measured to be at $z>4$ with the default template set, have $z<4$ when the maximally red template is introduced. This is graphically shown in panel d) of Figure \ref{fig:z_mstar}. The grey filled circles mark the sample of pure $z>4$ galaxies, from the previous section, while the blue circles identify the objects whose redshifts have been computed with the template set containing the maximally red template and whose stellar masses have been recomputed according to the new photometric redshifts. The fraction of objects which have $z>4$ after the introduction of the old and dusty template with respect to the default $z>4$ population is 81\% (61\% when considering the initial sample). The SED and $p(z)$ of two objects randomly chosen from the objects with $z>4$ without the dusty template but with $z<4$ when the old and dusty template is introduced, are shown in Figure \ref{fig:sed_dusty}. The introduction of the dusty template generally provides better fits to the data, in agreement with what found in e.g., \citet{muzzin2013a}.

\subsection{Effect of bayesian luminosity prior}
\label{sect:prior}
The core of most photometric redshifts codes is a $\chi^2$ minimization on the observed photometry of fluxes from a set of templates. However there are degenerate cases where the minimisation process alone can lead to an incorrect value. The most typical case is the inability, usually for faint objects, to distinguish if a break observed in the photometry is the result of the Balmer/4000\AA~break or of the Lyman break at 912\AA. In order to provide a way to remove such degeneracies, \citet{benitez2000} introduced the bayesian analysis to the measurements of photometric redshifts. The idea is to complement the flux measurements with information such as the distribution of apparent magnitude of objects with redshift. This technique has proven to be reliable at $z<4-5$. However, the prior is generally built on the basis of either properties of galaxies at lower redshift or from semi-analytic models (as is the case for EAzY). This makes its robustness questionable when it comes to measure photometric redshifts at higher z. For this reason, we considered the introduction of the bayesian prior in the measurements of the photometric redshift as a systematic effect. For the subsample selected from the UltraVISTA catalog, we needed to extend up to $z=10$ the original EAzY K-band prior, which reaches $z=7$. This was done by fitting the standard EAzY prior with a functional form $\Pi(z)\propto z^\gamma\times\exp[-(z/z_0)^\gamma]$, with $\gamma$ and $z_0$ free parameters \citep{benitez2000,brammer2008}  and extrapolating its values to $z=10$. 

By construction the sample recovered from the IRAC residual images is characterized by  very faint (when not absent) fluxes in the K-band, making unsuitable the adoption of the prior in the K-band.  We therefore built a prior for the IRAC $3.6\mu$m band. This was achieved  by fitting the same functional form as before to the distribution of redshifts in bins of  apparent magnitudes for objects from the simulation data by \citet{henriques2012}, based on the semi-analytic model presented in \citet{guo2011}, and accessed through the Virgo - Millennium database (\citealt{lemson2006}, \citealt{springel2005}).

\begin{figure} 
\hspace{-0.1cm}\includegraphics[width=8.5cm]{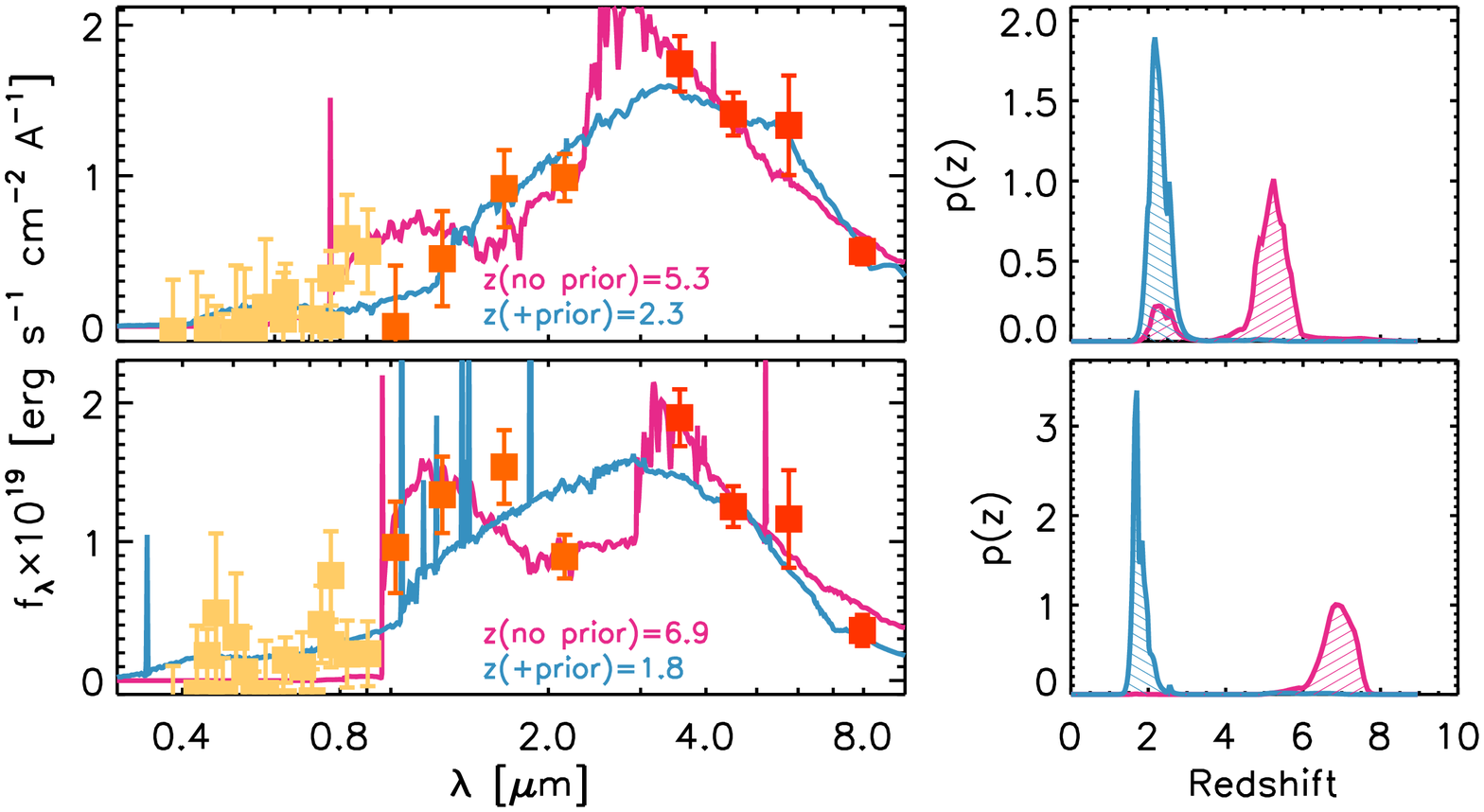} 
\caption{Similar to Figure \ref{fig:sed_dusty} but for the case of excluding (magenta curves and regions) or including (blue curves and regions) the bayesian luminosity prior. The bayesian luminosity prior favours the low-redshift solution. \label{fig:sed_prior} }
\end{figure}

The activation of the prior results in 52 residual objects at redshift $z>4$, i.e., 83\% of the objects from the sample at $z>4$ after the introduction of the old and dusty template have a redshift $z<4$ when the bayesian prior is also activated (the percentage becomes 90\% when comparing the 52 objects to the 502 objects in the initial sample). The introduction of the prior has then the largest systematic effect, among those considered in this work, on the sample selection. A graphical representation of this substantial selection effect is presented in panel e) of Figure \ref{fig:z_mstar}, while  the SED and $p(z)$ of two objects randomly chosen from the objects with $z>4$ without applying the bayesian  luminosity prior but with $z<4$ when the bayesian  luminosity prior is introduced, are shown in Figure \ref{fig:sed_prior}.

We would like to note here that, since the prior is based on semi-analytic models, which are still very uncertain in the redshift range considered in this work, the sample of $z>4$ galaxies selected using photometric redshifts obtained with the adoption of the prior should be considered as one of the sources of systematic effect.

\begin{figure*}
\hspace{-1.5cm}\includegraphics[width=19cm]{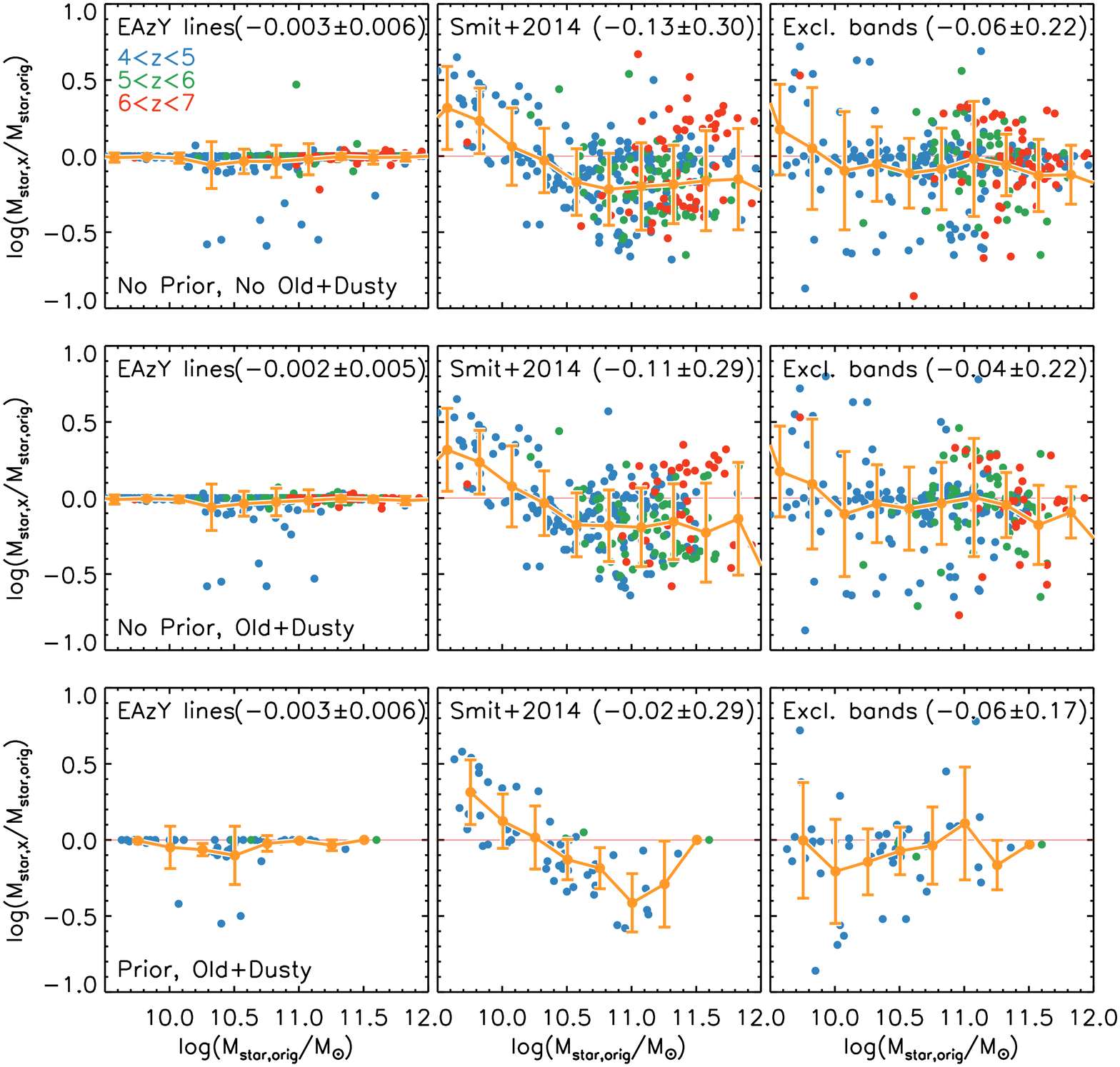} 
\caption{Excess in stellar mass measured after correcting the photometry for nebular emission lines. The top row refers to photometric redshifts computed without the old and dusty template and with no bayesian luminosity prior, the central row to the inclusion of the old and dusty template but not applying the luminosity prior while the bottom row refers to photometric redshifts obtained with the inclusion of the old and dusty template and with the adoption of the bayesian luminosity prior. Points are color coded according to their redshift, as specified by the legend in the top-left panel. These three configurations correspond to panels (c), (d) and (e) of Figure \ref{fig:z_mstar}. For each row, panels from left to right correspond to one of three methods implemented to correct the photometry from contamination by emission lines. Specifically, the panels on the left refer to the EW computed from EAzY templates; the central panels to the procedure presented in \citet{smit2014}, while the panels on the right refer to the exclusion of the fluxes in those bands possibly contaminated by the nebular lines. The yellow circles and error bars mark the average values together with the associated standard deviation. The robust mean and associated standard deviation are quoted in parentheses in each panel. The largest excess correction is found with \citet{smit2014} method, while the exclusion of the contaminated bands has the effect of introducing a scatter in the stellar mass measurement, although the average excess is nearly zero. The same behavior is observed for the case of including the old and dusty template and with the bayesian luminosity prior, although the lower number ob objects in the sample affects the statistics measurement. \label{fig:mstar_comp}}
\end{figure*}

\subsection{Nebular line contamination}
\label{sect:neblines}
Recently, a number of works have investigated  how the contamination to broad- and narrow-band photometry by nebular emission lines may introduce a systematic excess in the measurements of the stellar masses (and consequently reduce the sSFRs) of galaxies (see e.g., \citealt{atek2011,shim2011,debarros2014,stark2013,bowler2014,straatman2014,labbe2013b,oesch2013,ellis2013,schenker2013,gonzalez2014}), although a consensus on the effectiveness of this correction, in particular for galaxies at $z>6$, has not yet been reached (see e.g., \citealt{labbe2013b,bowler2014}). Given the above uncertainties in the potential contamination from nebular lines at high redshift, we can not adopt a single and straightforward approach to correct for the nebular line contamination. Instead, the systematic effects that the contamination by nebular lines emission can have on the measurement of the stellar population parameters (in particular of the stellar mass) have been analyzed implementing three independent recipes to correct the observed photometry from potential contamination, with each recipe applied on a per-objects basis. Stellar masses and population parameters were then re-computed, adopting the same configuration used for the default sample.  The result of this process are four different measurements of the stellar mass and population parameters associated to each sample of $z>4$ galaxies: one sample obtained with the original photometry and three new samples each one associated to one of the distinct methods of the nebular line contamination correction.

\begin{figure*}
\begin{tabular}{c}
\hspace{-0.1cm}\includegraphics[width=17cm]{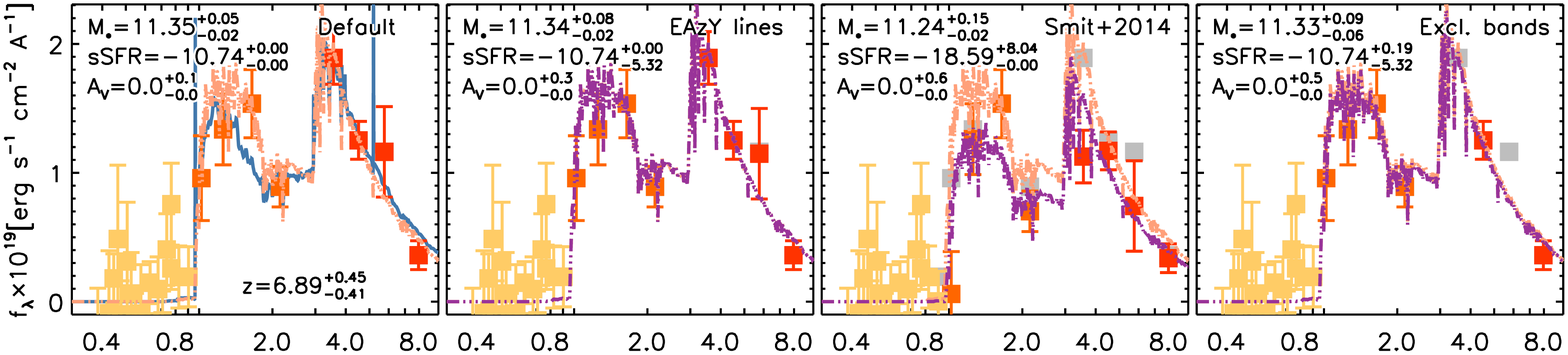}  \\
 \hspace{-0.1cm}\includegraphics[width=17cm]{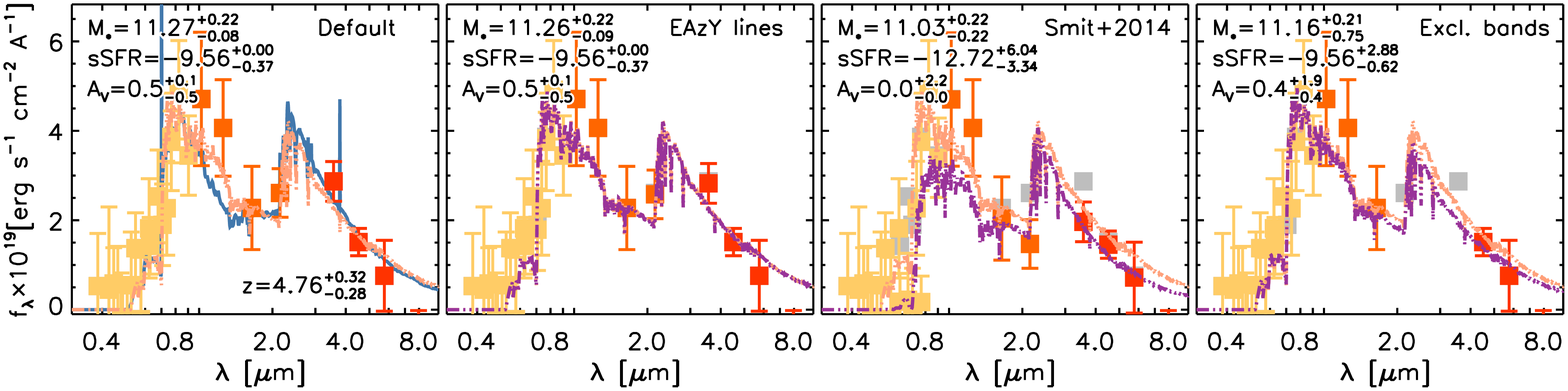}
\end{tabular}
\caption{Examples of stellar population parameters recovered from SED fitting on the photometric catalog after correction from nebular lines emission. The two rows refer each to a different object. In each panel, the photometry (in the observer frame) adopted for the fitting is represented by the colored squares; grey squares mark the photometry before applying the corrections. The pink curves represent the best-fit templates from FAST without applying any correction for emission line contamination; the violet curves in the remaining panels represent the best-fit template from FAST for each method implemented to correct for emission lines contamination. The best-fit template for the no-correction scenario is reported in each panel for comparison (pink curve). The blue curve in the leftmost panel marks the best-fit SED from EAzY. The main physical properties are listed at the top-left corner, together with the 68\% confidence level uncertainties (see Figure \ref{fig:sed_AGN} for further details). Left to right, the panels refer to the cases of original data, photometry corrected from the EW of lines in the best-fit EAzY template, photometry corrected following the procedure in \citet{smit2014}; and excluding from the photometry those bands being potentially contaminated by nebular emission. \label{fig:sed_neblines} }
\end{figure*}

In the first method, for each object we identified those bands which could be contaminated by the (redshifted) main nebular emission lines (Ly${\alpha}$, H$\alpha$, H$\beta$, [O \textsc{II}], [O \textsc{III}]). The potential contribution was computed from the line equivalent width (EW) recovered from the best-fit EAzY template, and the corresponding flux was then rescaled by the factor $EW\times R(\lambda_{obs})/(\lambda_{obs}\int R(\lambda)/\lambda d\lambda)$, where $R(\lambda)$ is the filter efficiency and $\lambda_{obs}$ is the redshifted wavelength of the emission line (see e.g., Eq. 1 in \citealt{smit2014}). The emission lines in the EAzY templates are tuned for objects at redshifts $z< 3$, resulting in EW smaller than those observed at higher redshift. This method reflects then an optimistic scenario, as the contamination by nebular lines is likely larger than the values recovered through this configuration.

\begin{figure*} 
\begin{tabular}{cc}
\hspace{-0.1cm}\includegraphics[width=8.5cm]{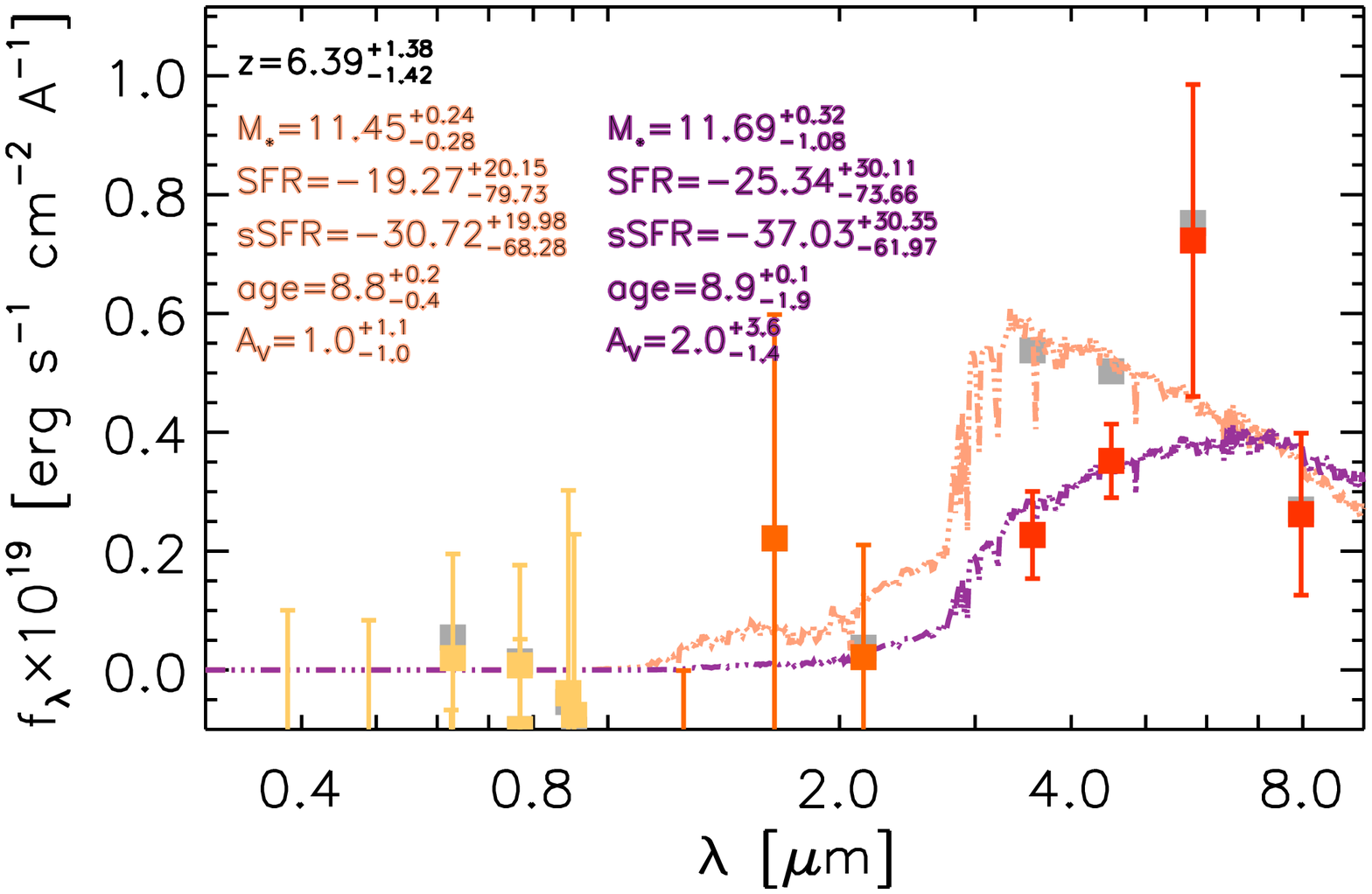} & \hspace{-0.1cm}\includegraphics[width=8.5cm]{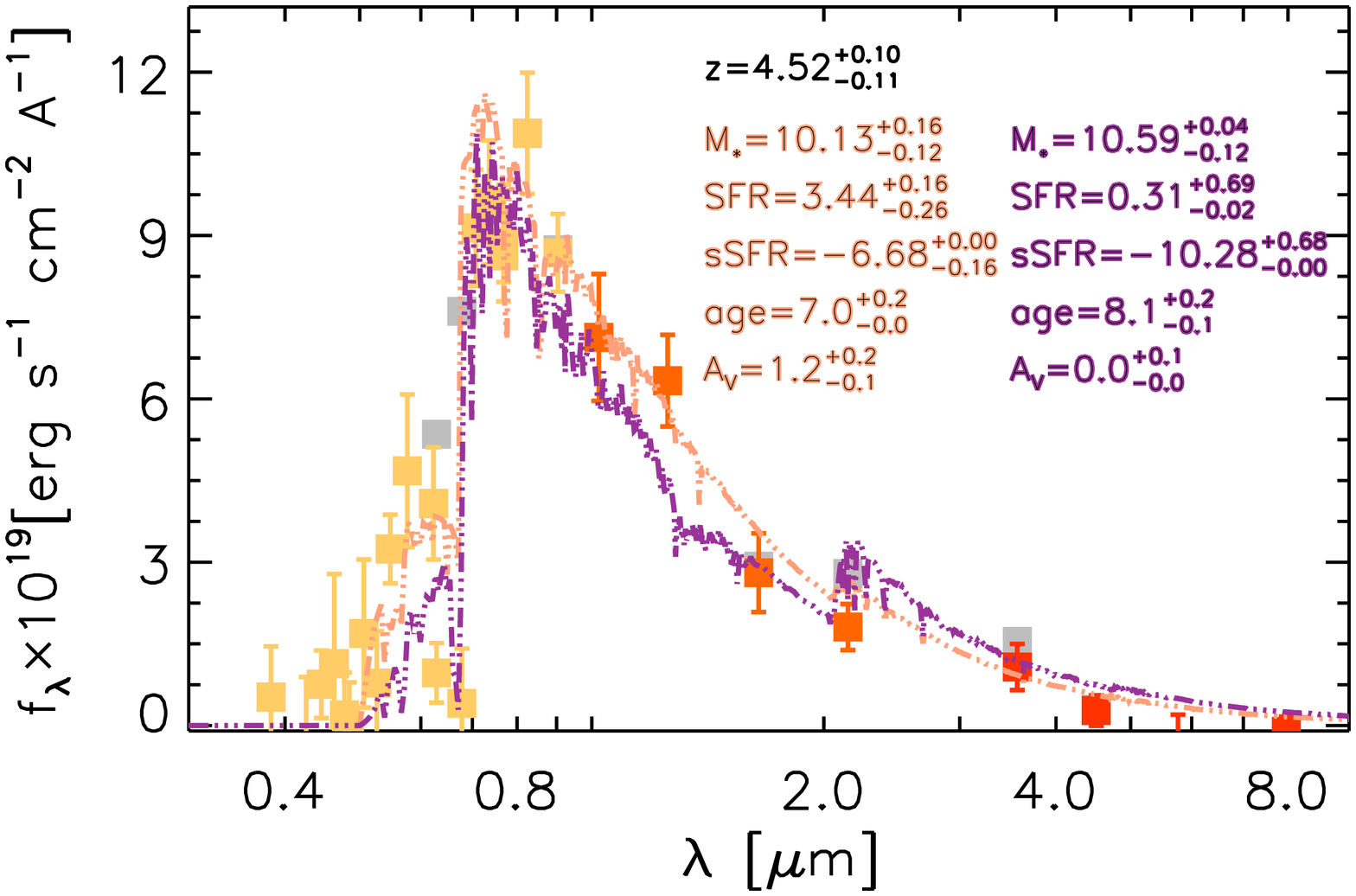}
\end{tabular}
\caption{ Left panel: example of $\log(M_*/M_\odot)>11$ object for which the \citet{smit2014} method for the correction of nebular emission contamination introduces an \emph{increase} in stellar mass. The colored points correspond to the photometry (in the observer frame)  after applying the correction of contamination. The original measurements are shown as filled grey squares. The best-fit FAST solution adopting the original photometry is shown by the pink curve, while the best-fit solution with the corrected photometry is represented by the violet curve. The main stellar population parameters are also reported, with the text colour matching the model they refer to. Specifically, top to bottom they are: the photometric redshift ($z$), the  $\log(M_*/M_\odot)$, the $\log($SFR) in units of $\log(M_\odot/\mathrm{yr})$, the $\log(\mathrm{sSFR/yr}^{-1})$, the $\log(\mathrm{age/yr})$ and the extinction expressed in magnitudes. Quoted errors refer to the 68\% confidence intervals.  Right panel: example of $\log(M_*/M_\odot)<10.5$ object for which the \citet{smit2014} method for the correction of nebular emission contamination introduces an increase in stellar mass. Same plotting conventions as for the left panel. The increase in stellar mass is due to a significant increase in either the dust extinction or the age. \label{fig:mstar_excess} }
\end{figure*}

Recent works (see e.g., \citealt{holden2014,smit2014}) have shown that $z>3$ star-forming galaxies are commonly characterized by high line ratios and large equivalent width which can even more bias the measurements of stellar masses. Specifically, \citet{smit2014} measured the rest-frame EW([O \textsc{III}]+H$\beta)$ for a sample of galaxies at $z\sim6.8$, finding a lower limit value of 637\AA~for the full sample, and an EW=1582\AA~for the bluest sources. These values were shown to be consistent with the extrapolation to $z\sim 6.8$ of the $z\sim 2.5$ EW(H$\alpha)$ measured   by \citet{fumagalli2012}, assuming an evolution of the EW as $(1+z)^{1.8}$ and converting the EW(H$\alpha)$ into EW([O \textsc{III}]+H$\beta)$ using the line intensity ratios in \citet{anders2003} for $Z=0.02Z\odot$. Following \citet{smit2014}, our second method was then implemented as follows. At first we computed the rest-frame EW([O \textsc{III}]+H$\beta)$ at the redshift observed for each galaxy  applying the $(1+z)^{1.8}$ evolution to the EW([O \textsc{III}]+H$\beta)$ measured by \citet{smit2014} at $z\sim6.8$. We adopted for the rest-frame EW of [O \textsc{III}]+H$\beta$ at $z=6.8$ the value EW([O \textsc{III}]+H$\beta)$=1582\AA, which coincides with the higher EW from \citet{smit2014}, as a way to consider the highest nebular emission contamination still consistent with observations. The EW of all the nebular emission lines in Table 1 of \citet{anders2003} where then computed adopting the line intensity ratios corresponding to $Z=0.02Z\odot$. The new flux was finally computed from Eq. (1) in \citet{smit2014}.

Our third method consisted in removing from the photometric catalog the fluxes in those bands which could be contaminated by the same nebular emission lines considered for the first method. This method is driven by the idea of not imposing any constraint on the contribution of the nebular lines in each photometric band.

The effects of the above three recipes on the stellar mass measurements are summarized in Figure \ref{fig:mstar_comp}, while in Figure \ref{fig:sed_neblines} we present examples of SED fitting to recover stellar population parameters before and after correcting the photometry for nebular emission contamination. For the default sample, with the first approach we find an excess in stellar mass  with respect to the stellar masses obtained without applying any correction for nebular line contamination, with biweight mean value of $0.003\pm0.006$ dex for the full $z>4$ sample ($0.003\pm0.006$ dex, $0.002\pm0.004$ dex  and $0.004\pm0.11$ dex for the stellar masses in the $z\sim4.5$, $z\sim5.4$ and $z\sim6.5$ bins respectively). These values are approximately one order of magnitude smaller than previous determinations. Specifically, \citet{schenker2013} report an excess of 0.2~dex for galaxies at $z\sim3.5$; at $z\sim4$ \citet{gonzalez2014} find a \emph{marginal correction}, while \citet{stark2013} report an excess of 0.04~dex; at $z\sim5$ the excess by \citet{gonzalez2014} is marginal, while \citet{stark2013} find 0.1~dex; at $z\sim6$ both \citet{stark2013} and \citet{gonzalez2014} find an excess of about 0.26~dex; finally an excess of 0.48~dex is found in \citet{labbe2013b} for $7<z<8$ galaxies. However, the small values found for the excesses in stellar mass are consistent with the working hypothesis that the EW of nebular emission lines in the EAzY templates are smaller than those observed for high-redshift galaxies.

The biweight mean excess in stellar mass from the \citet{smit2014} method is $0.13\pm0.30$~dex for sources at $z>4$ ($0.06\pm0.26$~dex, $0.19\pm0.19$~dex and $0.05\pm0.29$~dex in the three redshift bin respectively) which appear to be more consistent with the literature, although they are also consistent with no excess. We note that there is a group of galaxies which experienced an \emph{increase} in stellar mass, rather than a decrease. For stellar mass $\log(M_*/M_\odot)>11$, these are almost entirely objects that were detected on IRAC residual images and with redshift $6<z<7$. An example, randomly extracted from the sample, in shown in the left panel of Figure \ref{fig:mstar_excess}. The SEDs of these objects are characterized by absence of flux in the optical bands and, by construction, little to no flux also in the UltraVISTA NIR bands. At these redshifts, the H$\alpha$+N \textsc{II} and H$\beta$+[O \textsc{III}] enter the IRAC $3.6\mu$m and $4.5\mu$m bands. Under our working hypotheses, the EW associated to these nebular lines are large, resulting in a large correction factors in these two bands. Since the redshift is not re-computed after the correction is applied, the stellar population parameters are best fitted by a template with higher dust extinction, which translates into higher stellar masses.

A different effect seems to be responsible most often for the larger stellar masses estimated when adopting the \citet{smit2014} method to correct for emission line contamination in galaxies with $\log(M_*/M_\odot)<10.5$. An example is shown in the right panel of Figure \ref{fig:mstar_excess}. In this case, the galaxies are at $4<z<5$ and are characterized by extreme SFR (SFR $>1000$) and an extinction of about 1~mag. Here, the strong contribution from the Lyman-$\alpha$ line in our model assumptions translates into large EW for the intermediate optical bands, resulting in low flux (generally consistent with 0 erg s$^{-1}$ cm$^{-2}$ \AA$^{-1}$) in the bands potentially contaminated by Lyman-$\alpha$ emission, after the correction is applied. The resulting best-fit template is then characterized by a strong Ly$\alpha$ absorption line, which excludes the high-star-formation solution. The observed red color is then explained by the resulting older population of stars, rather than the effect of dust extinction, which translates into lower SFR and $A_V$ and higher age. 

\begin{figure}
\hspace{-0.1cm}\includegraphics[width=8.5cm]{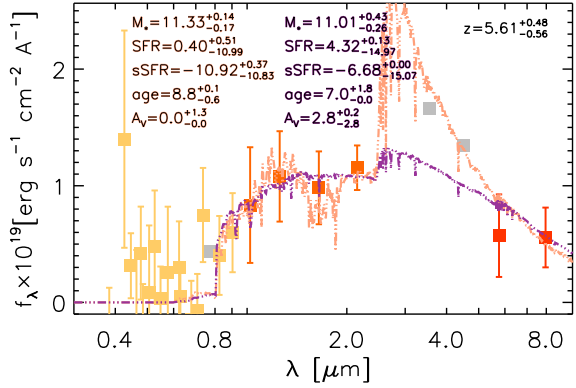}
\caption{Example of the effect of measuring the stellar population parameters excluding the bands possibly contaminated by nebular emission. Same plotting conventions as for Figure \ref{fig:mstar_excess}. The exclusion of the fluxes in the IRAC $3.6\mu$m and $4.5\mu$m bands and the fact that the flux in K-band is higher than the flux in IRAC $5.8\mu$m disfavors solutions with pronounced Balmer break. This results in best-fit templates with extremely high SFR.\label{fig:nl99} }
\end{figure}

The major effect in the measurements of the stellar mass introduced by our third method is to increase the spread in stellar masses, likely the result of the  lower number of bands available for the fit. Indeed the measured excesses are $0.06\pm0.22$~dex for $z>4$ and $0.04\pm0.12$~dex, $0.03\pm0.20$~dex and $0.02\pm24$~dex in the three redshift bins. The lack of flux information during the fit is particularly critical for objects at $5.4<z<6.5$. In this range of redshift H$\alpha$ and H$\beta$ enter IRAC $3.6\mu$m and $4.5\mu$m bands. The simultaneous exclusion of these two bands from the fitting process introduces a higher degree of freedom in the choice of the best-fitting template. For those sources with a K-band flux value larger  (in $f_\lambda$) than the flux value in IRAC $5.8\mu$m and a plateau in the observer-frame NIR wavelength shorter than $2.5\mu$m, the fitting code favors a solution without a strong Balmer break and with high values of dust extinction. The combination of these two effects generates best-fit templates characterized by a very unlikely high SFRs, with common values about  few$\times10^4M_\odot$yr$^{-1}$ (see Figure \ref{fig:nl99} for an example). If, instead, the observer-frame NIR region presents flux decreasing with wavelength, the best-fit template does not show such extreme values of SFR and $A_V$  (see e.g., the lower-right panel of Figure \ref{fig:sed_neblines}). The effectiveness of the results from this third prescription were then considered on a per-case basis.

When the three methods are applied to the sample obtained from the introduction of the old and dusty template and to the sample with both the old and dusty template and applying the bayesian luminosity prior, we find average excesses of $0.002\pm0.005$, $0.11\pm0.29$ and $0.04\pm0.22$ and $0.003\pm0.006$, $0.02\pm0.29$ and $0.06\pm0.17$ respectively for the three methods, in qualitative agreement with the results found for the default sample. The method introducing the largest excess in stellar mass is that from \citet{smit2014}.

The contribution of the nebular emission lines to the stellar mass measurements derived with the above scenarios are to be intended in a statistical sense. A quantitative and accurate determination of the contamination by emission lines for galaxies at $z>3$ would require spectroscopic studies on large samples, which is currently missing.

\subsection{Star formation histories}
\label{sect:sfh}
\begin{figure}
\includegraphics[width=8.5cm]{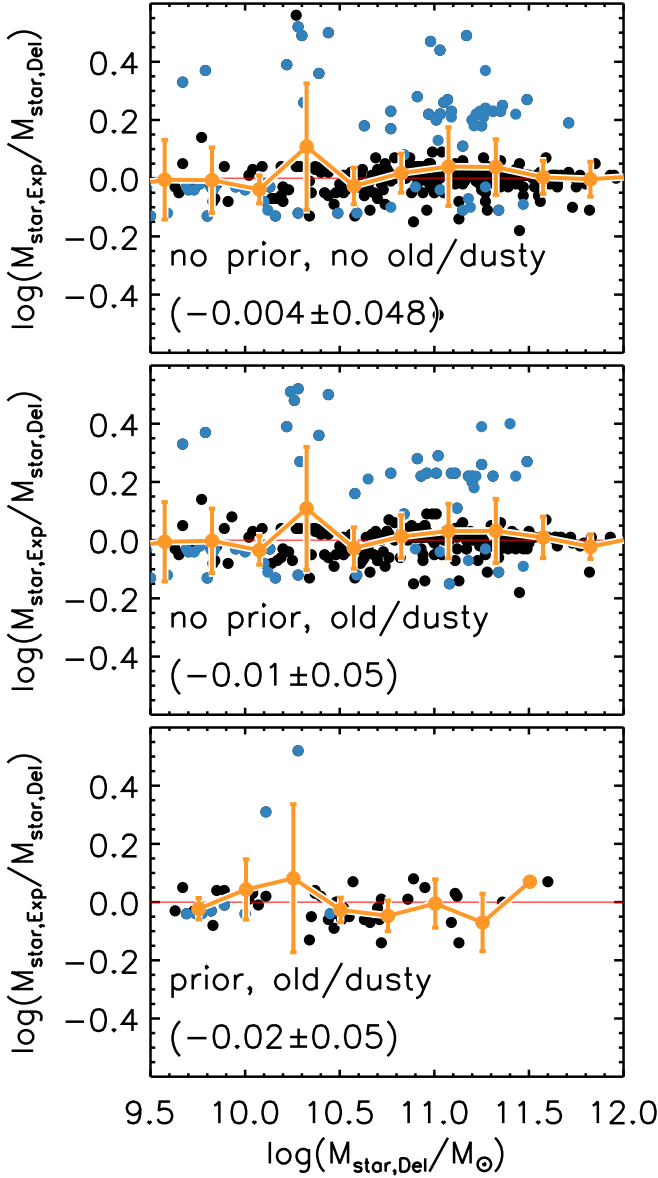}
\caption{Excess in stellar mass adopting the exponential SFH compared to the stellar mass from the delayed-exponential SFH. The top panel refers to photometric redshifts computed without the old and dusty template and with no bayesian luminosity prior, the central panel to the inclusion of the old and dusty template but not applying the prior while the bottom panel refers to photometric redshifts obtained with the inclusion of the old and dusty template and with the adoption of the bayesian luminosity prior. The yellow circles and error bars mark the average values together with the associated standard deviation. The blue points identify those galaxies with sSFR$>10^{-7}$yr$^{-1}$. The robust mean and associated standard deviation are quoted in parentheses in each panel. For most of the galaxies and configurations, the different SFHs do not alter significantly the measurement of the stellar mass. In the cases without prior, a few massive galaxies have stellar masses larger by $\sim0.2$dex than the corresponding from the delayed exponential case. \label{fig:sfh_comp}}
\end{figure}

\begin{figure}
\hspace{-0.1cm}\includegraphics[width=8.5cm]{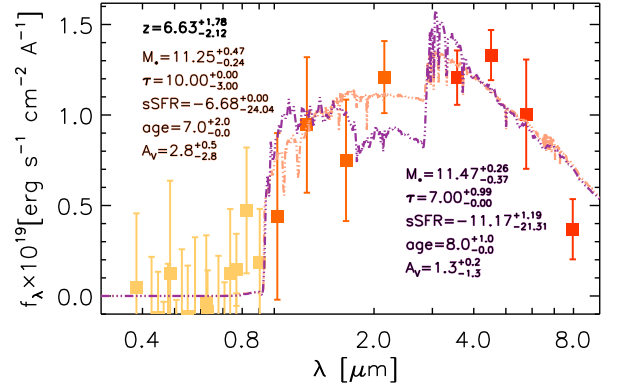}
\caption{Example of objects whose stellar mass obtained assuming an exponential SFH has an excess of $\sim0.2$~dex compared to the stellar mass from the delayed-exponential SFH. The colored points mark the observed photometry in the observer frame; the pink curve marks the best-fit template from the delayed-exponential SFH, while the violet curve represents the bet-fit template from the exponential SFH. Stellar population parameters are also reported in the labels, coded by the color of the corresponding best-fit template. Other plotting conventions as in Figure \ref{fig:mstar_excess}.  The exponential SFH provides an SED with older age and hence a higher stellar mass. \label{fig:exp-delexp} }
\end{figure}

We finally analyzed the systematic effects of adopting different SFHs in the measurement of the stellar population parameters; specifically, in addition to the delayed-exponential SFH, which constitutes out default configuration,stellar population parameters were also computed adopting an exponential SFH. The comparison between the stellar masses from the two SFHs is shown in Figure \ref{fig:sfh_comp}. For most of the galaxies, we observe no systematic offset in the stellar mass measurements. We can identify, however, a small sample of galaxies characterized by a stellar mass larger by $\gtrsim0.2$~dex when using the exponential SFH. The observed SED and best-fit templates for both SFHs for one object randomly chosen among those showing the increase in stellar mass when adopting the exponential SFH is shown in Figure \ref{fig:exp-delexp}. The observed SED of these objects are characterized by a plateau in the NIR bands shorter of $\sim2.5\mu$m.  During the fitting process, this plateau can be described by either a very young, highly star-forming galaxy with strong extinction by dust, or by an object with lower dust content and SFR, but older stellar population. The best fit SED template from the delayed-exponential SFH is characterized by a slowly increasing  SFH ($\log(\tau$/yr$)\sim10$) with young age ($\log$(age/yr)$\sim7$), high SFR (SFR$\sim$few$\times10^3 M_\odot/$yr$^{-1}$) and high dust extinction ($A_V\gtrsim2$~mag). The exponential SFH, instead, provides a best fit SED with older age ($\log$(age/yr)$\sim8-9$), and, correspondingly, lower dust extinction ($A_V\gtrsim1$~mag), resulting in a larger stellar mass. Although the delayed exponential SFH can, in principle, mimic the same best fit SED of the exponential SFH, as it can be seen from Figure \ref{fig:exp-delexp}, the solution with large $\tau$ is finally preferred as it provides a slightly better fit to the data in the NIR wavelength region.

\subsection{Stellar mass completeness}
\label{sect:smf_completeness}

\begin{figure}
\includegraphics[width=8cm]{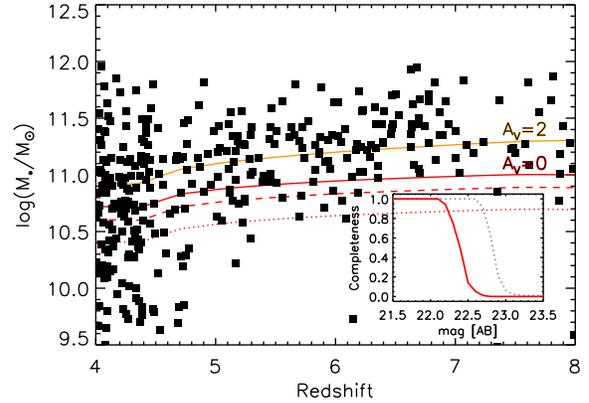} 
\caption{Stellar mass completeness as a function of redshift: the black points represent the $z>4$ galaxies from the sample obtained without bayesian luminosity prior. The red solid, dashed and dotted lines respectively mark the 90\%, 50\% and 5\% completeness limits for a passively evolving galaxy with $A_V=0$, obtained considering the corresponding detection level in IRAC $3.6\mu$m, as described in the text, while the solid yellow line marks the 90\% completeness limit in IRAC $4.5\mu$m with an additional absorption of $A_V=2$~mag, and corresponds to the limit in stellar mass we adopted in this work. The inset shows the detection completeness as a function of apparent magnitude for a point source in IRAC $3.6\mu$m (grey dotted curve) and IRAC $4.5\mu$m (red solid curve), with 90\% confidence levels equal to 23.4~mag and 22.9~mag respectively. The depth of the IRAC maps allows us to consider only galaxies with $\log(M_*/M_\odot)\gtrsim 11.3$. \label{fig:mstar_compl}}
\end{figure}

In Figure \ref{fig:mstar_compl} we mark with the solid red curve the completeness in stellar mass corresponding to the evolution from $z=20$ of a simple stellar population (SSP) model with stellar mass corresponding to that computed from the 90\% completeness limit in IRAC $3.6\mu$m. However, as shown by the inset, the 90\% completeness limit in IRAC $4.5\mu$m is fainter by 0.5~mag than in IRAC $3.6\mu$m (IRAC $3.6\mu$m and $4.5\mu$m 90\% completeness limits are 23.4~mag and 22.9~mag respectively). Furthermore, recently there has been evidence of significant dust extinction at the high-mass end of $z\gtrsim2$ galaxies (see e.g., \citealt{whitaker2010,marchesini2010,marchesini2014}). Because of these two factors, we adopt for the completeness limit the orange solid curve in Figure \ref{fig:mstar_compl}, which includes the effect of 2~mag of extinction in the V-band to the completeness from IRAC $4.5\mu$m.  As it is shown in Figure \ref{fig:mstar_compl}, the current depth of the IRAC coverage to the COSMOS field only allows for $\log(M_{*}/M_\odot) > 11.10, 11.20$ and $11.26$ galaxies at  $4<z<5, 5<z<6$ and $6<z<7$ respectively, corresponding to the high-mass end of the SMF.

\begin{figure*}
\begin{tabular}{c}
\hspace{-0.6cm} \includegraphics[width=17cm]{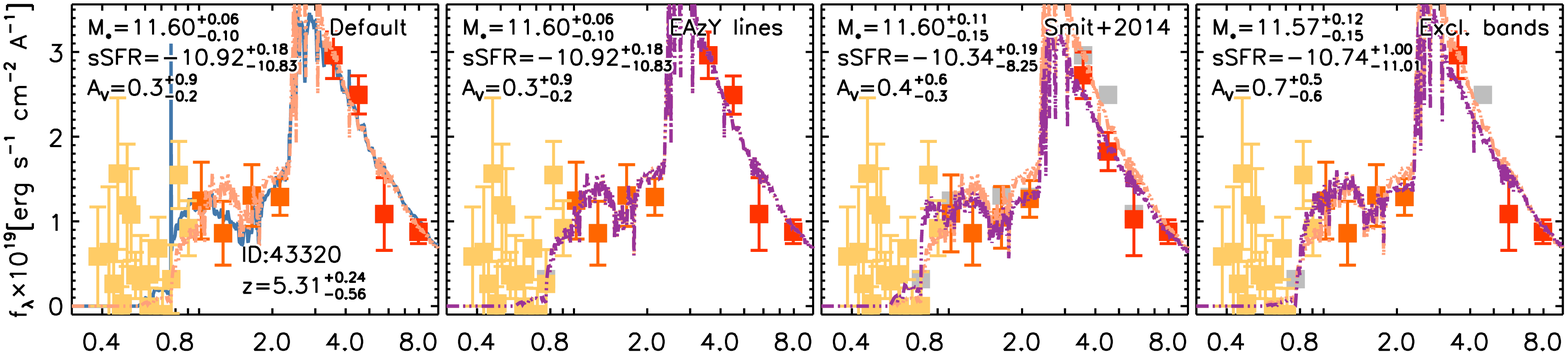} \\
\hspace{-0.6cm}\includegraphics[width=17cm]{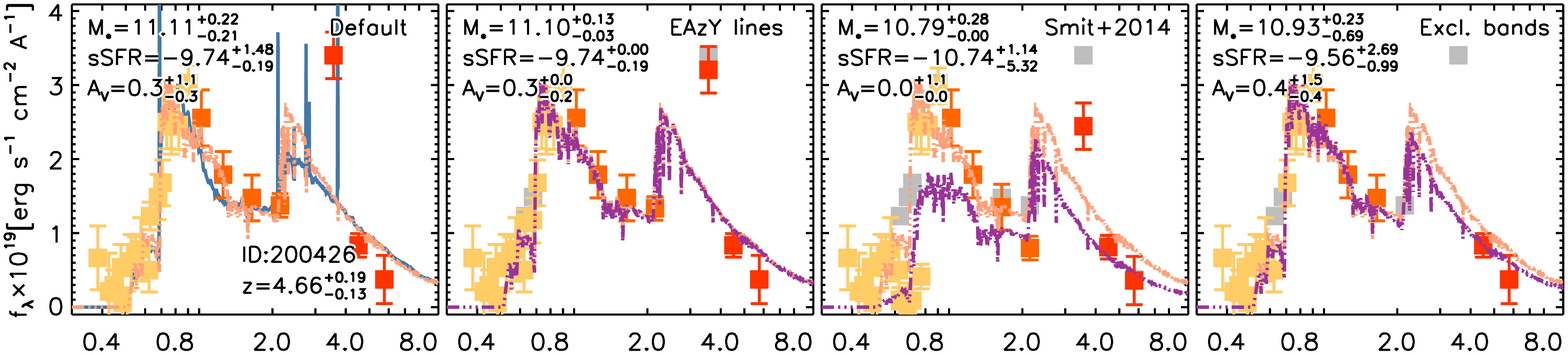} \\
\hspace{-0.6cm}\includegraphics[width=17cm]{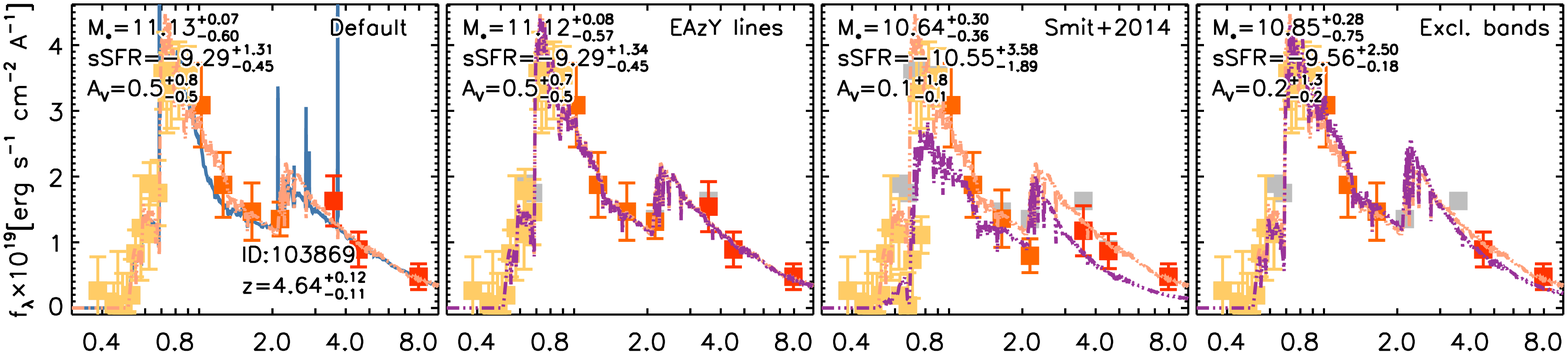} \\
\hspace{-0.6cm}\includegraphics[width=17cm]{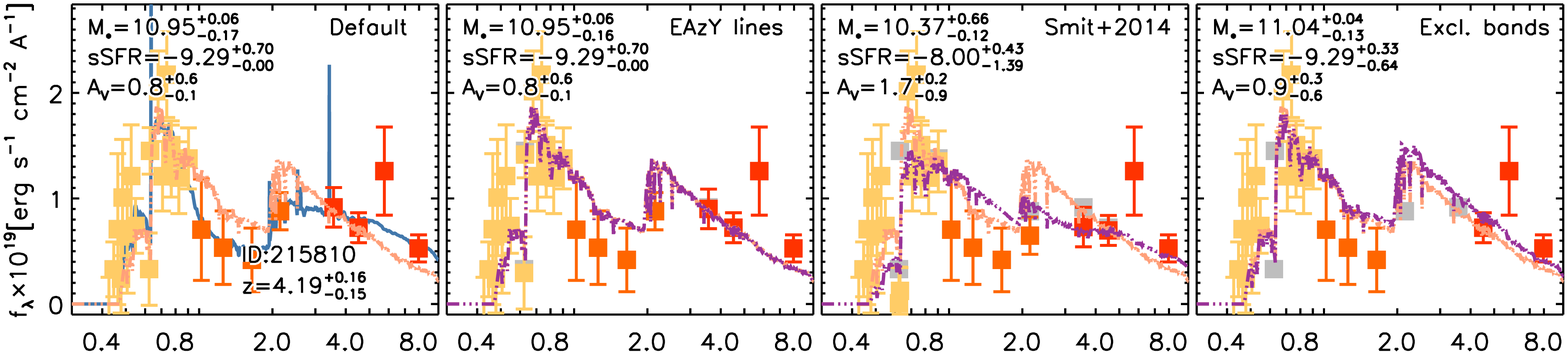} \\
\hspace{-0.6cm}\includegraphics[width=17cm]{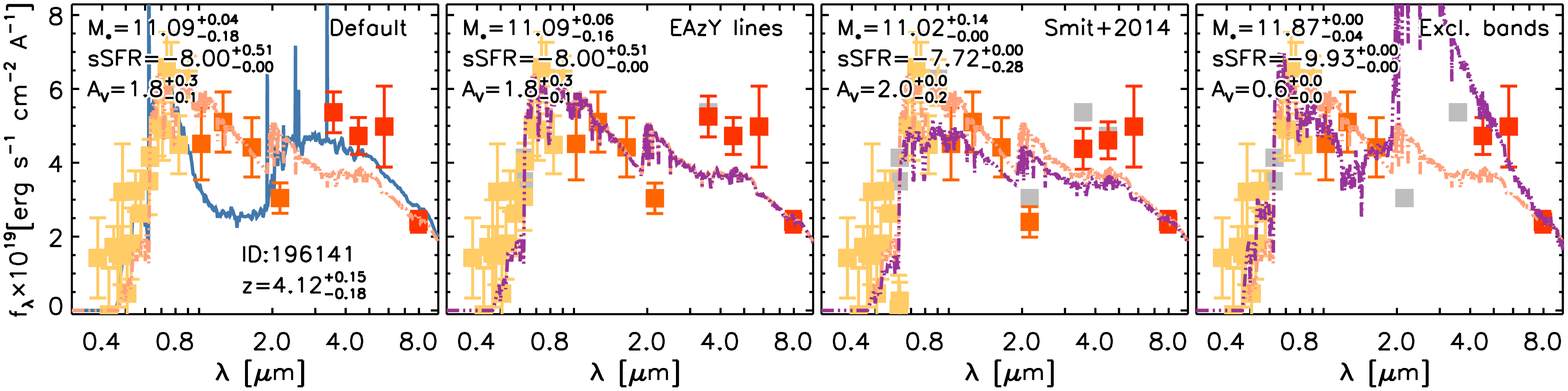} \\
\end{tabular}
\caption{These plots show the SED of the stellar mass complete sample (one object per row) obtained adopting the most conservative configuration, i.e., with the inclusion of the old and dusty template in the set of templates used for the measurement of photometric redshifts, and with the application of the bayesian luminosity prior in the measurement of photometric redshifts. Each panel refers to different fluxes adopted for the measurement of the stellar population parameters, as explained in Figure \ref{fig:sed_neblines}.   \label{fig:z4_SEDs}}
\end{figure*}

\addtocounter{figure}{-1}

\begin{figure*} 
%\ContinuedFloat
\begin{tabular}{c}
\hspace{-0.6cm}\includegraphics[width=17cm]{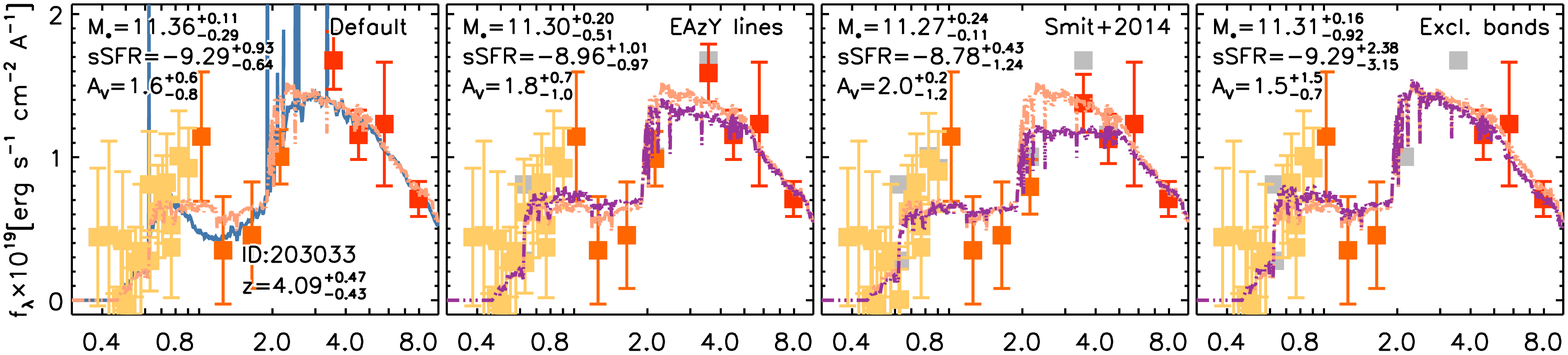} \\
\hspace{-0.6cm}\includegraphics[width=17cm]{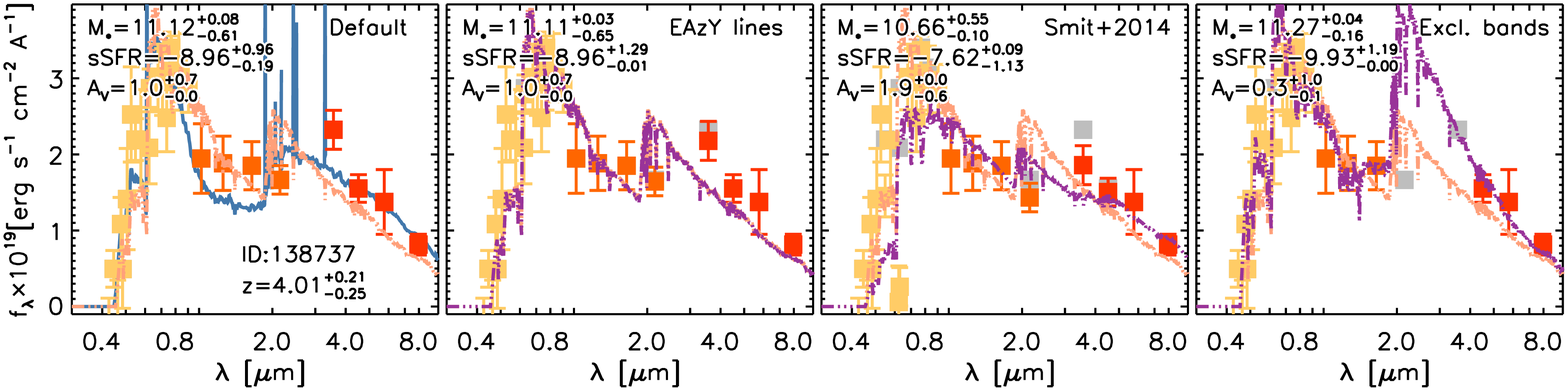} \\
\end{tabular}
\caption{ - Continued}
\end{figure*}

\subsection{Cosmic Variance}
\label{sect:cv}
The total area covered by COSMOS/UltraVISTA is approximately $1.5$ square degrees in one single field, making the effects of cosmic variance not negligible for very massive galaxies.  The contribution of cosmic variance to the total error budget was estimated through the recipe of \citet{moster2011}. The average uncertainties due to this effect are 21\%, 28\% and 37\%  for the $4<z<5$, $5<z<6$ and $6<z<7$ redshift bins respectively for stellar masses $\log(M_*/M_\odot)=11.25$. These values were added in quadrature to the poissonian error of the SMFs.

\begin{figure*}
\includegraphics[width=17cm]{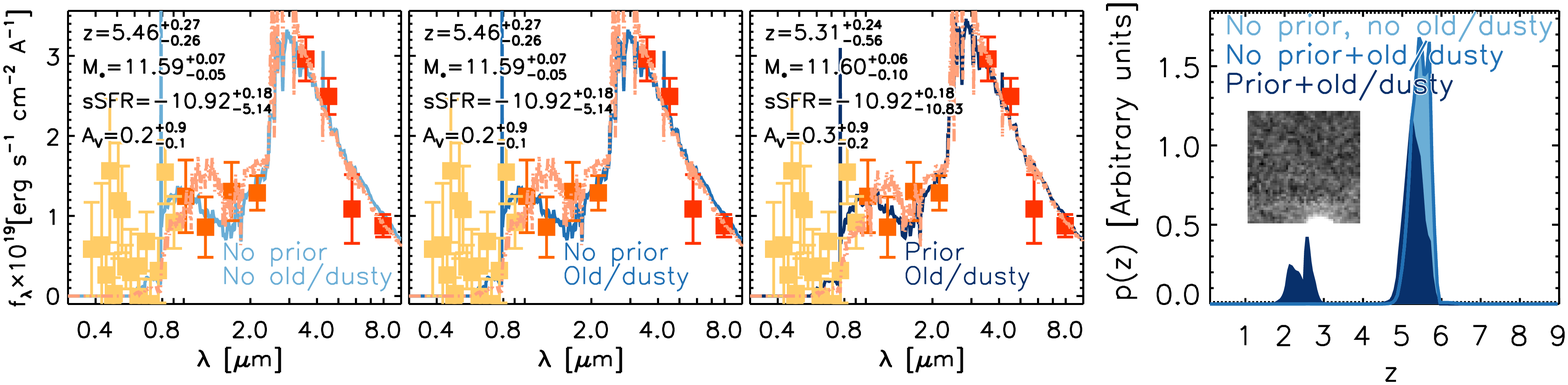}
\vspace{0.4cm}
\caption{The robust massive $z\sim5.4$ galaxy (corresponding to ID 43320 in Figure \ref{fig:z4_SEDs}). The three panels on the left show the SED from the measured photometry in the observer frame (colored points with error bars), together with the best-fit template from EAzY (blue curve) and from FAST (pink curve), this latter obtained without applying any correction for nebular emission contamination to the photometry. The main physical properties are listed at the top-left corner (see Figure \ref{fig:sed_AGN} for further details). Each one of the three panels refers to a different configuration adopted for the measurement of the photometric redshift. Left to right, these represent the cases of: excluding the bayesian luminosity prior and the old and dusty template; excluding the bayesian luminosity prior, but introducing the old and dusty template among the set of SED templates adopted for the photometric redshift measurements;  activating the bayesian luminosity prior and including the old and dusty template. The panel on the right shows the redshift probability distribution $p(z)$ for the three cases. The inset ($7.5''$ wide on each side), centred at the position of the object, shows the results from stacking  the filters bluer than the Lyman limit. The best-fit templates well describe the photometric data.  The $p(z)$ are characterized by a narrow peak at $z\sim5.4$, with a secondary peak at $z\sim2.5$ which appears when the luminosity prior is introduced, but whose probability is $p\sim21\%$. \label{fig:43320}}
\end{figure*}

\section{The population of $\MakeLowercase{z}>4$ galaxies}

\subsection{Robust massive galaxies at $z>4$}
\label{sect:robust_massive}
In this work, we consider as massive a galaxy with $M_*>10^{11}M_\odot$, while quiescent a galaxy whose specific star formation rate (sSFR) is smaller than $1/[3t_H(z)]$, with $t_H(z)$ the Hubble time at redshift $z$ \citep{damen2009,lundgren2014}. Our stellar mass complete sample is then composed by massive galaxies only. However, the different configurations adopted for the computation of redshift and stellar masses produce samples which, in general, do not always contain the same galaxies. As we discussed in Section \ref{sect:photoz}, systematic effects can arise in the measurements of photometric redshifts and/or stellar population parameters depending on the adopted set of SED templates, the inclusion of the bayesian luminosity prior, and the inclusion of the potential contamination from nebular lines.

Our analysis allows us to identify seven robust massive galaxies with redshift $z>4$ measured adopting the most restrictive configuration, i.e., with the adoption of both the bayesian luminosity prior and the maximally red SED template. Their SEDs are presented in Figure \ref{fig:z4_SEDs}. For each object, the panel on the left presents the original photometric points together with the best-fit SED template from EAzY (solid blue curve) and the best-fit FAST template with the default assumptions (dash-dotted pink curve). The three panels on the right show the results of the FAST run on the photometric catalogs obtained after applying the correction from nebular emission lines contamination. We note that, out of the seven objects, only three of them (ID 43320, 196141 and 203033) would be included in the stellar-mass complete sample irrespectively of the assumptions adopted for the measurement of photometric redshifts and stellar population parameters.

The sample is characterized by stellar masses in the range  $10.95<\log(M_*/M_\odot)<11.60 $, specific star-formation rates in the range $-10.92 < \log(\mbox{sSFR/yr}^{-1}) < -8.0$,  extinction in the range $0.3<A_V<1.8$~mag and ages $7.6<\log($age/yr$)<9.0$ (obtained with the default assumptions: no luminosity prior, no old/dusty template and delayed exponential SFH), with median of the log-values of $10^{11.11}M_\odot$, $10^{-9.29}$~yr$^{-1}$ and $10^{9.0}$yr for the stellar mass, sSFR and age, respectively and consistent with what found by \citet{bowler2014}, while the median of the dust extinction is 0.8~mag. Specifically, the high value of the dust extinction is qualitatively in agreement with the trend of increasing $A_V$ with redshift found by e.g., \citet{whitaker2010} and \citet{marchesini2014}. Most of the SEDs are characteristic of star-forming or post-starburst galaxies, and their redshift probability distributions show very pronounced peak, with a small  (where non absent) secondary peak at $z<1$. The redshift probability distribution are well constrained because most of these objects show both the Lyman break and the Balmer/4000\AA~break (although, given the young ages, this second break is mostly originated by the Balmer break).

\subsection{Massive and quiescent galaxies at $z>4$}

The adoption of a set of different configurations for the measurements of photometric redshifts and stellar population parameters, as described in Sect. \ref{sect:photoz}, in principle can produce different values of the physical properties from the same photometry. Specifically, the effect of both the old/dusty SED template and the introduction of the bayesian luminosity prior result in photometric redshifts generally lower than those obtained when excluding the luminosity prior and the old/dusty SED template. 

Despite the above indetermination, we are able to identify one robust candidate for a massive galaxy at $z>4$ irrespectively of the configuration adopted for the measurements of photometric redshift and stellar population parameters, whose sSFR from SED fitting is consistent with being quiescent. Its SED is presented in the top row of Figure \ref{fig:z4_SEDs} labeled as ID 43320; in Figure \ref{fig:43320} we show the best-fit SEDs from EAzY for the three cases of no luminosity prior and no old/dusty template, no  luminosity prior and old/dusty template, and luminosity prior and old/dusty template (blue curves) and corresponding FAST best-fit templates (pink curves), together with the redshift probability distributions for the three above cases (filled regions in the panel on the right).

In the following sections we analyze in more detail its main physical parameters.

% 0) photometry
\subsubsection{Observed SED}
\label{sect:43320_sed}
The photometric SED for this object is presented in the top row of Figure \ref{fig:z4_SEDs}, labeled as ID 43320. The flux in the IA527 filter presents an excess compared to that in the adjacent filters. Although this would be consistent with a Lyman $\alpha$ emission, a visual inspection of the science frame did not show any clear evidence for emission from an object. Instead, the same region is crossed by an horizontal band, possibly an instrumental defect, few pixels wide and characterized by an emission slightly above the local background, which is likely the origin of the measured flux excess.

The  MIPS 24$\mu$m  map shows a clear detection at $12\sigma$ level, indicating either that a solution at $2<z<3$ could be more appropriate or that the rest-frame emission at $\sim3.8\mu$m originates from the torus hot dust heated by a hidden AGN.

As we show in Section \ref{sect:43320_sfr}, the MIPS $24\mu$m emission measured for this source corresponds to high infra-red luminosity. If this excess originated from obscured star formation, the high IR luminosity would make the galaxy detected in Herschel maps. The HerMES SPIRE 250$\mu$m map (\citealt{oliver2012,viero2013}) does not show any significant source centered at the position of this galaxy; according to the HerMES catalog (\citealt{smith2012,roseboom2010,wang2014}), the closest source lays at an apparent distance of about $3''$. However, the large FWHM of SPIRE $250\mu$m (FWHM$_{S250}=18''$) does not allow us to completely rule out the absence of SPIRE 250$\mu$m flux for this galaxy.

% 1) redshift
\subsubsection{Redshift}
In Figure \ref{fig:43320} we show the best-fit SEDs from EAzY for the three cases of no luminosity prior and no old/dusty template, no luminosity prior and old/dusty template, and luminosity prior with old/dusty template (blue curves); the panel on the right shows the redshift probability distributions for the three above cases (filled regions). The galaxy has a photometric redshift consistent with $z=5.39\pm0.08$, depending on the configuration adopted for the measurement. The best-fit SEDs well describe the photometric points, supporting the redshift measurement. In the right panel of Figure \ref{fig:43320} we also show the stack of those bands bluer than the Lyman limit under the assumption that $z=5.4$. The resulting image does not present any clear evidence of flux excess, increasing the confidence on the measured value for the redshift. According to the distribution of redshift resulting from the SED fitting, the probability of this galaxy to be at $2<z<3$ is smaller than 21\% (this upper value corresponds to the introduction of the bayesian  luminosity prior and of the dusty template). However, forcing the redshift to be $z<3$ produces a best-fit solution which is worse than the $z\sim5.4$ one.

% 2) stellar mass
\subsubsection{Stellar mass}

The recovered stellar population parameters vary both because of the different redshift measurement and because of the different recipes we adopted to take into account the potential contamination by nebular lines. Despite this, the stellar mass measurements for this object are all consistent with the value of $\log(M_*/M_\odot)=11.53\pm0.07$. The FAST best-fit SEDs are marked by the pink curves in Figure \ref{fig:43320}. The stellar mass changes only marginally  even when assuming that IRAC $3.6\mu$m and $4.5\mu$m are contaminated by strong emission lines, as from \citet{smit2014} recipe for which the rest-frame EW([O \textsc{III}]+$H\beta$)$\sim$1230\AA~at $z\sim5.4$. We note that the measurements of the stellar population parameters obtained excluding from the fit those bands potentially contaminated by nebular emission when no luminosity prior is adopted for the measurement of photometric redshifts suffer from the issue presented in Section \ref{sect:neblines}  (see also Figure \ref{fig:nl99}): the exclusion of the IRAC $3.6\mu$m and $4.5\mu$m fluxes, together with the fact that the $K$-band flux is higher than the IRAC $5.8\mu$m disfavors a solution with pronounced Balmer/4000\AA~break; the best-fit SED is instead characterized by extremely high SFR (SFR$\sim10^4M_\odot$yr$^{-1}$). The corresponding stellar population parameters were then excluded during the selection process. A definitive assessment of the intrinsic physical properties of this object will necessarily require spectroscopic observations.

%3) SFR + MIPS
\subsubsection{sSFR}
\label{sect:43320_sfr}

The values for the sSFR recovered from the SED  analyses are all consistent with log(sSFR/yr$^{-1}$)=$-10.26\pm0.6$, with $\log(\mbox{age/yr})=8.5\pm0.4$ and an extinction $A_V=0.5\pm0.2$~mag. Similarly to the stellar mass measurement, also the sSFR change marginally under the assumption that IRAC $3.6\mu$m and $4.5\mu$m are contaminated by strong emission lines. At redshift $z=5.4$ the sSFR values satisfies the criterion of sSFR$< 1/[3t_H(z)]$ for the identification of quiescent galaxies. However, the value of the sSFR from the SED fitting analysis is potentially in contrast with the observed MIPS $24\mu$m flux.

Assuming that the observed MIPS 24$\mu$m flux comes entirely from the dust-enshrouded star formation, at the measured redshift, it corresponds to a luminosity $\log(L_{\mbox{\tiny IR}}/L_\odot)=15.0 \pm 0.3$ (adopting the recipe in \citet{wuyts2008} with \citet{dale2002} template set; no significant discrepancy was obtained using the \citet{chary2001} recipe). Using this value of the infrared luminosity, we estimate the star-formation rate to be SFR$_{\mbox{\tiny IR}}= 0.98 \times 10^{-10} L_{\mbox{\tiny IR}} \approx 10^4 M_\odot$yr$^{-1}$ for a \cite{kroupa2001} IMF (\citealt{kennicutt1998,bell2005,muzzin2013a}), an unlikely high physical value. If instead we assume for this galaxy a redshift $z\sim2.5$, roughly corresponding to the secondary peak in the $p(z)$ distribution, its total infra-red luminosity ($L_{\tiny IR} \equiv L_{8\mu m-1000\mu m}$) and the SFR would be typical of hyper-luminous infra-red galaxies (HLIRGs).  Indeed, using the same prescriptions used above, the luminosity recovered from the MIPS 24$\mu$m flux would be $\log(L_{\mbox{\tiny IR}}/L_\odot)=13.5 \pm 0.3$, with a star-formation rate SFR$_{\mbox{\tiny IR}}= 0.98 \times 10^{-10} L_{\mbox{\tiny IR}} \approx 3000 M_\odot$yr$^{-1}$ for a \cite{kroupa2001} IMF.

If the MIPS emission of this galaxy originated from obscured star formation, the high infra-red luminosity at $z\sim2.5$, and most likely even that at $z=5.4$, would make the galaxy detected in Herschel data. However, as discussed in Section \ref{sect:43320_sed}, current HerMES data do not show any clear evidence of flux which could be associated to this object. This fact, together with the unlikely high SFR$_{\mbox{\tiny IR}}$ recovered from the observed MIPS $24\mu$m when the galaxy is considered to be at redshift  $z\sim5.4$, suggest that the observed MIPS flux is likely not originated by an intense SFR, but, instead, by the torus of hot dust heated by a hidden AGN, supporting the SFR values recovered from the SED fitting analysis. However, a robust assessment of the presence of an AGN (and hence of the origin of the MIPS excess) requires spectroscopy, which is still lacking.

\subsection{Stellar Mass Functions}
\label{sect:smf}
We computed the SMFs in the three redshift bins  $4<z<5$ , $5<z<6$ and  $6<z<7$ using the 1/Vmax formalism \citep{schmidt1968,avni1980}, which provides both the shape and the normalization of the SMF at the same time. Upper and lower poissonian uncertainties were computed using the recipe of \citet{gehrels1986}, valid for small samples. Uncertainties due to cosmic variance computed  as described in Sect. \ref{sect:cv} were added in quadrature.  Estimating how the uncertainties in the stellar mass measurements propagate into the computation of the SMF is a non-trivial task since the stellar mass is not directly observed, but it is instead recovered through SED template fitting on multi-wavelength photometry. Similarly to what done in \citet{muzzin2013b}, we then estimated the effects that flux uncertainties have on the measurement of the SMF using Monte Carlo simulations.  We implemented 100 Monte Carlo realizations of the multi-wavelength photometric catalog randomly perturbing the measured fluxes according to the flux uncertainties. Photometric redshifts and stellar population parameters were then re-computed using the same methods applied to the original unperturbed sample. For each one of the 100 catalogs, the SMFs were finally measured. The dispersion of the density values of the SMF in each stellar mass bin then gives an estimate of the uncertainties in the SMF from photometry errors propagating to photometric redshifts and stellar mass measurements. The uncertainties ranged from 8\% to 40\%. These uncertainties were added in quadrature to the uncertainties from poissonian noise and cosmic variance. Furthermore, given the high uncertainties in the dust content of the SED templates adopted for the recovery of the photometric redshift, we caution the reader that the lower stellar mass bin in each redshift range may still suffer from incompleteness.

\begin{figure*}
\begin{tabular}{cc}
\hspace{-0.5cm}\vspace{-0.6cm}\includegraphics[width=9cm]{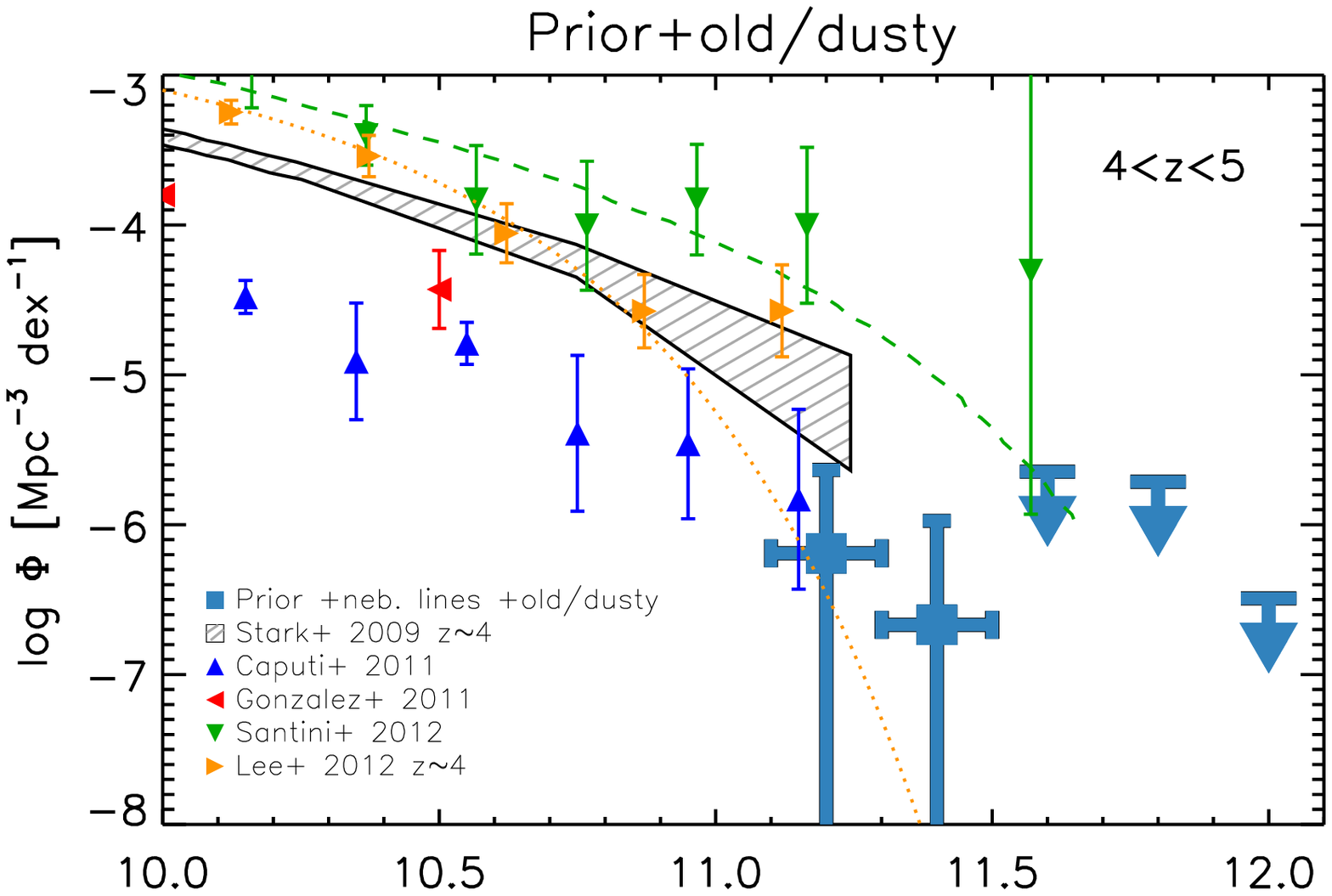} &  \hspace{-0.6cm}\vspace{-0.5cm}\includegraphics[width=9cm]{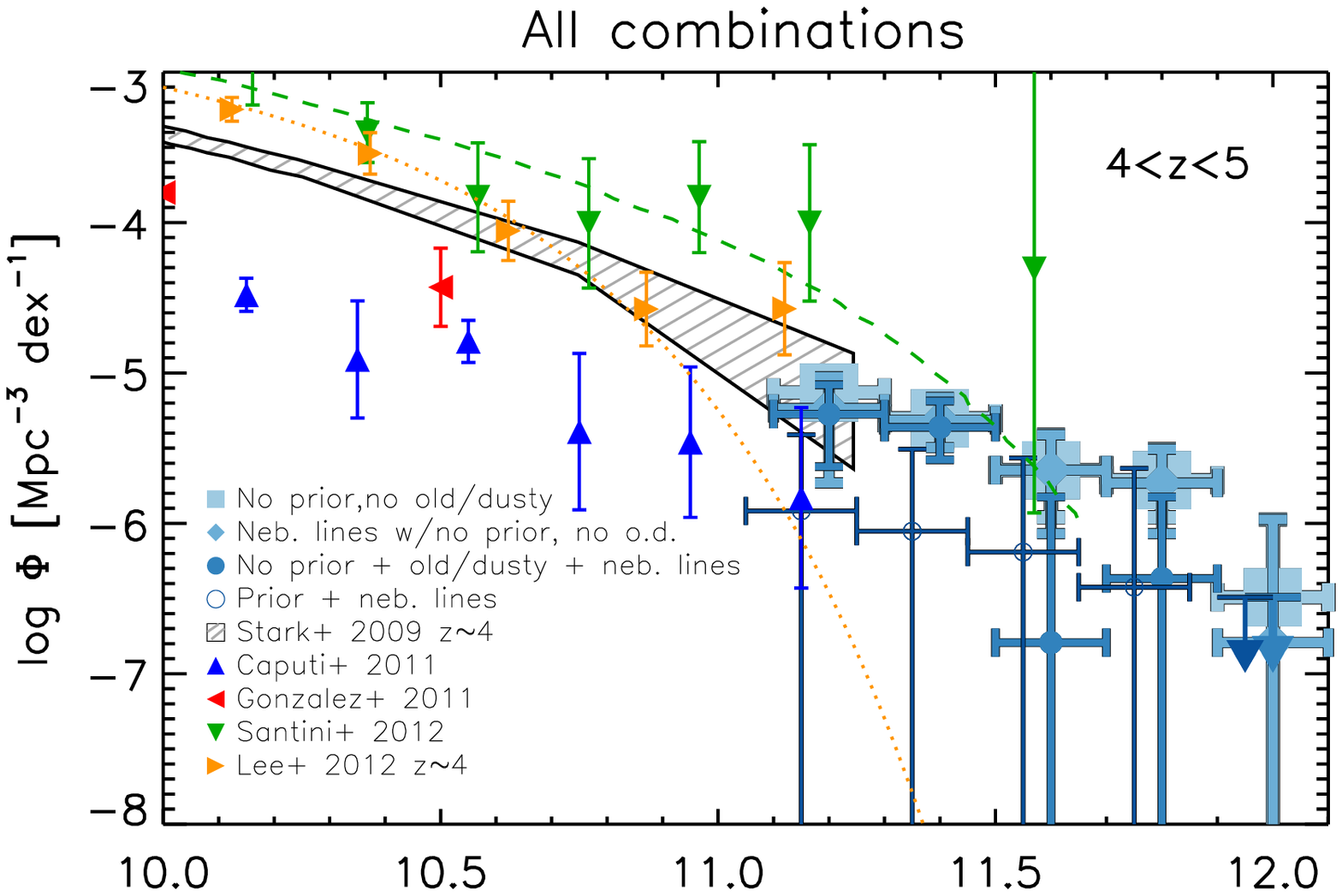} \\
\hspace{-0.5cm}\vspace{-0.6cm}\includegraphics[width=9cm]{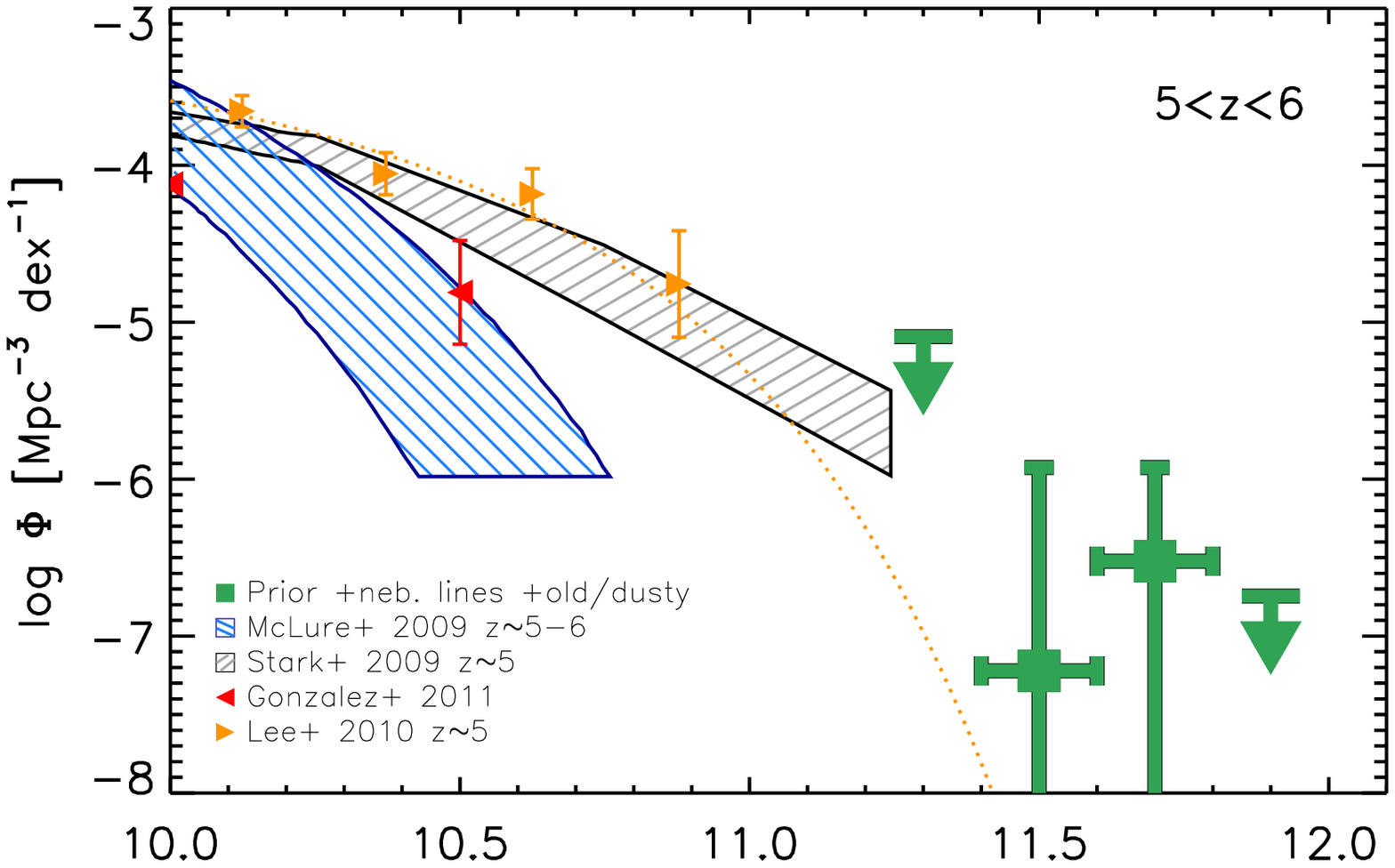} &  \hspace{-0.6cm}\vspace{-0.5cm}\includegraphics[width=9cm]{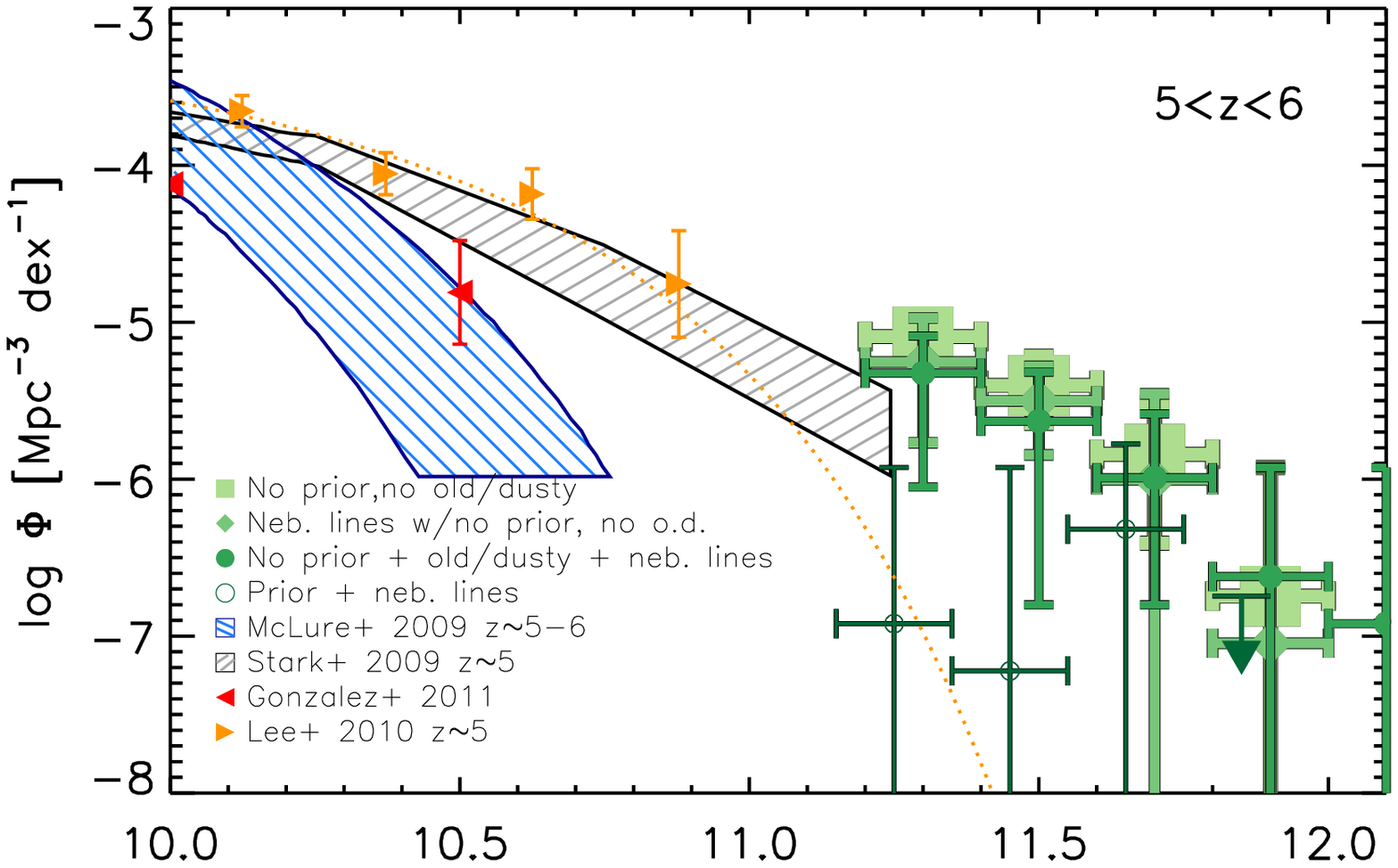} \\ 
\hspace{-0.5cm}\vspace{-0.1cm}\includegraphics[width=9cm]{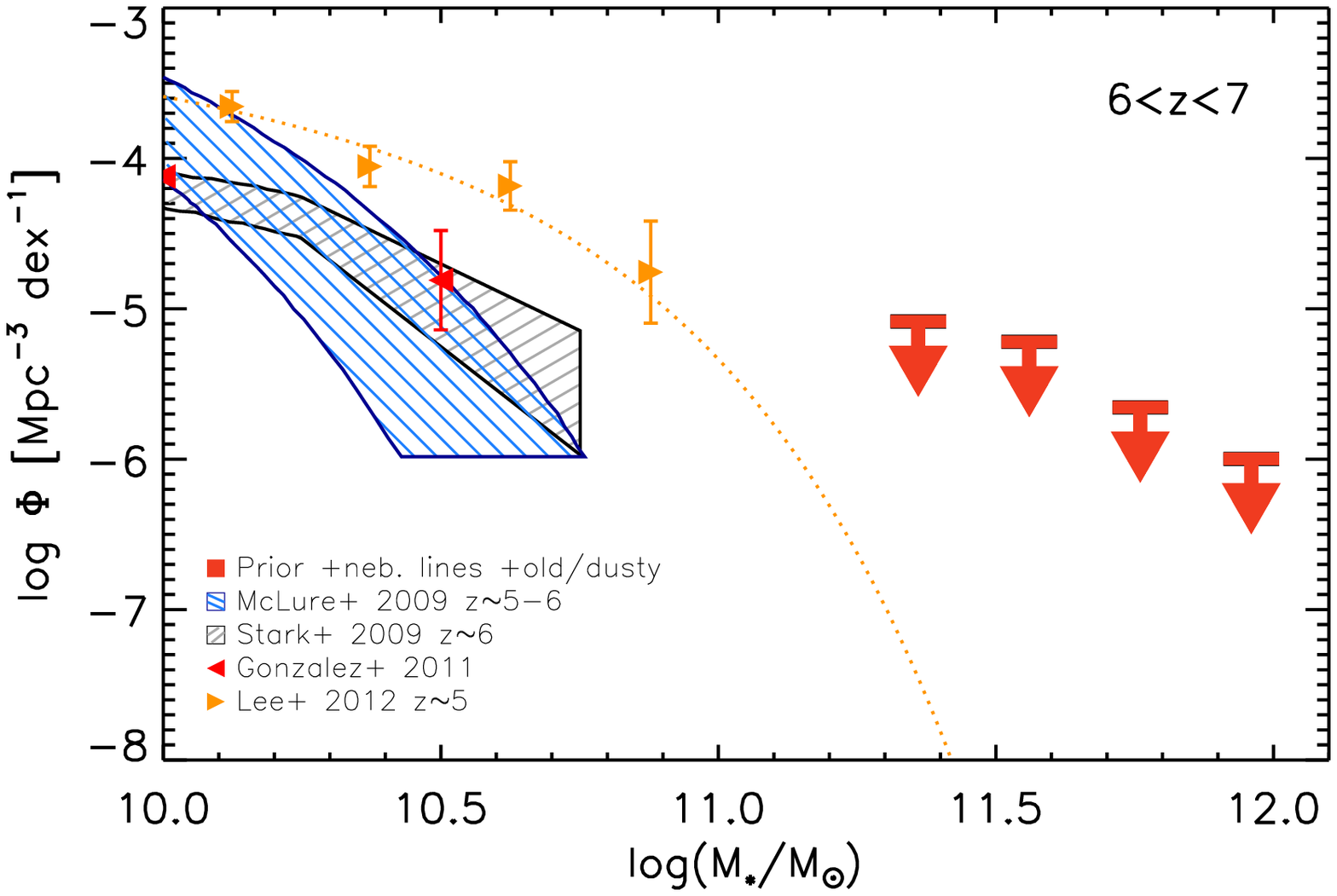} &  \hspace{-0.1cm}\includegraphics[width=9cm]{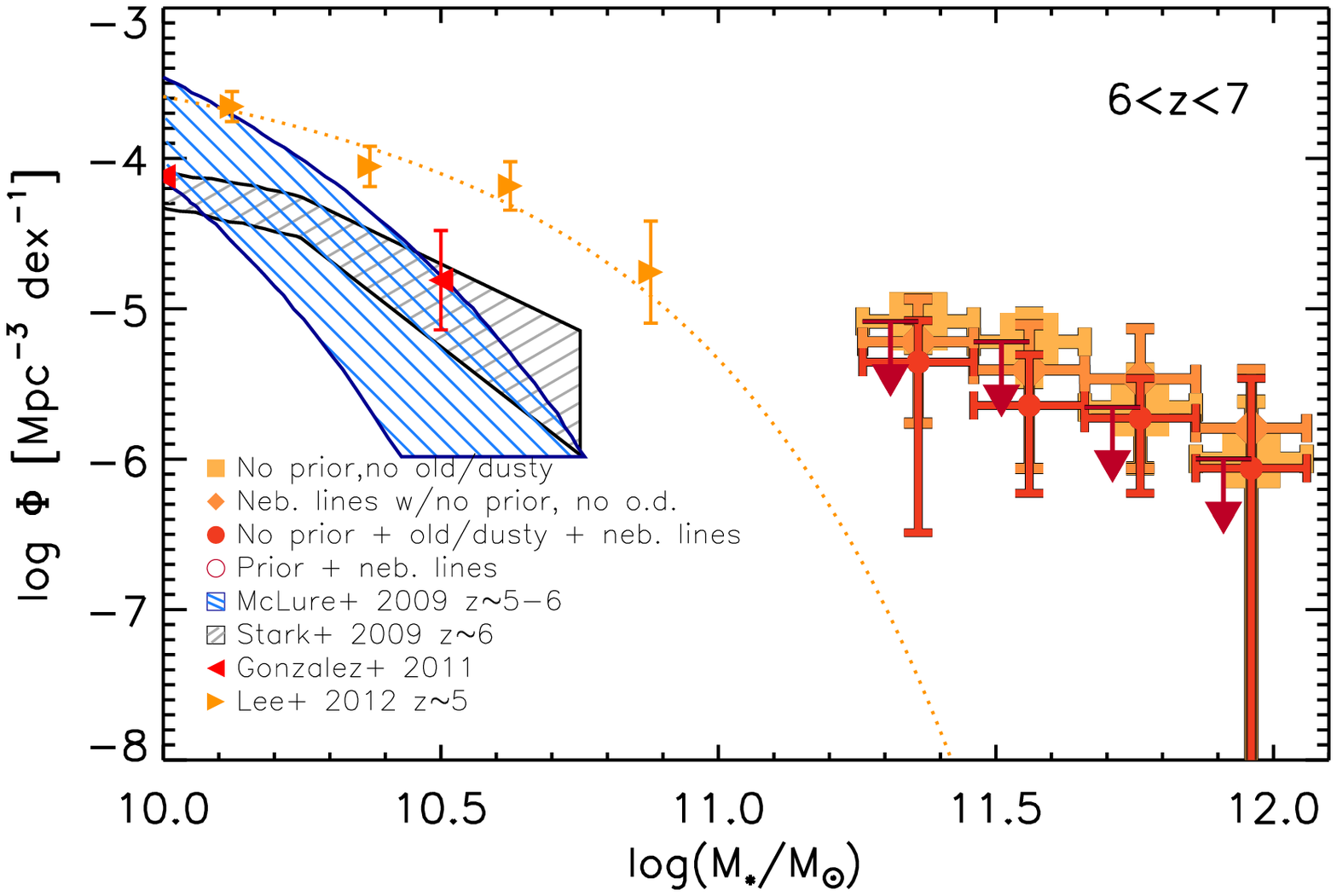} 
\end{tabular}
\caption{Stellar mass functions at $4<z<5$ (top panels), $5<z<6$ (middle panels), $6<z<7$ (bottom panels). The plots in the left column refer to the SMF of the robust sample of massive galaxies presented in Section \ref{sect:robust_massive}, while in the column on the right we present the SMF measurements for different configurations adopted for the measurement of photometric redshifts and stellar population parameters. For the measurements in our work, the vertical error bars include the effect of cosmic variance and the systematic uncertainties arising from the adopted SFHs, the correction of the nebular lines in the photometry, the inclusion of the old and dusty template and the application of the bayesian luminosity prior for the measurements of the photometric redshifts.  The horizontal error bars reflect the bin size adopted for the computation of the SMF. When the bayesian  luminosity prior is introduced in the computation of the photometric redshifts, most of the resulting SMFs measurements turn into upper limits. The measurements from the luminosity prior case have been arbitrarily offset by -0.05~dex to help visualisation. The SMF at $4<z<5$ is consistent with previous measurements at $2\sigma$ level up to $M_*\sim10^{11.6}M_\odot$. \label{fig:SMF}}
\end{figure*}

\begin{figure}
\begin{tabular}{c}
\hspace{-0.5cm}\vspace{-0.5cm}\includegraphics[width=9cm, bb=0 0 504 330,clip=true]{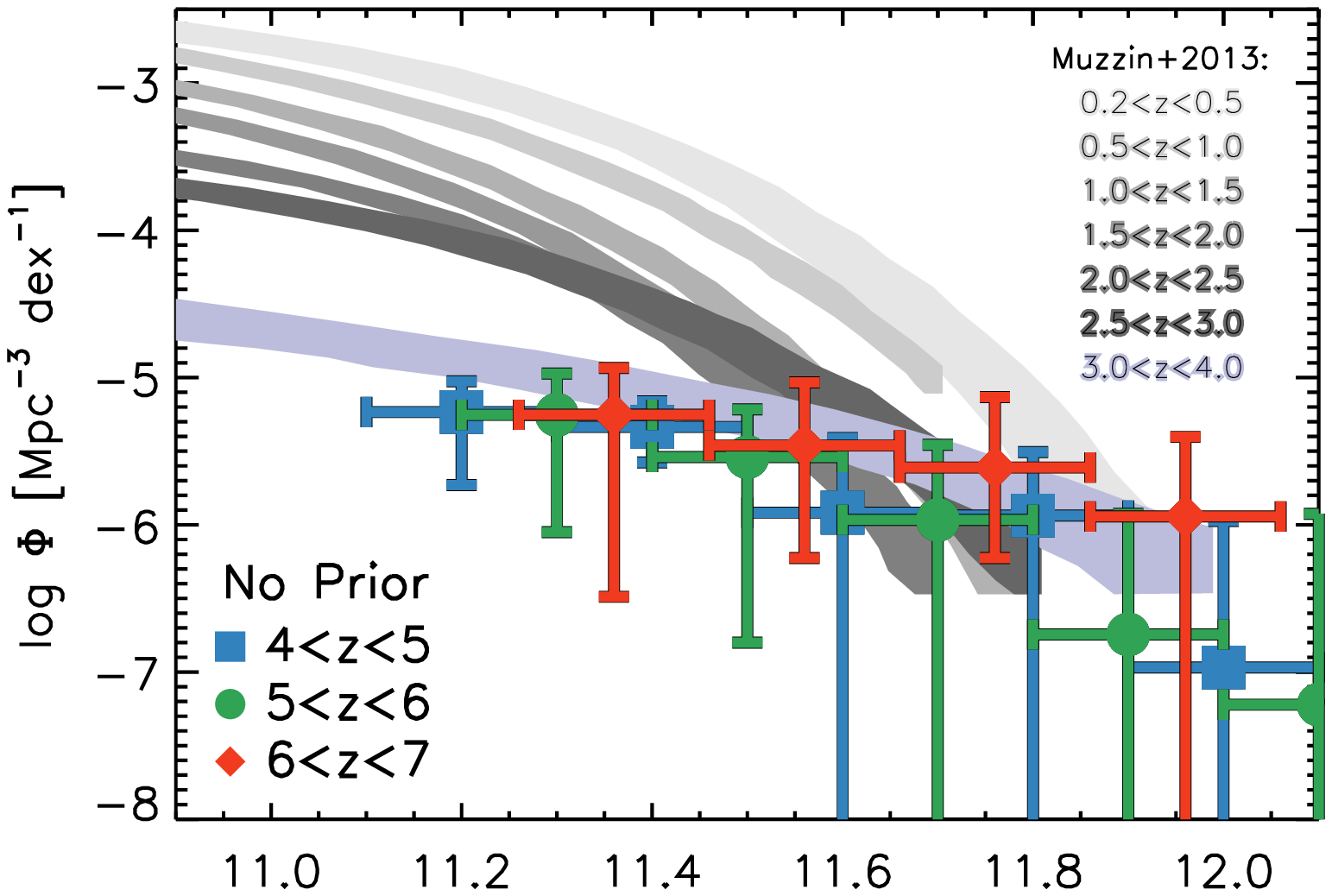} \\
\hspace{-0.5cm}\vspace{-0.5cm}\includegraphics[width=9cm, bb=0 0 504 330,clip=true]{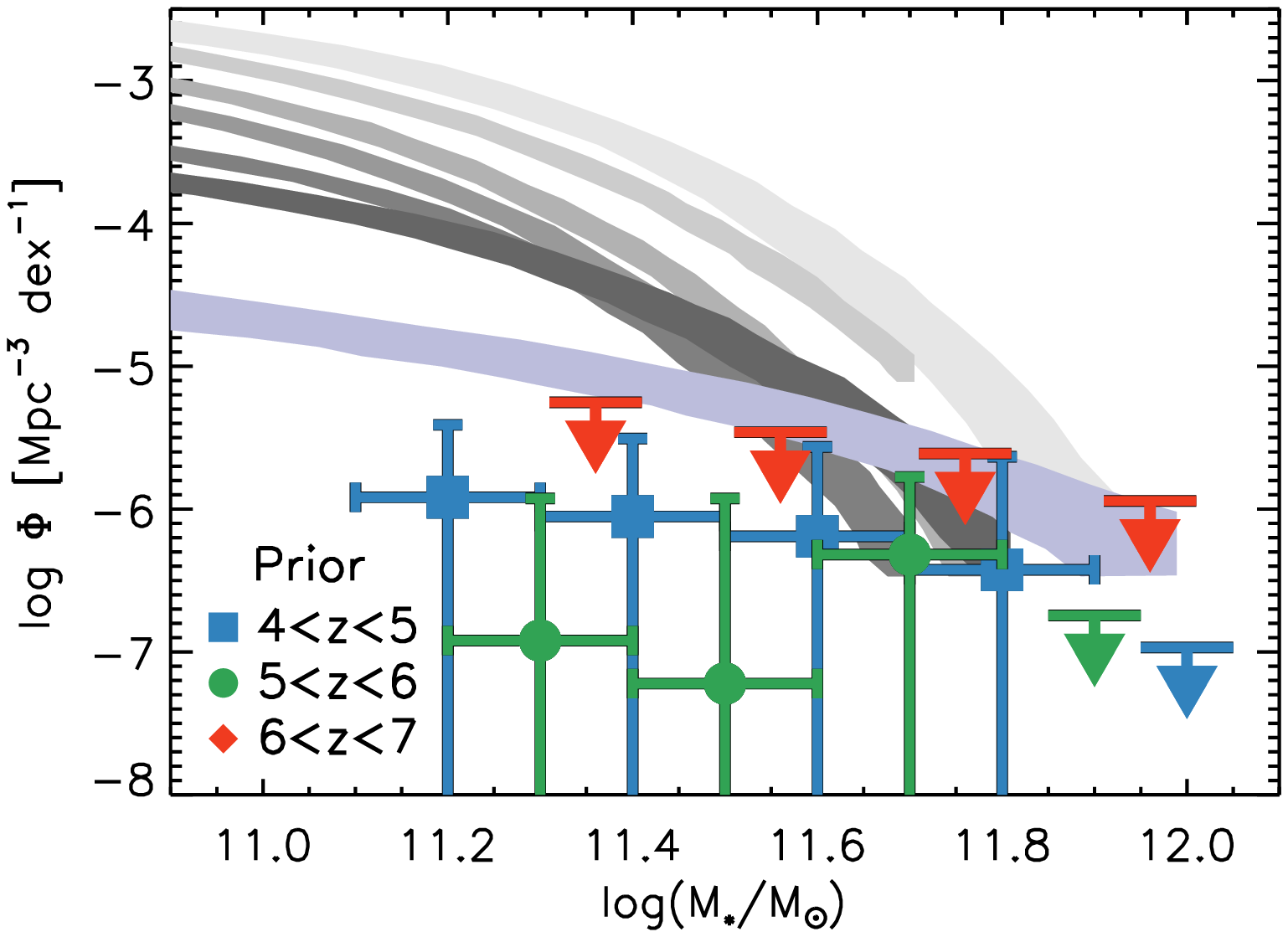} 
\end{tabular}
\caption{Comparison of the SMF in the three redshift bins. The top panel refers to photometric redshifts obtained without the bayesian luminosity prior, while the lower panel refers to the case with the bayesian luminosity prior. In each panel, the points refer to the median value of the SMF measurements obtained with all the different configurations adopted to compute the stellar masses (i.e., using the measured flux as well as after applying the correction for nebular emission contamination, as described in Section \ref{sect:neblines}.  For comparison, the SMF measurements from \citet{muzzin2013b} for $0.5<z<4$ are also plotted. Specifically, the blue shaded region corresponds to the $3<z<4$ SMF. No evidence for evolution is found in the SMFs from $z\sim6.5$ to $z\sim3.5$. \label{fig:SMF_comp}}
\end{figure}

In Table \ref{tab:SMF_del} and \ref{tab:SMF_exp} we report the measurements with the total errors of the SMFs for all the combinations for the measurements of photometric redshifts and stellar population parameters adopted in this work for the delayed-exponential and exponential SFH, respectively. A graphical representation of the SMFs is shown in Figure \ref{fig:SMF}. The column on the left shows the SMF of the robust sample of massive galaxies presented in Section \ref{sect:robust_massive}, while in the right column we present the SMF measurements from different configurations adopted for the measurement of photometric redshifts and stellar population parameters. For each stellar mass bin, points correspond to the median of the measurements from the different configurations adopted for the measurement of photometric redshift and/or for the correction of nebular emission contamination; upper (lower) error bars correspond to the maximum (minimum) $1\sigma$ value. A direct comparison among the SMFs in the three redshift bins is presented in Figure \ref{fig:SMF_comp}. We however note that, given the large uncertainties in the SED fitting arising particularly at $z>5.4$ from the exclusion of the flux in  those bands potentially contaminated by nebular emission (see discussion in Sect. \ref{sect:neblines}), the values reported in Figures \ref{fig:SMF} and \ref{fig:SMF_comp} do not include the measurements of the SMF obtained with this configuration, although, for completeness, these values are included in Table \ref{tab:SMF_del} and \ref{tab:SMF_exp}.

\begin{table*}
\caption{SMF measurements in the three redshift bins, corresponding to the delayed-exponential SFH and for the four different configurations adopted for the measurements of the photometric redshifts, namely without luminosity prior and old/dusty template, without  luminosity prior but including the old/dusty template, with luminosity prior but without old/dusty template, and with luminosity prior and the old/dusty template. For each photometric redshift measurement, the SMF corresponding to each of the four configurations adopted for the measurement of the stellar population parameters and the correction of nebular line contamination are reported. Uncertainties include poisson noise and cosmic variance. \label{tab:SMF_del}}
\begin{tabular}{c|c|cccc|cccc}

$z $ range & Central $M_*$ & \multicolumn{8}{c}{ $\Phi$ $[ 10^{-6}$ Mpc$^{-3}$ dex$^{-1} ] $} \\
\hline
$4<z<5$ &  & \multicolumn{4}{|c|}{No prior, No old/dusty} & \multicolumn{4}{|c}{No prior+old/dusty}\\

 & & Default & EAzY lines & Smit+2014 & Excl. bands & Default & EAzY lines & Smit+2014 & Excl. bands \\

% copy/paste from here
 & $ 11.20\pm 0.1 $ & $  7.42^{+2.02}_{-1.74} $ & $  6.77^{+1.94}_{-1.66} $ & $  2.90^{+1.37}_{-1.05} $ & $  4.51^{+1.63}_{-1.33} $ & $  6.13^{+1.86}_{-1.57} $ & $  4.84^{+1.68}_{-1.38} $ & $  3.55^{+1.48}_{-1.17} $ & $  4.84^{+1.68}_{-1.38} $\\
 & $ 11.40\pm 0.1 $ & $  4.84^{+1.68}_{-1.38} $ & $  5.16^{+1.72}_{-1.43} $ & $  4.84^{+1.68}_{-1.38} $ & $  5.16^{+1.72}_{-1.43} $ & $  4.51^{+1.63}_{-1.33} $ & $  4.84^{+1.68}_{-1.38} $ & $  4.19^{+1.58}_{-1.28} $ & $  3.87^{+1.53}_{-1.22} $\\
 & $ 11.60\pm 0.1 $ & $  1.93^{+1.18}_{-0.84} $ & $  1.61^{+1.11}_{-0.76} $ & $  2.26^{+1.25}_{-0.91} $ & $  2.26^{+1.25}_{-0.91} $ & $ < 1.93 $ & $ < 1.93 $ & $  0.32^{+0.74}_{-0.32} $ & $  0.32^{+0.74}_{-0.32} $\\
 & $ 11.80\pm 0.1 $ & $  1.93^{+1.18}_{-0.84} $ & $  1.93^{+1.18}_{-0.84} $ & $  1.61^{+1.11}_{-0.76} $ & $  2.26^{+1.25}_{-0.91} $ & $  0.32^{+0.74}_{-0.32} $ & $  0.32^{+0.74}_{-0.32} $ & $  0.32^{+0.74}_{-0.32} $ & $  0.64^{+0.85}_{-0.46} $\\
 & $ 12.00\pm 0.1 $ & $  0.32^{+0.74}_{-0.32} $ & $  0.32^{+0.74}_{-0.32} $ & $ < 0.32 $ & $ < 0.32 $ & $ < 0.32 $ & $ < 0.32 $ & $ < 0.32 $ & $ < 0.32 $\\

 & & & & & & & & & \\
 &  & \multicolumn{4}{|c|}{Prior, No old/dusty} & \multicolumn{4}{|c}{Prior+old/dusty}\\

 & & Default & EAzY lines & Smit+2014 & Excl. bands & Default & EAzY lines & Smit+2014 & Excl. bands \\

 & $ 11.20\pm 0.1 $ & $  2.58^{+1.31}_{-0.98} $ & $  2.58^{+1.31}_{-0.98} $ & $  1.29^{+1.03}_{-0.67} $ & $  1.29^{+1.03}_{-0.67} $ & $  0.97^{+0.95}_{-0.57} $ & $  1.29^{+1.03}_{-0.67} $ & $  0.32^{+0.74}_{-0.32} $ & $  0.32^{+0.74}_{-0.32} $\\
 & $ 11.40\pm 0.1 $ & $  1.61^{+1.11}_{-0.76} $ & $  1.61^{+1.11}_{-0.76} $ & $  1.29^{+1.03}_{-0.67} $ & $  1.61^{+1.11}_{-0.76} $ & $  0.32^{+0.74}_{-0.32} $ & $ < 4.84 $ & $ < 4.84 $ & $  0.64^{+0.85}_{-0.46} $\\
 & $ 11.60\pm 0.1 $ & $  1.61^{+1.11}_{-0.76} $ & $  0.97^{+0.95}_{-0.57} $ & $  1.29^{+1.03}_{-0.67} $ & $  1.29^{+1.03}_{-0.67} $ & $ < 1.93 $ & $ < 1.93 $ & $ < 1.93 $ & $ < 1.93 $\\
 & $ 11.80\pm 0.1 $ & $  0.32^{+0.74}_{-0.32} $ & $  0.64^{+0.85}_{-0.46} $ & $  0.64^{+0.85}_{-0.46} $ & $  0.64^{+0.85}_{-0.46} $ & $ < 1.93 $ & $ < 1.93 $ & $ < 1.93 $ & $  0.32^{+0.74}_{-0.32} $\\
 & $ 12.00\pm 0.1 $ & $ < 0.32 $ & $ < 0.32 $ & $ < 0.32 $ & $ < 0.32 $ & $ < 0.32 $ & $ < 0.32 $ & $ < 0.32 $ & $ < 0.32 $\\
% to here

 & & & & & & & & & \\
\hline
 & & & & & & & & & \\
$5<z<6$ &  & \multicolumn{4}{|c|}{No prior, No old/dusty} & \multicolumn{4}{|c}{No prior+old/dusty}\\

 & & Default & EAzY lines & Smit+2014 & Excl. bands & Default & EAzY lines & Smit+2014 & Excl. bands \\

% copy/paste from here
 & $ 11.30\pm 0.1 $ & $  8.28^{+2.37}_{-2.14} $ & $  8.28^{+2.37}_{-2.14} $ & $  3.24^{+1.56}_{-1.26} $ & $  7.92^{+2.32}_{-2.08} $ & $  6.12^{+2.05}_{-1.80} $ & $  5.40^{+1.94}_{-1.67} $ & $  2.88^{+1.49}_{-1.18} $ & $  5.40^{+1.94}_{-1.67} $\\
 & $ 11.50\pm 0.1 $ & $  3.60^{+1.63}_{-1.33} $ & $  3.60^{+1.63}_{-1.33} $ & $  2.52^{+1.42}_{-1.09} $ & $  2.88^{+1.49}_{-1.18} $ & $  3.24^{+1.56}_{-1.26} $ & $  3.24^{+1.56}_{-1.26} $ & $  0.72^{+0.96}_{-0.56} $ & $  3.60^{+1.63}_{-1.33} $\\
 & $ 11.70\pm 0.1 $ & $  1.08^{+1.07}_{-0.69} $ & $  1.08^{+1.07}_{-0.69} $ & $  0.36^{+0.83}_{-0.36} $ & $  1.44^{+1.17}_{-0.80} $ & $  0.72^{+0.96}_{-0.56} $ & $  0.72^{+0.96}_{-0.56} $ & $  0.72^{+0.96}_{-0.56} $ & $ < 1.08 $\\
 & $ 11.90\pm 0.1 $ & $  0.36^{+0.83}_{-0.36} $ & $  0.36^{+0.83}_{-0.36} $ & $ < 0.36 $ & $ < 0.36 $ & $  0.36^{+0.83}_{-0.36} $ & $  0.36^{+0.83}_{-0.36} $ & $ < 0.36 $ & $  0.72^{+0.96}_{-0.56} $\\

 & & & & & & & & & \\
 &  & \multicolumn{4}{|c|}{Prior, No old/dusty} & \multicolumn{4}{|c}{Prior+old/dusty}\\

 & & Default & EAzY lines & Smit+2014 & Excl. bands & Default & EAzY lines & Smit+2014 & Excl. bands \\

 & $ 11.30\pm 0.1 $ & $  0.36^{+0.83}_{-0.36} $ & $  0.36^{+0.83}_{-0.36} $ & $ < 8.28 $ & $ < 8.28 $ & $ < 8.28 $ & $ < 8.28 $ & $ < 8.28 $ & $ < 8.28 $\\
 & $ 11.50\pm 0.1 $ & $ < 3.60 $ & $ < 3.60 $ & $ < 3.60 $ & $  0.72^{+0.96}_{-0.56} $ & $ < 3.60 $ & $ < 3.60 $ & $ < 3.60 $ & $  0.36^{+0.83}_{-0.36} $\\
 & $ 11.70\pm 0.1 $ & $  0.72^{+0.96}_{-0.56} $ & $  0.72^{+0.96}_{-0.56} $ & $  0.72^{+0.96}_{-0.56} $ & $ < 1.08 $ & $  0.36^{+0.83}_{-0.36} $ & $  0.36^{+0.83}_{-0.36} $ & $  0.36^{+0.83}_{-0.36} $ & $ < 1.08 $\\
 & $ 11.90\pm 0.1 $ & $ < 0.36 $ & $ < 0.36 $ & $ < 0.36 $ & $ < 0.36 $ & $ < 0.36 $ & $ < 0.36 $ & $ < 0.36 $ & $ < 0.36 $\\
% to here

 & & & & & & & & & \\
\hline
 & & & & & & & & & \\

$6<z<7$ &  & \multicolumn{4}{|c|}{No prior, No old/dusty} & \multicolumn{4}{|c}{No prior+old/dusty}\\

 & & Default & EAzY lines & Smit+2014 & Excl. bands & Default & EAzY lines & Smit+2014 & Excl. bands \\

% copy/paste from here
 & $ 11.36\pm 0.1 $ & $  8.45^{+2.72}_{-2.61} $ & $  8.45^{+2.72}_{-2.61} $ & $  3.22^{+1.73}_{-1.48} $ & $  7.64^{+2.58}_{-2.46} $ & $  5.63^{+2.22}_{-2.05} $ & $  5.23^{+2.15}_{-1.96} $ & $  2.41^{+1.54}_{-1.26} $ & $  4.83^{+2.07}_{-1.87} $\\
 & $ 11.56\pm 0.1 $ & $  5.23^{+2.15}_{-1.96} $ & $  5.63^{+2.22}_{-2.05} $ & $  2.82^{+1.64}_{-1.37} $ & $  6.84^{+2.44}_{-2.30} $ & $  1.61^{+1.33}_{-1.01} $ & $  1.61^{+1.33}_{-1.01} $ & $  3.22^{+1.73}_{-1.48} $ & $  2.41^{+1.54}_{-1.26} $\\
 & $ 11.76\pm 0.1 $ & $  2.41^{+1.54}_{-1.26} $ & $  2.01^{+1.44}_{-1.14} $ & $  4.02^{+1.90}_{-1.68} $ & $  1.21^{+1.21}_{-0.88} $ & $  2.01^{+1.44}_{-1.14} $ & $  2.01^{+1.44}_{-1.14} $ & $  1.61^{+1.33}_{-1.01} $ & $  1.61^{+1.33}_{-1.01} $\\
 & $ 11.96\pm 0.1 $ & $  0.80^{+1.08}_{-0.80} $ & $  0.80^{+1.08}_{-0.80} $ & $  2.01^{+1.44}_{-1.14} $ & $  1.61^{+1.33}_{-1.01} $ & $  0.40^{+0.93}_{-0.40} $ & $  0.40^{+0.93}_{-0.40} $ & $  1.61^{+1.33}_{-1.01} $ & $  0.80^{+1.08}_{-0.80} $\\

 & & & & & & & & & \\
 &  & \multicolumn{4}{|c|}{Prior, No old/dusty} & \multicolumn{4}{|c}{Prior+old/dusty}\\

 & & Default & EAzY lines & Smit+2014 & Excl. bands & Default & EAzY lines & Smit+2014 & Excl. bands \\

 & $ 11.36\pm 0.1 $ & $ < 8.45 $ & $ < 8.45 $ & $ < 8.45 $ & $ < 8.45 $ & $ < 8.45 $ & $ < 8.45 $ & $ < 8.45 $ & $ < 8.45 $\\
 & $ 11.56\pm 0.1 $ & $ < 5.23 $ & $ < 5.23 $ & $ < 5.23 $ & $ < 5.23 $ & $ < 5.23 $ & $ < 5.23 $ & $ < 5.23 $ & $ < 5.23 $\\
 & $ 11.76\pm 0.1 $ & $ < 2.41 $ & $ < 2.41 $ & $ < 2.41 $ & $ < 2.41 $ & $ < 2.41 $ & $ < 2.41 $ & $ < 2.41 $ & $ < 2.41 $\\
 & $ 11.96\pm 0.1 $ & $ < 0.80 $ & $ < 0.80 $ & $ < 0.80 $ & $ < 0.80 $ & $ < 0.80 $ & $ < 0.80 $ & $ < 0.80 $ & $ < 0.80 $\\
% to here

\end{tabular}
\end{table*}

%%%%%%%%%%%%%%%%%%%
%% Exponential

\begin{table*}
\caption{SMF measurements in the three redshift bins, corresponding to the exponential SFH and for the four different configurations adopted for the measurements of the photometric redshifts, namely without luminosity prior and old/dusty template, without luminosity prior but including the old/dusty template, with prior but without old/dusty template, and with luminosity prior and the old/dusty template. For each photometric redshift measurement, the SMF corresponding to each of the four configurations adopted for the measurement of the stellar population parameters and the correction of nebular line contamination are reported. Uncertainties include poisson noise and cosmic variance. \label{tab:SMF_exp}}
\begin{tabular}{c|c|cccc|cccc}

$z $ range & Central $M_*$ & \multicolumn{8}{c}{ $\Phi$ $[ 10^{-6}$ Mpc$^{-3}$ dex$^{-1} ] $} \\
\hline
$4<z<5$ &  & \multicolumn{4}{|c|}{No prior, No old/dusty} & \multicolumn{4}{|c}{No prior+old/dusty}\\

 & & Default & EAzY lines & Smit+2014 & Excl. bands & Default & EAzY lines & Smit+2014 & Excl. bands \\

% copy/paste from here
 & $ 11.20\pm 0.1 $ & $  7.42^{+2.02}_{-1.74} $ & $  7.42^{+2.02}_{-1.74} $ & $  6.13^{+1.86}_{-1.57} $ & $  6.13^{+1.86}_{-1.57} $ & $  6.45^{+1.90}_{-1.61} $ & $  6.45^{+1.90}_{-1.61} $ & $  4.51^{+1.63}_{-1.33} $ & $  6.13^{+1.86}_{-1.57} $\\
 & $ 11.40\pm 0.1 $ & $  5.16^{+1.72}_{-1.43} $ & $  4.84^{+1.68}_{-1.38} $ & $  4.51^{+1.63}_{-1.33} $ & $  5.16^{+1.72}_{-1.43} $ & $  3.87^{+1.53}_{-1.22} $ & $  3.87^{+1.53}_{-1.22} $ & $  4.84^{+1.68}_{-1.38} $ & $  4.19^{+1.58}_{-1.28} $\\
 & $ 11.60\pm 0.1 $ & $  2.58^{+1.31}_{-0.98} $ & $  2.58^{+1.31}_{-0.98} $ & $  2.58^{+1.31}_{-0.98} $ & $  1.93^{+1.18}_{-0.84} $ & $ < 2.58 $ & $ < 2.58 $ & $  0.64^{+0.85}_{-0.46} $ & $ < 2.58 $\\
 & $ 11.80\pm 0.1 $ & $  1.93^{+1.18}_{-0.84} $ & $  1.93^{+1.18}_{-0.84} $ & $  1.93^{+1.18}_{-0.84} $ & $  2.26^{+1.25}_{-0.91} $ & $  0.64^{+0.85}_{-0.46} $ & $  0.64^{+0.85}_{-0.46} $ & $  0.32^{+0.74}_{-0.32} $ & $  0.64^{+0.85}_{-0.46} $\\
 & $ 12.00\pm 0.1 $ & $  0.32^{+0.74}_{-0.32} $ & $  0.32^{+0.74}_{-0.32} $ & $ < 0.32 $ & $ < 0.32 $ & $ < 0.32 $ & $ < 0.32 $ & $ < 0.32 $ & $ < 0.32 $\\

 & & & & & & & & & \\
 &  & \multicolumn{4}{|c|}{Prior, No old/dusty} & \multicolumn{4}{|c}{Prior+old/dusty}\\

 & & Default & EAzY lines & Smit+2014 & Excl. bands & Default & EAzY lines & Smit+2014 & Excl. bands \\

 & $ 11.20\pm 0.1 $ & $  1.61^{+1.11}_{-0.76} $ & $  1.61^{+1.11}_{-0.76} $ & $  0.97^{+0.95}_{-0.57} $ & $  1.93^{+1.18}_{-0.84} $ & $  0.64^{+0.85}_{-0.46} $ & $  0.32^{+0.74}_{-0.32} $ & $  0.32^{+0.74}_{-0.32} $ & $  0.64^{+0.85}_{-0.46} $\\
 & $ 11.40\pm 0.1 $ & $  1.61^{+1.11}_{-0.76} $ & $  1.29^{+1.03}_{-0.67} $ & $  1.93^{+1.18}_{-0.84} $ & $  1.61^{+1.11}_{-0.76} $ & $  0.32^{+0.74}_{-0.32} $ & $  0.32^{+0.74}_{-0.32} $ & $  0.32^{+0.74}_{-0.32} $ & $  0.64^{+0.85}_{-0.46} $\\
 & $ 11.60\pm 0.1 $ & $  1.29^{+1.03}_{-0.67} $ & $  0.97^{+0.95}_{-0.57} $ & $  1.61^{+1.11}_{-0.76} $ & $  1.29^{+1.03}_{-0.67} $ & $ < 2.58 $ & $ < 2.58 $ & $ < 2.58 $ & $ < 2.58 $\\
 & $ 11.80\pm 0.1 $ & $  0.97^{+0.95}_{-0.57} $ & $  1.29^{+1.03}_{-0.67} $ & $  0.64^{+0.85}_{-0.46} $ & $  0.64^{+0.85}_{-0.46} $ & $ < 1.93 $ & $ < 1.93 $ & $ < 1.93 $ & $  0.32^{+0.74}_{-0.32} $\\
 & $ 12.00\pm 0.1 $ & $ < 0.32 $ & $ < 0.32 $ & $ < 0.32 $ & $ < 0.32 $ & $ < 0.32 $ & $ < 0.32 $ & $ < 0.32 $ & $ < 0.32 $\\
% to here

 & & & & & & & & & \\
\hline
 & & & & & & & & & \\
$5<z<6$ &  & \multicolumn{4}{|c|}{No prior, No old/dusty} & \multicolumn{4}{|c}{No prior+old/dusty}\\

 & & Default & EAzY lines & Smit+2014 & Excl. bands & Default & EAzY lines & Smit+2014 & Excl. bands \\

% copy/paste from here
 & $ 11.30\pm 0.1 $ & $  7.92^{+2.32}_{-2.08} $ & $  7.92^{+2.32}_{-2.08} $ & $  2.88^{+1.49}_{-1.18} $ & $  7.56^{+2.27}_{-2.03} $ & $  6.12^{+2.05}_{-1.80} $ & $  6.12^{+2.05}_{-1.80} $ & $  1.80^{+1.26}_{-0.91} $ & $  5.76^{+2.00}_{-1.73} $\\
 & $ 11.50\pm 0.1 $ & $  4.32^{+1.76}_{-1.47} $ & $  3.60^{+1.63}_{-1.33} $ & $  2.88^{+1.49}_{-1.18} $ & $  4.32^{+1.76}_{-1.47} $ & $  3.24^{+1.56}_{-1.26} $ & $  2.52^{+1.42}_{-1.09} $ & $  1.08^{+1.07}_{-0.69} $ & $  5.40^{+1.94}_{-1.67} $\\
 & $ 11.70\pm 0.1 $ & $  1.80^{+1.26}_{-0.91} $ & $  2.16^{+1.34}_{-1.00} $ & $  0.36^{+0.83}_{-0.36} $ & $  1.08^{+1.07}_{-0.69} $ & $  1.44^{+1.17}_{-0.80} $ & $  1.44^{+1.17}_{-0.80} $ & $  1.08^{+1.07}_{-0.69} $ & $  0.36^{+0.83}_{-0.36} $\\
 & $ 11.90\pm 0.1 $ & $ 0.36^{+0.83}_{-0.36} $ & $ 0.36^{+0.83}_{-0.36} $ & $ < 0.36 $ & $  0.72^{+0.96}_{-0.56} $ & $  0.36^{+0.83}_{-0.36} $ & $  0.36^{+0.83}_{-0.36} $ & $ < 0.36 $ & $  0.72^{+0.96}_{-0.56} $\\

 & & & & & & & & & \\
 &  & \multicolumn{4}{|c|}{Prior, No old/dusty} & \multicolumn{4}{|c}{Prior+old/dusty}\\

 & & Default & EAzY lines & Smit+2014 & Excl. bands & Default & EAzY lines & Smit+2014 & Excl. bands \\

 & $ 11.30\pm 0.1 $ & $  0.36^{+0.83}_{-0.36} $ & $  0.36^{+0.83}_{-0.36} $ & $ < 7.92 $ & $ < 7.92 $ & $ < 7.92 $ & $ < 7.92 $ & $ < 7.92 $ & $ < 7.92 $\\
 & $ 11.50\pm 0.1 $ & $ < 4.32 $ & $ < 4.32 $ & $  0.36^{+0.83}_{-0.36} $ & $  0.36^{+0.83}_{-0.36} $ & $ < 4.32 $ & $ < 4.32 $ & $  0.36^{+0.83}_{-0.36} $ & $ < 4.32 $\\
 & $ 11.70\pm 0.1 $ & $  0.72^{+0.96}_{-0.56} $ & $  0.72^{+0.96}_{-0.56} $ & $  0.36^{+0.83}_{-0.36} $ & $  0.36^{+0.83}_{-0.36} $ & $  0.36^{+0.83}_{-0.36} $ & $  0.36^{+0.83}_{-0.36} $ & $ < 1.80 $ & $  0.36^{+0.83}_{-0.36} $\\
 & $ 11.90\pm 0.1 $ & $ < 0.36 $ & $ < 0.36 $ & $ < 0.36 $ & $ < 0.36 $ & $ < 0.36 $ & $ < 0.36 $ & $ < 0.36 $ & $ < 0.36 $\\
 % to here

 & & & & & & & & & \\
\hline
 & & & & & & & & & \\

$6<z<7$ &  & \multicolumn{4}{|c|}{No prior, No old/dusty} & \multicolumn{4}{|c}{No prior+old/dusty}\\

 & & Default & EAzY lines & Smit+2014 & Excl. bands & Default & EAzY lines & Smit+2014 & Excl. bands \\

% copy/paste from here
 & $ 11.36\pm 0.1 $ & $  8.04^{+2.65}_{-2.53} $ & $  8.85^{+2.79}_{-2.69} $ & $  3.62^{+1.82}_{-1.58} $ & $  8.45^{+2.72}_{-2.61} $ & $  6.03^{+2.30}_{-2.13} $ & $  6.03^{+2.30}_{-2.13} $ & $  1.21^{+1.21}_{-0.88} $ & $  6.43^{+2.37}_{-2.22} $\\
 & $ 11.56\pm 0.1 $ & $  6.84^{+2.44}_{-2.30} $ & $  5.23^{+2.15}_{-1.96} $ & $  2.01^{+1.44}_{-1.14} $ & $  8.04^{+2.65}_{-2.53} $ & $  2.01^{+1.44}_{-1.14} $ & $  2.01^{+1.44}_{-1.14} $ & $  3.22^{+1.73}_{-1.48} $ & $  2.41^{+1.54}_{-1.26} $\\
 & $ 11.76\pm 0.1 $ & $  2.01^{+1.44}_{-1.14} $ & $  2.41^{+1.54}_{-1.26} $ & $  5.23^{+2.15}_{-1.96} $ & $  1.21^{+1.21}_{-0.88} $ & $  2.01^{+1.44}_{-1.14} $ & $  2.01^{+1.44}_{-1.14} $ & $  1.61^{+1.33}_{-1.01} $ & $  1.61^{+1.33}_{-1.01} $\\
 & $ 11.96\pm 0.1 $ & $  1.21^{+1.21}_{-0.88} $ & $  1.21^{+1.21}_{-0.88} $ & $  2.41^{+1.54}_{-1.26} $ & $  1.61^{+1.33}_{-1.01} $ & $  0.40^{+0.93}_{-0.40} $ & $  0.40^{+0.93}_{-0.40} $ & $  2.01^{+1.44}_{-1.14} $ & $  0.80^{+1.08}_{-0.80} $\\

 & & & & & & & & & \\
 &  & \multicolumn{4}{|c|}{Prior, No old/dusty} & \multicolumn{4}{|c}{Prior+old/dusty}\\

 & & Default & EAzY lines & Smit+2014 & Excl. bands & Default & EAzY lines & Smit+2014 & Excl. bands \\

 & $ 11.36\pm 0.1 $ & $ < 8.04 $ & $ < 8.04 $ & $ < 8.04 $ & $ < 8.04 $ & $ < 8.04 $ & $ < 8.04 $ & $ < 8.04 $ & $ < 8.04 $\\
 & $ 11.56\pm 0.1 $ & $ < 6.84 $ & $ < 6.84 $ & $ < 6.84 $ & $ < 6.84 $ & $ < 6.84 $ & $ < 6.84 $ & $ < 6.84 $ & $ < 6.84 $\\
 & $ 11.76\pm 0.1 $ & $ < 2.01 $ & $ < 2.01 $ & $ < 2.01 $ & $ < 2.01 $ & $ < 2.01 $ & $ < 2.01 $ & $ < 2.01 $ & $ < 2.01 $\\
 & $ 11.96\pm 0.1 $ & $ < 1.21 $ & $ < 1.21 $ & $ < 1.21 $ & $ < 1.21 $ & $ < 1.21 $ & $ < 1.21 $ & $ < 1.21 $ & $ < 1.21 $\\
% to here

\end{tabular}
\end{table*}

%%%%%%%%%%%%%%%%%%%

The configurations we adopted for the measurements of redshift and stellar population parameters, described in Section \ref{sect:photoz}, introduce a dispersion in our SMF measurements. The considered SFHs introduce the smaller dispersion in the SMF measurements; the dispersion of the stellar mass measurements from the inclusion of the nebular emission lines is smaller than the uncertainty from the combination of poissonian noise and cosmic variance. On the contrary, the inclusion or not of the bayesian  luminosity prior in the measurements of the photometric redshifts influences the values of photometric redshifts themselves for a non-negligible fraction of objects, and, as a consequence, of the stellar masses. Specifically, the adoption of the bayesian luminosity prior transforms many measurements into upper limits.

In the following subsections we compare our SMF estimates with measurements from the literature.

\subsubsection{The SMF at $4<z<5$}

In the redshift range $4<z<5$, SMF measurements have already been published by \citet{stark2009,caputi2011,gonzalez2011,santini2012,lee2012}. Of these, the SMFs of \citet{stark2009}, \citet{gonzalez2011} and of \citet{lee2012} are based on dropouts selections, while those of \citet{caputi2011} and \citet{santini2012} are based on photometric redshift measurements. The measurements by \citet{caputi2011} are based on observations over relatively wide fields (0.6 deg$^2$), while the other works over fields about one order of magnitude narrower than UltraVISTA.  Since the total exposure time is roughly of the same order, this translates into detection of fainter sources, and allowing for a lower stellar mass completeness limit for the same stellar mass-to-light ratio. The smaller fields together with the selection effects introduced by the dropout technique, have reduced so far the chances of detecting the most massive objects. This is evident from the top-left plot of Figure \ref{fig:SMF}. Points from the literature reach an upper limit in stellar mass which is at most $\log(M_*/M_\odot)\sim11.3$, the only exception here being the measurements at $\log(M_*/M_\odot)\sim11.6$ by \citet{santini2012}, although the error bar is quite large. On the opposite side, \citet{gonzalez2011} measure the SMF up to $z\sim7$, but their stellar mass range never overlaps with ours.

The stellar mass range over which our SMF measurements is defined overlaps with those by  \citet{stark2009}, \citet{caputi2011} and \citet{lee2012} in the lowest stellar mass bin, while it overlaps with \citet{santini2012} up to $\log(M_*/M_\odot)\sim11.6$. The lowest stellar mass bin measurements with no luminosity prior is consistent with the measurements by \citet{stark2009} and with \citet{caputi2011} and at $2\sigma$-level with the measurements by \citet{lee2012}, while it is consistent with \citet{santini2012} at 3$\sigma$. We however note that \citet{santini2012} do not include the effects of cosmic variance in the error budget, which for $3.5<z<4.5$ we estimate using the recipe of \citet{moster2011} to be as high as 50\% and 70\% for $\log(M_*/M_\odot)=10.5$  and 11 respectively.  At higher stellar mass, our SMF becomes consistent with both the Schechter parameterization and Vmax measurements by \citet{santini2012} at 1$\sigma$ level. Under the caveat that extrapolations are characterized by a high degree of uncertainty and should be considered with care, we finally compare our measurements to the extrapolation of the Schecher fit of \citet{lee2012}. While the measurements obtained without the bayesian  luminosity prior are larger than the extrapolation of the Schechter fit by \citet{lee2012} (marked by the yellow dotted curve  in the top panels of Figure \ref{fig:SMF}) by at least a factor of ten, the measurements obtained with the adoption of the luminosity prior are consistent at $1\sigma$ level. However, this agreement is mostly the result of the large uncertainties from both the systematic effects discussed in Section 3 and from the limited  size of the sample.

\subsubsection{The SMF at $5<z<6$}

In this range of redshift there are fewer previous determination of the SMF than at $4<z<5$, namely the works by \citet{mclure2009,stark2009,gonzalez2011} and \citet{lee2012}. The SMF by \citet{stark2009} is the only measurements whose stellar mass range partly overlaps with ours. In the overlapping range, the two measurements are consistent within 1$\sigma$.

As for the $4<z<5$ case,  our SMF is consistent with the extrapolation of the Schechter fit by \citet{lee2012} when the measurements obtained with the bayesian luminosity prior are considered. As for the $4<z<5$ SMF, also in this case we caution about over interpreting the comparison between our measurements and the extrapolation of the Schechter fit by \citet{lee2012}.

\subsubsection{The SMF at $6<z<7$}
In this range of redshift current measurements of the SMF include the works by \citet{stark2009,gonzalez2011} and \citet{lee2012}. However, if we do not consider the extrapolation of the Schechter fit by \citet{lee2012}, none of them is defined over a stellar mass range which is overlapping with ours. Our SMF is consistent with the extrapolation of the Schecher only considering our upper limit measurements.

\subsubsection{The evolution of the SMF}
In this section we consider the evolution of the SMF from $z\sim6.5$ to $z\sim4.5$. In the top panel of Figure \ref{fig:SMF_comp} we report the median of the SMF measurements without luminosity prior (i.e., with and without the dusty template) from each redshift bin, while the bottom panel presents the median of the SMF from the  luminosity prior configurations. In the same plot, we also report the SMF measurements at $z<4$ from \citet{muzzin2013b} as filled regions (light blue for the SMF at $z\sim 3.5$; the other filled regions corresponding to the SMF at $z\sim2.75,2.25,1.75,1.25,0.75$ from darker to lighter grey). 

For the case without luminosity prior, any potential evolution at $\log(M_*/M_\odot)>11.4$ is hidden by the large error bars. Nonetheless, the points at $4<z<5$ and $\log(M_*/M_\odot)<11.4$ are a factor of 3 lower than the SMF at $3<z<4$. An even larger evolution of about one full order of magnitude is observed when considering the SMF measurements from the luminosity prior configurations. At higher redshift, the large uncertainties associated to different sources of uncertainty in the measurement of the stellar mass convert the SMF measurements obtained with the luminosity prior into upper limits, implying significant evolution of the high-mass end of the SMF in the first 1.5 Gyr of cosmic history.

\section{Discussion}

Massive galaxies in the redshift range $4<z<7$ constitute an important test for the models of galaxy formation and evolution.
To date, massive and massive, quiescent galaxies have been found using multi-wavelength photometric surveys up to $z\lesssim4$ \citep{fontana2009,kriek2009b,marchesini2010,whitacker2012,muzzin2012,stefanon2013, 
straatman2014,fan2013,muzzin2013b,lundgren2014,marchesini2014}, while at $z>4$ their identification is still uncertain (e.g., \citealt{dunlop2007}). 

Recently, \citet{bowler2014} found a population of galaxies at $6<z<7$ using data from UltraVISTA DR2, with the detection in a $Y+J$ stacked frame, weighted by the inverse of the variance. The galaxies are characterized  by stellar masses $M_*\lesssim10^{10.5}M_\odot$ and sSFR$\sim10^{-7}-10^{-9}$yr$^{-1}$, consistent with the sample presented here.  

\citet{lundgren2014}, analyzing the SED of $z<6$ galaxies in the HUDF, found that the fraction of  quiescent galaxies with $\log(M_*/M_\odot)\sim10$ at $z>4$ is $<50\%$ and it is consistent with 0. The small field characterizing the HUDF, however, translates into a high cosmic variance ($>50\%$) for that range of redshifts and stellar masses.

Attempts to search for $\log(M_*/M_\odot)> 11$ galaxies at $z>4$ have been made so far by \citealt{mobasher2005,mclure2006,dunlop2007} and \citep{wiklind2008}. \citet{mobasher2005} reported a detection of a  $\log(M_*/M_\odot)\sim 11.6$ post-starburst galaxy at $z=6.5$ in the Hubble Ultra Deep Field (HUDF), while neither \citet{mclure2006} nor \citet{dunlop2007} were able to detect any in larger fields. \citet{wiklind2008}, on the other side, detected  in the GOODS-S field  five galaxies at $5<z<6.5$ with stellar mass $\log(M_*/M_\odot)> 11$ (after correcting for the different IMF).

In this work we detect galaxies with stellar mass $M_*>10^{11}M_\odot$ up to $z\sim7$ using a multi-wavelength photometric catalog. Within this sample, we are also able to identify seven robust candidates of very massive galaxies at $4<z<7$. Our photometric redshift measurements are based on two analyses, one introducing bayesian  luminosity priors, the other excluding the luminosity prior from the analysis. The two approaches  
produce redshift distributions  which do not fully overlap. Indeed,  this uncertainty in the determination of photometric redshifts could be mitigated by higher S/N data in the $Y, J, H, K$ bands. The difference in photometric redshift measurements is then reflected at the time of measuring the SMFs. Specifically, the effect of introducing the luminosity priors in the measurements of the photometric redshifts can qualitatively be 
 identified with the conversion of the SMF measurements into upper limits, suggesting a strong evolution of the co-moving number density of galaxies during the $\sim1$Gyr occurred from $z\sim6.5$ to $z\sim 3.5$. However, given the still large uncertainties in the knowledge of the galaxy populations at $z>4$, we caution the reader 
about the reliability of the SMF analysis based on the sample obtained with luminosity priors. Indeed, bayesian luminosity priors are currently derived either from low-redshift luminosity functions/SMFs or from semi-analytic models, and no observationally tested prior is available over the redshift interval probed by our work. We adopted three different approaches to account for nebular line contamination to the photometry. The additional dispersion in the stellar mass measurements is smaller than the uncertainties from Poisson noise and cosmic variance for the adopted dataset.

\begin{figure}
\hspace{-1cm}\includegraphics[width=10cm]{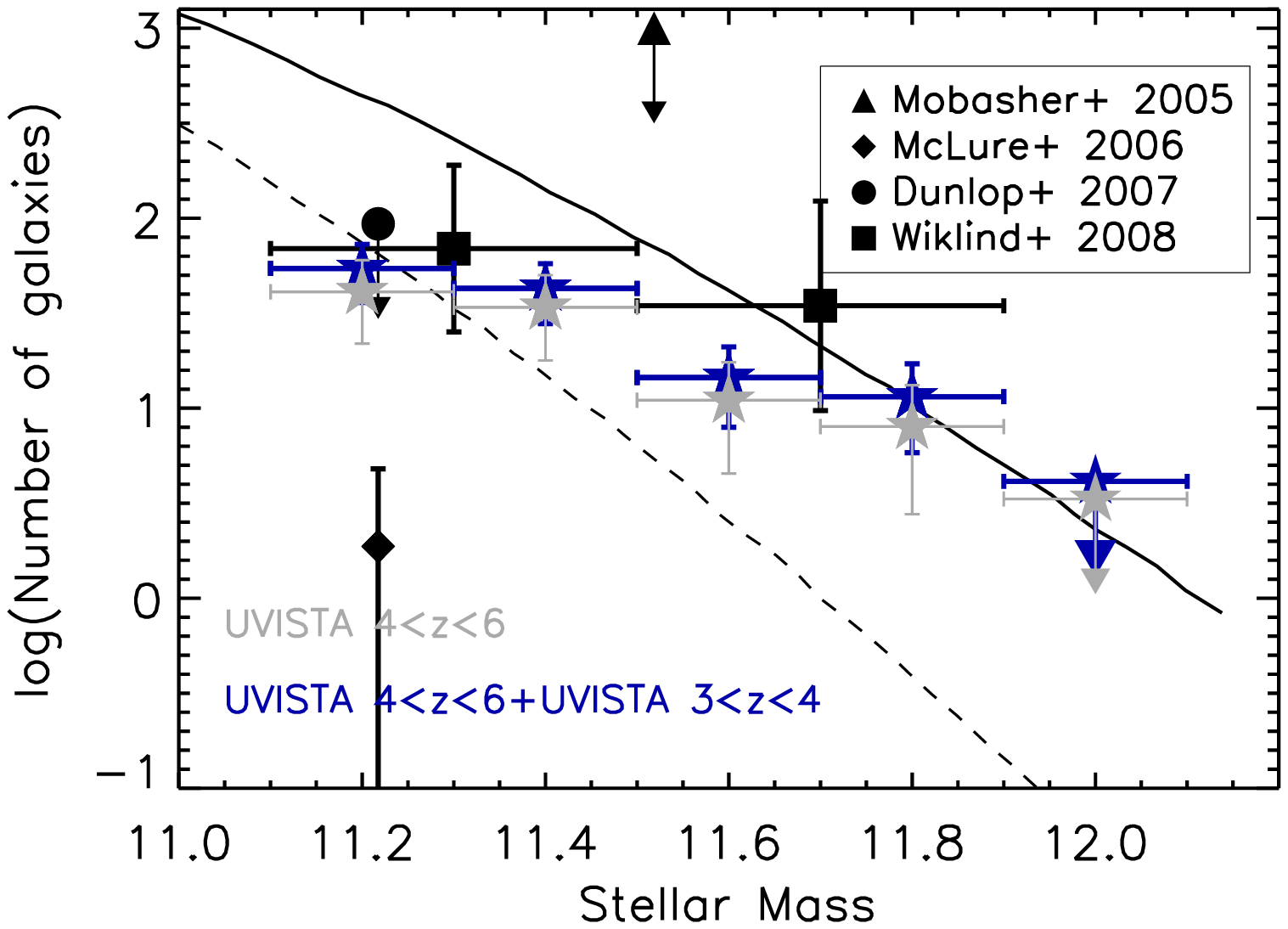}
\caption{Distribution of the number of high-redshift galaxies according to their stellar mass in an area equivalent to that of UltraVISTA. The curves represent the expected number of $3<z<6$ galaxies observed (solid) and intrinsic (dashed), taken from \citet{davies2013} rescaled to the actual UltraVISTA area, and are based on \citet{behroozi2013}. The points mark the number of massive galaxies found in this work at $4<z<6$ (grey points) and at $3<z<6$. The horizontal error bars identify each bin in stellar mass, while the vertical error bars include the effects of poisson statistics and cosmic variance. The number of objects observed in UltraVISTA is consistent with the predicted number counts on the whole stellar mass range .\label{fig:num_comp}}
\end{figure}

As a check of consistency for galaxies with stellar mass $\log(M_*/M_\odot)> 11$ at $z>4$, we reproduce in Figure \ref{fig:num_comp} the actual number of galaxies we recover in the redshift range $4<z<6$ for our default sample and at $3<z<6$ after adding the number of massive galaxies at $3<z<4$ from the UltraVISTA catalog of \citet{muzzin2013a}. The associated uncertainties include poisson noise and cosmic variance. The solid and dashed lines mark the expected observed and intrinsic number of galaxies, respectively, for the redshift range $3<z<6$ taken from \citet{davies2013} after they have been rescaled to match the area covered by UltraVISTA. Specifically, the expected number of intrinsic galaxies represents the number of galaxies which are truly at $3<z<6$ in \citet{behroozi2013} simulation, while the expected observed number of galaxies refers to the number of galaxies which can be detected in real data, taking into account effects such as the Eddington bias. Our measurements fall roughly in between the two curves, revealing a good agreement with the theoretical expectations.

In Figure \ref{fig:num_comp}, we also display the number of detected objects from  \citet{mobasher2005,mclure2006,dunlop2007} and \citet{wiklind2008}, rescaled to the UltraVISTA area and with poissonian and cosmic variance uncertainties computed in the same way as in our analysis. Since \citet{mclure2006} and \citet{dunlop2007} do not report any detection, the corresponding value is taken as the upper limit from the detection of a single galaxy. Specifically, \citet{wiklind2008} identified a sample of five galaxies at $z>5$ with stellar masses  $M_*>10^{11}M_\odot$ (after applying an offset of 0.2~dex to convert them to \citet{chabrier2003} IMF) in the GOODS-S region over an area of 156 arcmin$^2$. When rescaled to the $\approx1.5$ degree$^2$ of the UltraVISTA field, this would translate into $\sim150$ massive $z>5$ galaxies. In our analysis, we find 138 galaxies at $5<z<7$, consistent with the number of galaxies in the sample of \citet{wiklind2008}. However, out of the 138 galaxies, only one has a sufficiently robust redshift determination such that it is still massive and at $z>5$ when applying bayesian luminosity prior and adding the old and dusty template to the set of templates used for the measurements of the photometric redshifts.

Given the small area of the fields (11.5arcmin$^2$, 125arcmin$^2$ 0.6degrees$^2$ and 156arcmin$^2$ ), the corresponding cosmic variance is quite high, ranging from 35\% for the UDS field to $\sim200\%$ for HUDF, which definitely converts most of the measurements into upper limits. Despite this, the number counts by \citet{mobasher2005}, \citet{dunlop2007} and \citet{wiklind2008} are consistent with the model, while the equivalent number from \citet{mclure2006} is consistently lower than the theoretical expectations. However, we note that \citet{wiklind2008} sample refers to the redshift range $5<z<6$, which implies a higher number of galaxies when the full $3<z<6$ range is considered.

Among the $M_*\gtrsim10^{11}M_\odot$ galaxies in our sample,  we identify one robust candidate for a massive galaxy at $z>4$, irrespective of the configuration adopted to measure the photometric redshifts and 
stellar population parameters. Specifically, the galaxy, at $z\sim 5.4$, is characterized by a stellar mass $\log(M_*/M_\odot)\sim11.5$, a log(sSFR/yr$^{-1}$)$\sim-10.3$ and a log(age/yr) of 8.5. All the considered SED fitting indicate that this galaxy is also quiescent; although we detect MIPS $24\mu$m flux. The absence of detection in Herschel data and the physically unlikely high value of the SFR associated to the MIPS flux support the SFR values from the SED fitting analysis, suggesting that the MIPS flux is originated by an AGN.

If this result were confirmed through e.g., NIR spectroscopy, this finding would shift by about 0.5~Gyr back in time the appearance of the first massive, post-starburst galaxies (see e.g., \citealt{marsan2014}), corresponding to a 
Universe 1~Gyr old. According to the delayed-exponential model, the age of this galaxy at the time of observation is $10^{8.5}$yr $\sim320$Myr, meaning that it started its formation less than 680Myr after the Big Bang (or $z\sim7.5$); the peak of star formation occurred at extremely early stages of its formation, when the galaxy was just about 65Myr old (corresponding to $z\sim7$) and reaching a SFR of $\sim 3800M_\odot$yr$^{-1}$, typical of HyLIRGS, in line with the downsizing scenario inferred from the fossil records in local most massive galaxies (e.g., \citealt{thomas2010}).

Assuming a constant (or average) SFH, with a SFR=400 $M_
\odot/$yr$^{-1}$, typical of a highly star-forming galaxy, this would imply that a $10^{11.5}M_\odot$ galaxy would take $0.8$Gyr to fully assemble its mass in stars. The requirement for these galaxies to be already dead by $z\sim 4.5$ - then - constrains the quenching time 
to no more than 0.5Gyr. If, instead, we assume an initial phase with star formation similar to those found in ULIRGs, and corresponding to few thousands solar masses per year, the same galaxy 
would form in $\approx0.1$ Gyr, with most of the time available for the quenching of star formation.

\section{Conclusions}

In this work we complemented the UltraVISTA multi-wavelength photometric catalog by \citet{muzzin2013a} with detections on the residual images resulting from the template-fitting photometry adopted to include IRAC $3.6\mu$m and $4.5\mu$m bands to the multi-wavelength catalog, effectively constructing an IRAC-selected catalog. This new catalog allowed us to decrease the stellar mass limit at $4<z<7$ to $\gtrsim10^{11.3}M_\odot$. Using this new catalog:

\begin{itemize}
\item We found a candidate for a massive ($M_*\sim10^{11.6}M_\odot$), quiescent (sSFR$\sim10^{-10.3}$yr$^{-1}$) galaxy at $z\sim5.4$. The photometric redshift and the stellar mass measurement showed to be robust using different recipes for photometric redshift and stellar population parameters, including the effects of nebular emission lines. However, the clear detection in MIPS $24\mu$m suggests that this galaxy could be hosting an AGN. A secondary solution exist at $z\sim2.5$, although the cumulative probability of being at $z<3$ is at most 21\%. Even at this low redshift, the galaxy would still be highly forming stars. 
\item We detected a sample of seven robust very massive galaxies with redshift $z>4$. Their SEDs are typical of star-forming or post-starburst galaxies.
\item We finally presented our measurements of the massive end of the SMF at $4<z<7$. These measurements are mostly affected by the systematic effects in the measurement of photometric redshifts from the introduction of the old and dusty template and from the adoption of the bayesian prior on the observed flux (see Section \ref{sect:prior}). This, together with the large uncertainties associated with the low number of galaxies in the sample and with the cosmic variance, prevents us from constraining the evolution of the high-mass end of the SMF of galaxies over the redshift range $4<z<7$  detecting any possible evolution in the range of redshift between $z\sim6.5$ and $z\sim4.5$.
\item The population of massive galaxies presented in this work is numerically consistent with theoretical expectations of the number of high-redshift massive galaxies.
\end{itemize}

The recent release of UltraVISTA DR2, reaching almost one magnitude deeper in the NIR bands than the previous release used in our work, will allow us to improve the observational constraints of the SEDs used in this work. However, as we have shown, systematic uncertainties from the still very limited knowledge of the intrinsic SEDs of massive galaxies at $z>4$ dominate the total error budget. Rest-frame optical spectroscopy of massive galaxies at $z>4$ is necessary to measure the amount of dust extinction and the level of contamination to the photometry by nebular lines.  This however will require the instrumental capabilities of the James Webb Space Telescope. Given the dusty and star-forming nature of most of the found candidates of massive galaxies at$ z>4$, ALMA provides a unique opportunity to spectroscopically measure their redshifts until the advent of JWST.

\acknowledgments
We warmly thank the referee for her/his constructive report. MS would like to thank Gregory Rudnick and Haojing Yan for useful discussions. He also acknowledges the support from NASA grants HST-GO-12286.11 and 12060.95. DM acknowledges the support of the Research Corporation for Science Advancement's Cottrell Scholarship. The Dark Cosmology Centre is funded by the DNRF.  JPUF and BMJ acknowledge support from the ERC-StG grant EGGS-278202. This study is based on a K$_{s}$-selected catalog of the COSMOS/UltraVISTA field from \citet{muzzin2013a}.  The catalog contains PSF-matched photometry in 30 photometric bands covering the wavelength range 0.15$\micron$ $\rightarrow$ 24$\micron$ and includes the available $GALEX$ \citep{martin2005} CFHT/Subaru \citep{capak2007}, UltraVISTA \citep{mccracken2012}, S-COSMOS \citep{sanders2007}, and zCOSMOS \citep{lilly2009} datasets.Ó  Based on data products from observations made with ESO Telescopes at the La Silla Paranal Observatory under ESO programme ID 179.A-2005 and on data products produced by TERAPIX and the Cambridge Astronomy Survey Unit on behalf of the UltraVISTA consortium. Based on observations obtained with MegaPrime/MegaCam, a joint project of CFHT and CEA/IRFU, at the Canada-France-Hawaii Telescope (CFHT) which is operated by the National Research Council (NRC) of Canada, the Institut National des Science de l'Univers of the Centre National de la Recherche Scientifique (CNRS) of France, and the University of Hawaii. This work is based in part on data products produced at Terapix available at the Canadian Astronomy Data Centre as part of the Canada-France-Hawaii Telescope Legacy Survey, a collaborative project of NRC and CNRS. This research has made use of data from HerMES project (http://hermes.sussex.ac.uk/). HerMES is a Herschel Key Programme utilising Guaranteed Time from the SPIRE instrument team, ESAC scientists and a mission scientist.
The HerMES data was accessed through the Herschel Database in Marseille (HeDaM - http://hedam.lam.fr) operated by CeSAM and hosted by the Laboratoire d'Astrophysique de Marseille. The Millennium Simulation databases used in this paper and the web application providing online access to them were constructed as part of the activities of the German Astrophysical Virtual Observatory (GAVO).

\bibliographystyle{apj}
\bibliography{uvista_gsmf_irac}

%\begin{thebibliography}{76}
%\expandafter\ifx\csname natexlab\endcsname\relax\def\natexlab#1{#1}\fi

%\end{thebibliography}

\end{document}